\documentclass[journal,12pt,onecolumn,draftclsnofoot]{IEEEtran}
\usepackage[T1]{fontenc}
\usepackage{times}
\usepackage{color}
\usepackage{amsmath}
\usepackage{amssymb}
\usepackage{stmaryrd}
\usepackage{stackrel}
\usepackage{graphicx}
\usepackage{cite}
\usepackage{cases}
\usepackage{stackrel}
\usepackage{multicol}
\usepackage[USenglish]{babel}
\usepackage{array}
\usepackage{multirow}
\usepackage{amsfonts}
\usepackage{float}
\usepackage{balance}
\usepackage{ragged2e}
\usepackage{subfigure}

\renewcommand{\raggedright}{\leftskip=0pt \rightskip=0pt plus 0cm}
\raggedright

\newtheorem{thm}{\textbf{\textit{Theorem}}}
\newtheorem{remrk}{\textbf{\textit{Remark}}}
\newtheorem{lemma}{\textbf{\textit{Lemma}}}
\newtheorem{cor}{\textbf{\textit{Corollary}}}

\begin{document}
\title{Harvest-and-Opportunistically-Relay: Analyses on Transmission Outage and Covertness}
\author{Yuanjian~Li, Rui~Zhao,~\IEEEmembership{Member,~IEEE,} Zhiqiao~Nie and A. Hamid~Aghvami,~\IEEEmembership{Fellow,~IEEE}\thanks{Yuanjian Li and A. Hamid Aghvami are with Centre for Telecommunications Research (CTR), King's College London, London WC2R 2LS, U.K. (e-mail: yuanjian.li@kcl.ac.uk; hamid.aghvami@kcl.ac.uk).}
\thanks{Rui Zhao and Zhiqiao Nie are with Xiamen Key Laboratory of Mobile Multimedia Communications, Huaqiao University, Xiamen 361021, China. (e-mail: rzhao@hqu.edu.cn;  zqnie@hqu.edu.cn).}
\thanks{This work has been submitted to the IEEE for possible publication.  Copyright may be transferred without notice, after which this version may no longer be accessible.}}
\maketitle
\begin{abstract}
For enhancing transmission performance, privacy level and energy manipulating efficiency
of wireless networks, this paper initiates a novel simultaneous wireless
information and power transfer (SWIPT) full-duplex (FD) relaying protocol,
termed harvest-and-opportunistically-relay (HOR). In the proposed
HOR protocol, the relay can work opportunistically in either
pure energy harvesting (PEH) or the FD SWIPT mode. Due to the FD characteristics,
the dynamic fluctuation of R's residual energy is
difficult to quantify and track. To solve this problem, we apply a
novel discrete-state Markov Chain (MC) method in which the practical
finite-capacity energy storage is considered. Furthermore, to improve the privacy level of the proposed HOR
relaying system, covert transmission performance analysis is developed and investigated, where closed-form expressions of optimal detection
threshold and minimum detection error probability are derived. Last
but not least, with the aid of stationary distribution of the MC,
closed-form expression of transmission outage probability is calculated,
based on which transmission outage performance is analyzed. Numerical results have validated the correctness
of analyses on transmission outage and covert communication. The impacts
of key system parameters on the performance of transmission outage
and covert communication are given and discussed. Based on mathematical
analysis and numerical results, it is fair to say that the proposed
HOR model is able to not only reliably enhance the transmission performance
via smartly managing residual energy but also efficiently improve the
privacy level of the legitimate transmission party via dynamically
adjust the optimal detection threshold.
\end{abstract}


\section{Introduction}

\subsection{Background}

\IEEEPARstart{Conventionally}, wireless communication systems are basically powered by rechargeable battery or electrical grid,
such as cellular, Bluetooth, Wi-Fi and sensor networks. There are several distinguish physical or/and economic disadvantages rooted
in these traditional wireless communication power supply methods, which has been the bottleneck restricting the ubiquitous applications of
wireless communication \cite{bi2016accumulate}. More precisely stated,
grid-powered wireless communication systems, e.g., cellular networks,
require solid support of electrical grid infrastructure, which
may not only need much more construction resources but also lead to
enormous energy consumption; The operational lifetime of battery-powered wireless networks
is usually limited, for finite battery capacity in practical
applications, leading to periodic battery
replacement or recharging. To prolong the wireless networks lifetime
and improve the energy efficiency, the research of energy-aware architectures
and transmission strategies has been a hotspot in recent years.

Energy harvesting (EH) technique is able to scavenge energy from natural
resources (e.g., solar power, piezoelectric energy, wind and mechanical
vibrations), which is known as a promising candidate to overcome the aforementioned disadvantages of the traditional power supply strategies. Unfortunately, the amount of energy
harvested from natural resources highly depends on several uncontrollable
factors, such as the weather condition, resulting in EH unreliablility. To aid this, a promising method
scavenging energy from man-made radio frequency (RF) radiation has gained lots of research concentrations \cite{chu2018resource}.
Inspired by the fact that the RF signals can carry the intended information
and energy at the same time, the concept of simultaneous wireless
information and power transfer (SWIPT) was coined in \cite{Varshney2008Transporting}.
Thereafter, two practical SWIPT strategies were introduced
in \cite{Zhang2013MIMO}, i.e., time-switching (TS) and power-splitting
(PS) based SWIPT, in which the missions of information decoding (ID) and EH are conducted
respectively in time or power domain. Specifically, the TS-based method
allocates part of the time slot to decode information and the remaining
to harvest energy, whereas one potion of the received signal power
is utilized for ID and the other potion is used for EH in the PS-based
strategy \cite{krikidis2014simultaneous}. Based on these practical
SWIPT strategies, various essential issues about SWIPT were studied
in different wireless transmission systems, e.g., maximizing the ergodic
rate for a dynamic SWIPT approach in the cooperative cognitive radio
network (CCRN) \cite{yan2017dynamic}, a non-cooperative game theoretic
approach for the resource optimization in SWIPT enabled heterogeneous
small cell network (HetSNet) \cite{zhang2018incomplete}, optimizing
the energy efficiency (EE) by delicately designing the precoders at the
transceivers in the multiple-input multiple-output (MIMO) two-way
wireless networks \cite{rostampoor2017energy}. 

Full-duplex (FD) technology which allows transceivers emit and receive
information simultaneously, can potentially achieve efficient utilization
of wireless resources (say, time and frequency), and thus it is expected
to overcome the shortcomings of half-duplex (HD) counterpart on spectral
efficiency (SE) \cite{chen2017spectral}. However, the theoretical
performance of FD nodes is significantly limited by the harmful self-interference
(SI) which represents that the emitted signals may be directly/indirectly
received by their own receivers at the FD nodes \cite{xing2017multipair}.
Fortunately, thanks to recent advances in SI cancellation (SIC) techniques
(e.g., passive and active SIC approaches), it is possible to suppress
the SI to noise level, which makes the FD technology more practical
and feasible in practice \cite{li2018antenna,ahmed2015all}. Nevertheless,
due to the RF impairments, SI cannot be restrained perfectly so that
the FD networks are still impacted by the so-called residual SI (RSI)
\cite{li2017outage}. 

Thereafter, FD technique has drawn attention from both academic
and industrial communities. Among various wireless FD transmission
applications, one popular candidate is the FD  relaying (FDR) technique,
which can not only extend the transmission coverage and combat the
severe fading in wireless communications but also enhance the utilization
efficiency of wireless resources \cite{Sohaib2017Full}. Some of the
corresponding works have coped with the performance of various FDR
network backgrounds, in the presence of RSI. For example, two buffer-aided
relaying approaches with adaptive transmission-reception at the FD
relay in the absence of direct link between the source and the destination
were proposed and studied in \cite{Razlighi2017Buffer}; whereas the
outage probability of a amplify-and-forward (AF) FDR network with
direct source-destination link was investigated in \cite{Wang2015Outage}.
Moreover, in the case of decode-and-forward (DF) relaying protocol,
\cite{li2017Secrecy} researched the ergodic achievable secrecy rate
issue in the FDR wiretap channels.

With rapid development of the fifth-generation (5G) wireless networks
and Internet of Things (IoT), rocketing sorts and amounts of private
information (e.g., location data, control orders, social identity
information, e-health indexes) are needed to be shared wirelessly
among tranceivers. Consequently, growing concerns have been posing
on security and privacy (low detection probability by the third party)
of wireless transmissions. In the existing literature, security issues
of wireless communications are much more concerned and investigated
than its privacy counterpart. To enhance wireless information security,
lots of works have been developed, like, cryptography and information-theoretic
physical layer security techniques. However, the inherently public
and visible nature of wireless medium (electromagnetic wave) not only
leads to the security vulnerability but also privacy weakness. Recently, increasing research efforts have been pouring
into the field of low probability of detection on the existence of
wireless transmissions, namely, covert communications. For example,
communicating covert messages under the detection of legitimate party
who does not desire the leakage of sensitive information which is
supposed to be kept confidential within authorized tranceivers, belongs
to covert communication research regime. The famous Square Root Law
which indicates the fact that $\mathcal{O}\left(\sqrt{n}\right)$
bits of information can be transmitted reliably and covertly in $n$
channel uses over additive white Gaussian noise (AWGN) channels as
$n\rightarrow+\infty$, was coined in \cite{bash2013limits}. Afterwards,
covert transmissions have been researched and investigated in various
wireless communication scenarios \cite{shahzad2018achieving,zheng2019multi,zhou2019joint}.
In \cite{shahzad2018achieving}, the authors studied covert wireless
communications in the presence of a FD receiver which can generate
artificial noise to cause uncertainty at the adversary so that low
probability of detection can be achievable. Considering both centralized
and distributed antenna systems (CAS/DAS), multi-antenna-aided covert
communications coexisting with randomly located wardens and interferers
was studied in \cite{zheng2019multi}. Authors of \cite{zhou2019joint}
jointly optimized trajectory and transmit power for covert transmission
in unmanned aerial vehicle (UAV) networks, aiming to hide a UAV for
communicating critical informations.

\IEEEpubidadjcol

\subsection{Related Works and Motivation}

Hereby, we review the related works, point out the differences
and claim our motivation.

It has been a promising solution to meet the green communication and
the reliable transmission demand in the upcoming 5G and IoT era by
introducing SWIPT into FDR wireless communications. Particularly,
in the scenario which contains power-constrained relay node, the SWIPT
FDR has the potential to not only solve power supply problem but also enhance significantly key wireless transmission performances,
like, reliability, SE, valid coverage, quality of service (QoS), etc.
Besides, by delicately designing the covert communication detection
strategy, it is promising to improve privacy level of the SWIPT FDR
system.

To the best of the authors' knowledge, there already exist inspiring
related literature which investigated and studied SWIPT FDR in the
context of different wireless network setups. In \cite{Wang2017Relay},
the characteristics and performance of PS-based two-way SWIPT FDR
networks as well as the relay selection issue were researched.
In \cite{Wen2016Joint}, a joint optimization method finding the source
as well as the relay beamformers was proposed and the numerical results
for the mean squared error (MSE) and bit error rate (BER) showed that
the proposed method performed well in the MIMO SWIPT FDR systems.
In \cite{zeng2015full}, the throughput maximization problem for a
FDR wireless communication network with simultaneous down-link energy
transfer and up-link information transmission was investigated. 
In \cite{liu2016power}, outage probability and average throughput performances were investigated in a SWIPT FDR wireless network. Unfortunately, the aforementioned works and the
majority of existing literature on SWIPT FDR did not include the consideration
on covert communications. To bridge this research
gap, we investigate covert communication problems of SWIPT FDR systems
in this paper.

Regarding the related works of covert communications in the field
of wireless relaying networks, it is still in its infancy stage. In
\cite{hu2019covert}, Hu \textit{et al.} examined the possibility,
performance limits, and associated costs for a power-constrained HD relay transmitting covert information on top of forwarding
the source\textquoteright s information. Wang \textit{et al.} \cite{wang2018covert}
investigated how channel uncertainty can influence covert communication
performance in wireless relaying networks. A covert communication
scheme under fading channels was proposed and studied in \cite{shahzad2018relaying}
where the relay not only forwards source's information but also plays
the role as a cooperative jammer. However, so far, discrete EH technique 
has not been considered in the existing literature regarding covert
communications, which is a main concern of this paper.

Motivated by the aforementioned contents, we propose
a novel wireless relaying protocol in which discrete-energy-state
SWIPT FDR and covert communications are combined and considered, aiming
to enhance wireless transmission performance while improving its privacy
level.

\subsection{Our Contributions}

In this paper, a new transmission protocol termed harvest-and-opportunistically-relay
(HOR) is designed and analyzed. Specifically,
the FD relay which contains
no sustainable power supply but wireless EH system and rechargeable
energy storage is deployed to opportunistically help the source and
the destination complete their wireless communication. In the proposed HOR protocol, according to
the relay's energy status and channel condition
between the source and the destination, the relay works
dynamically in either pure energy harvesting (PEH) or the FD SWIPT
mode. Furthermore, to evaluate
the detection performance on potential covert communication, i.e., improving the privacy level of the proposed HOR protocol, performance analysis on covert transmission is developed and investigated. As far as the authors
know, we are the first to introduce both discrete EH and covert communications
into SWIPT FDR systems. The main contributions
of the paper are concluded in details as follows.

\begin{itemize}
\item \it{Protocal Design}\rm{: We systematically establish a novel HOR protocol from listing necessary hardware facilities to designing feasible transmission stragegy. The proposed HOR scheme can efficiently enhance wireless transmission performance between the source and the destination, via flexibly managing the relay's precious stored energy. Besides, through covert communication analysis, privacy level of the proposed HOR system can be improved. It is fair to claim that the proposed HOR scheme is able to not only enchance wireless transmission performance but also improve the system's privacy level.}
\item \it{Hybrid Energy Storage and Markov Chain}\rm{: We consider the practical energy storage model under the limitation of finite capacity at the relay. To enable the relay's FD functionality, a hybrid energy storage scheme is adopted, which consists of both primary and secondary energy storages. To track dynamic fluctuation of residual energy, energy discretization and discrete-state MC method is applied to model the complicated energy state transitions. It is worth noting that all the transition probabilities are calculated in closed-form. Then, the statinoary distribution of the MC is given.}
\item \it{Covert Commmunication Analysis}\rm{: Under the cover of forwarded source massages, there exists potential threat of critical information leakage. To improve the HOR protocol's privacy level, we provide covert communication analysis given channel uncertainty in this paper. The optimility of radiometer for covert massage detection is proved. Closed-form expressions of false alarm and missed detection probabilities are derived, based on which we calculate closed-form expressions of the optimal detection threshold and the coresponding minimum detection error probability. Furthermore, the impacts of imperfect channle estimation on the minimum detection error probability is also discussed.}
\item \it{Transmission Performance Analysis}\rm{: Invoking the MC's stationary distribution, closed-form expression of transmission outage probability is derived, then we provide transmission outage analysis of the proposed HOR scheme. Furthermore, the impacts of key system parameters on transmission outage performance are investigated via numerical results.} 
\end{itemize}

\subsection{Outline and Notation}

\textit{\textcolor{black}{Organization}}\textcolor{black}{: The paper
is organized as follows. Section \uppercase\expandafter{\romannumeral2}
presents the HOR model and its transmission strategy. Section \uppercase\expandafter{\romannumeral3}
describes the energy discretization, detailed derivation on the MC
and the stationary distribution. Section \uppercase\expandafter{\romannumeral4}
shows covert communication analysis. Section
\uppercase\expandafter{\romannumeral5} gives transmission outage
performance analysis. Simulation results
are presented in Section \uppercase\expandafter{\romannumeral6} and
conclusions are drawn in Section \uppercase\expandafter{\romannumeral7}.}

\textit{\textcolor{black}{Notation}}\textcolor{black}{: Bold lower
case letters denote vectors, e.g., $\mathbf{v}$. Bold upper case
letters denote matrices, e.g., $\mathbf{M}$. $(\cdot)^{T}$, $(\cdot)^{-1}$,
${\bf I}$ indicate transpose of matrix, inverse of matrix, unit matrix,
while $\mathbb{E}\left\{ \cdot\right\} $ and $|\cdot|$ mean statistical
expectation and modulo operators of a complex number, respectively.
$\mathcal{CN}\left(\mu,\sigma^{2}\right)$ stands for the complex
Gaussian distribution with mean $\mu$ and variance $\sigma^{2}$.
The intersection of two sets $A$ and $B$ is denoted by $A\cap B$.
$\Pr\left(\cdot\right)$ is the operator calculating probability of
a specific objective. Symbols $\sum$ and $\prod$ represent the summation
and product operations of a sequence of terms, respectively.}

\section{System model and transmission strategy}

A classical three-node wireless relaying network, which comprises
one source (S), one destination (D) and one relay (R), is considered
in this paper. Energy-constrained R is equipped with
dual antenna so that it can adopt the FD technique, whereas S and
D are both single-antenna node. The novel HOR protocol
is coined originally to assist wireless communication from S to D, with the ability of managing RF energy smartly, while improving the overall privacy level.

\subsection{Assumptions Regarding Wireless Channels}

In this paper, we assume that all wireless channels are modeled as
quasi-static Rayleigh fading channels, which means that these fading
channels remain static within each transmission slot, and vary independently
over different transmission slots. The Rayleigh fading distribution
that the self-interference (SI) channel at R follows is considered
because the line-of-sight (LoS) component can be largely eliminated
via antenna isolation and the scattering plays the principal role
herein. Note that the aforementioned slot is equivalent to a block
of time over which the intended massages are transmitted. Besides, this paper
considers the widely used infinite block-length model which means
that each transmission slot is composed of $n$ symbols and $n\rightarrow\infty$
is assumed. Moreover, the block boundaries in wireless links are
predefined to be synchronized perfectly throughout the whole system.
Without loss of generality, the block duration in the considered model
is normalized to one time unit so that the measures of power and energy
are identical and can be used interchangeably in this paper. Wireless
channels S$\shortrightarrow$D, S$\shortrightarrow$R, and R$\shortrightarrow$D
are denoted as $h_{\textnormal{SD}}$, $h_{\textnormal{SR}}$, and
$h_{\textnormal{RD}}$, respectively. Moreover, $h_{\textnormal{RR}}$
indicates the SI link caused by the FD characteristic at R. It is
worth noting that the channel coefficients $h_{\textnormal{SD}}$,
$h_{\textnormal{SR}}$, $h_{\textnormal{RD}}$ and $h_{\textnormal{RR}}$ are manipulated to
encompass the gains of transmit and receive antennas as well as the
path losses cased by propagation distances among the nodes in this paper. The aforementioned
wireless channel coefficients follow independently and identically
distributed (i.i.d.) complex Gaussian distribution with zero means
and variance $\mathbb{E}\left\{ \vert h_{\text{SD}}\vert^{2}\right\} =\Omega_{\text{SD}}$,
$\mathbb{E}\left\{ \vert h_{\text{SR}}\vert^{2}\right\} =\Omega_{\text{SR}}$,
$\mathbb{E}\left\{ \vert h_{\text{RD}}\vert^{2}\right\} =\Omega_{\text{RD}}$
and $\mathbb{E}\left\{ \vert h_{\text{RR}}\vert^{2}\right\} =\Omega_{\text{RR}}$.

Regarding the availability of global CSIs, the instantaneous CSI of
channel between S and D are assumed to be available at S via channel
estimation, but D can only gain the imperfect instantaneous CSI estimation
of wireless channel between R and D. We note hereby that the availability
of instantaneous S$\shortrightarrow$R and R$\shortrightarrow$R CSIs
poses no influence on the considered performance analyses so that we do not rise any assumption on their availabilities.

\subsection{Relay Model}

In the considered system model, R is known publicly to be energy-limited,
leading to rigorous power supply problem which is expected to be solved
by the promising SWIPT technique. Different from the traditional relay
strategy, in this paper, a novel relay protocol named HOR is proposed,
which allows R to work in either the PEH mode or the FD SWIPT mode opportunistically.
In specific, when performing the FD SWIPT, R receives and forwards information
simultaneously to assist the wireless transmission between S and D,
while the PS-based EH solution is applied to harvest the RF energy. In the case of adopting the PEH mode, R concentrates on
capturing wireless energy from the RF signals without any information
processing. Apart from assisting wireless transmission, R is considered
as potential leaker who intends to leak vital information regarding
the source signals to the third party, which should keep covert from
the legitimate party, i.e., S and D. The malicious intention of R
keeps secrecy and the legitimate party cannot make sure whether R
is innocent or not, the legitimate party treads R as an innocent and
friendly node initially but keeps an eye on R detecting the
potential covert communication generated by R.

For achieving the proposed HOR functionality, R should equip the following
hardwares:

\begin{itemize}
\item[1)] Three RF chains, enabling the EH, information forwarding and covert massage emitting.
\item[2)] One rectifier utilized to transform the RF signals into direct currents (DC).
\item[3)] A battery serving as the principal energy carrier (PEC) with high energy capacity.
\item[4)] One minor battery (MB) for storing harvested energy temporarily, e.g., a capacitor.
\item[5)] A constant energy supply for sending covert massage, whose existence is unaware publicly.
\end{itemize}

In details, the receive antenna at R is permanently bounded with the
rectifier via one RF chain. One single battery cannot be charged and
discharged simultaneously so that the FD SWIPT mode may not be realized,
we herein apply both the PEC and the MB at R to crack this dilemma.
Note that the PEC is directly connected to the rectifier and the transmitting
RF chain for absorbing and releasing energy, respectively. In the PEH mode, the harvested energy are absorbed by the PEC directly.
In the FD SWIPT mode, the PEC releases its residual energy to support the transmitting RF chain. Meanwhile, the MB stores
the harvested energy temporarily and delivers all the stored energy
into the PEC when the FD SWIPT mode terminates. The hidden constant
energy supply which is connected to the other RF chain will release its power only when the relay decides to
leak the system's informations.

Alongside assisting signal transmission between S and D, R is a rapacious
node which intends to leak essential information (defined as the covert
massage herein) regarding the source signals, when the right opportunity
occurs. The legal destination D also plays the role as a\textit{ warden}
detecting the potential information leakage. Reducing the probability
of being detected by the legitimate party, R would like to emit the
covert massage under some solid covers. In this proposed system, the
forwarded version of the source signals is the only existing shield.
Reasonably, R would consider the worst case (D can gain perfect channel
estimation and know its own noise power) and intends to broadcast
the covert massage merely when itself works in the FD SWIPT mode.
Otherwise, the covert communication initiated by R will be detected
by D without any hesitation (i.e., probability one), which is definitely
an undesired circumstance R ever expects. This is because, in the
case of PEH, R is supposed to focus on EH without forwarding, any
additional transmit power initiated at R will be detected easily by D.

\subsection{Transmission Protocol}

In our proposed HOR protocol, before each transmission block is sent
out, S broadcasts pilot signal to estimate $h_{\textnormal{SD}}$
which will be utilized to calculate the received instantaneous signal-to-noise-ratio
(SNR) at D, denoted as $\gamma_{\textnormal{SD}}=P_{\textnormal{S}}\left|h_{\textnormal{SD}}\right|^{2}/\sigma_{\text{D}}^{2}$
where $P_{\textnormal{S}}$ represents average transmit power at S, $\text{\ensuremath{\sigma}}_{\textnormal{D}}^{2}$
is the power of additive white Gaussian noise (AWGN) at D. In the
case of $\gamma_{\textnormal{SD}}\geq\gamma_{th}$, D feeds back two
bits ``11'' to S through a feedback link, where $\gamma_{th}$ is
a predefined instantaneous SNR threshold. Otherwise, D feeds back
two bits ``00'' instead. When S receives the feedback bits ``11'',
S broadcasts two bits ``01'' to R. Otherwise (i.e., S receives \textquotedbl 00\textquotedbl ),
S sends out bits ``10'' alternatively. If R receives \textquotedbl 01\textquotedbl ,
it means the direct link between S and D is good enough so that R
is not necessarily needed to assist the transmission between S and
D, R keeps working in the PEH mode without forwarding any information
(of course, including the possible covert massage). If R receives
bits ``10'', which means the quality of received information at
D is poor, R is expected to help the transmission from S to D. Before
participating in transmission, R has to estimate its residual energy,
checking whether the available energy is sufficient to support the
transmission. If the energy state of R is greater than a given residual
energy threshold $E_{th}$, i.e., $E_{i}\geq E_{th},$ R feeds back
bit ``1'' to S, otherwise, feeds back bit ``0'' instead. Once
S receives the feedback bit ``1'' from R, S starts to broadcast
the intended information signal, and R turns into the FD SWIPT mode,
i.e., R helps S forward the information signal and harvests energy
simultaneously. If S receives the feedback bit ``0'' from R, S broadcasts
energy signal to charge the battery at R. At this very time, D ceases
signal processing, because the energy signal is randomly generated
by S and conveys no useful information.

The condition $\gamma_{\textnormal{SD}}\geq\gamma_{th}$ is referred
as the ``SNR requirement'' which is applied to guarantee the reliability
of communication from S to D. On the other hand, the condition $E_{i}\geq E_{th}$
is regarded as the ``energy requirement'', ensuring that the residual
energy at R is sufficient to support the relying work.

We would like to explain the PEH and the FD SWIPT modes thoroughly
in the following:

\subsubsection{The PEH Mode}

When R works in the PEH mode, R employs the reception antenna for
receiving RF signals. Note that the PEH mode will be enabled in the
case of either $\gamma_{\textnormal{SD}}\geq\gamma_{th}$ or $\left\{\gamma_{\textnormal{SD}}<\gamma_{th}\right\}\cap \left\{E_{i}<E_{th}\right\}$.
By ignoring the negligible energy harvested from the noise at the
receiver, the total amount of energy harvested at R in a transmission
slot can be given by 
\begin{equation}
E_{\text{PEH}}=\eta P_{\text{S}}\left|h_{\text{SR}}\right|^{2},\label{eq:EnerPEH}
\end{equation}
where $\eta\left(0<\eta<1\right)$ means the efficiency of energy
conversion, and the harvested energy in this stage will be straight
transferred into the PEC.

\subsubsection{The FD SWIPT Mode}

It is worth noting that the FD SWIPT mode will be invoked when the
case $\left\{\gamma_{\textnormal{SD}}<\gamma_{th}\right\}\cap \left\{E_{i}\geq E_{th}\right\}$
holds. Only in this circumstance, R gets chance to broadcast covert
massage under the cover of the forwarded version of legitimate signals.

When R does not emit covert massages, the received signals
at R and D can be expressed respectively as
\begin{equation}
\boldsymbol{y}_{\text{R}}\left[\omega\right]=\sqrt{P_{\text{S}}}h_{\text{SR}}\boldsymbol{x}_{\text{S}}\left[\omega\right]+\sqrt{kP_{\text{R}}}h_{\text{RR}}\boldsymbol{x}_{\text{R}}\left[\omega\right]+\boldsymbol{n}_{\text{R}}\left[\omega\right],
\end{equation}
\begin{equation}
\boldsymbol{y}_{\text{D}}\left[\omega\right]=\sqrt{P_{\text{S}}}h_{\text{SD}}\boldsymbol{x}_{\text{S}}\left[\omega\right]+\sqrt{P_{\text{R}}}h_{\text{RD}}\boldsymbol{x}_{\text{R}}\left[\omega\right]+\boldsymbol{n}_{\text{D}}\left[\omega\right],\label{eq:y_DH0}
\end{equation}
where $P_{\text{R}}$ means average transmit power at
R, $\boldsymbol{x}_{\text{S}}\left[\omega\right]\sim\mathcal{CN}\left(0,1\right)$
represents the intended signal emitted from S, $\omega\in\left\{ 1,2,...,n\right\} $
denotes the symbol index in a transmission block and $n$ measures
the block-length, i.e., the total number of channel uses in each specific
transmission slot. $\boldsymbol{x}_{\text{R}}\left[\omega\right]=\boldsymbol{x}_{\text{S}}\left[\omega-\eth\right]$
is the forwarded version of $\boldsymbol{x}_{\text{S}}\left[\omega-\eth\right]$
after decoding and recoding where $\boldsymbol{x}_{\text{R}}\left[\omega\right]\sim\mathcal{CN}\left(0,1\right)$,
and integer $\eth$ represents the number of delayed symbols due to
signal processing. The AWGNs received at R and D are respectively
marked as $\boldsymbol{n}_{\text{R}}$ and $\boldsymbol{n}_{\text{D}}$,
subjected to $\boldsymbol{n}_{\text{R}}\left[\omega\right]\sim\mathcal{CN}\left(0,\sigma_{\text{R}}^{2}\right)$
and $\boldsymbol{n}_{\text{D}}\left[\omega\right]\sim\mathcal{CN}\left(0,\sigma_{\text{D}}^{2}\right)$.
In the FD SWIPT mode, R suffers from the SI which will definitely
degrade R's reception quality. Thanks to the
promising SIC techniques, R can debilitate the SI up to a relatively
low degree. Practically, constrained by computation capacity and impanelment
complexity, the perfect SIC cannot be reached. Thus, we consider a
practical scenario where imperfect SIC assumption is adopted, and
variable $k\in\left(0,1\right]$ represents the SIC coefficient which
implies different SIC levels.

When R does decide to broadcast covert massage, the received signals
at R and D can be expressed respectively as
\begin{equation}
\boldsymbol{y}_{\text{R}}\left[\omega\right]=\sqrt{P_{\text{S}}}h_{\text{SR}}\boldsymbol{x}_{\text{S}}\left[\omega\right]+\sqrt{kP_{\text{R}}}h_{\text{RR}}\boldsymbol{x}_{\text{R}}\left[\omega\right]+
\sqrt{kP_{\Delta}}h_{\text{RR}}\boldsymbol{x}_{\text{c}}\left[\omega\right]+\boldsymbol{n}_{\text{R}}\left[\omega\right],
\end{equation}
\begin{equation}
\boldsymbol{y}_{\text{D}}\left[\omega\right]=\sqrt{P_{\text{S}}}h_{\text{SD}}\boldsymbol{x}_{\text{S}}\left[\omega\right]+\sqrt{P_{\text{R}}}h_{\text{RD}}\boldsymbol{x}_{\text{R}}\left[\omega\right]+
\sqrt{P_{\Delta}}h_{\text{RD}}\boldsymbol{x}_{\text{c}}\left[\omega\right]+\boldsymbol{n}_{\text{D}}\left[\omega\right],\label{eq:y_DH1}
\end{equation}
where $P_{\Delta}$ means average transmit power of covert massage $\boldsymbol{x}_{\text{c}}$
subjected to $\boldsymbol{x}_{\text{c}}\left[\omega\right]\sim\mathcal{CN}\left(0,1\right)$.
Note that $P_{\Delta}$ merely comes from the constant energy supply.

Enabling the FD SWIPT mode, the PS-based EH protocol is adopted in
this paper. Specifically, R splits the power of received signal into
$\rho:\left(1-\rho\right)$ proportions. The $\rho$ portion of the
received signal power is used to EH and the remaining $\left(1-\rho\right)$
portion is allocated to information processing. Therefore, after ignoring
the negligible energy harvested form the AWGN, the energy harvested
at R in each time slot can be respectively calculated as
\begin{equation}
E_{\text{FS0}}=\eta\rho\left(P_{s}\left|h_{\text{SR}}\right|^{2}+kP_{\text{R}}\left|h_{\text{RR}}\right|^{2}\right),\label{eq:EnerFS0}
\end{equation}
\begin{equation}
E_{\text{FS1}}=\eta\rho\left(P_{s}\left|h_{\text{SR}}\right|^{2}+kP_{\text{R}}\left|h_{\text{RR}}\right|^{2}+kP_{\text{\ensuremath{\Delta}}}\left|h_{\text{RR}}\right|^{2}\right),\label{eq:EnerFS1}
\end{equation}
where the lower suffix ``FS0'' refers to the circumstance in which
the FD SWIPT mode is invoked without sending covert massage, another
lower suffix ``FS1'' means that the FD SWIPT mode with covert massage
is adopted. Particularly, we constrain the total transmit power at
R in the FD SWIPT mode as $P_{\text{FS0}}=P_{\text{R}}$ and $P_{\text{FS1}}=P_{\text{R}}+P_{\Delta}$, respectively.
Hence, (\ref{eq:EnerFS0}) and (\ref{eq:EnerFS1}) can be reconstructed uniformly as
\begin{equation}
E_{\text{FS}}=\eta\rho\left(P_{s}\left|h_{\text{SR}}\right|^{2}+kP_{\text{FS}}\left|h_{\text{RR}}\right|^{2}\right),\label{eq:EnerFS}
\end{equation}
where $P_{\text{FS}}\in\left\{ P_{\text{FS0}},P_{\text{FS1}}\right\} $
and $E_{\text{FS}}\in\left\{ E_{\text{FS0}},E_{\text{FS1}}\right\} $.
Note that $P_{\text{R}}=E_{th}$ holds for each specific transmission
block, the lower suffixes ``FS'', ``FS0'' and ``FS1'' designed
in this paper is for concise expression. In any specific mathematical
expression, ``FS'' is applied to sorely invoke ``FS0'' or ``FS1'',
and no combinations of them will be used. It is
worth noting that the harvested energy is collected via the MB at
the first place, and then transferred into the PEC within ignorable
time duration when the FD SWIPT mode completes.

\section{Markov Chain and Stationary Distribution}

Enabling the FD SWIPT mode at R, the hybrid energy container
composed of the PEC and the MB is considered in our proposed model. This hybrid
energy container system makes R possible to absorb and release energy
at the same time, which plays the essential role of hardware foundation
in the FD SWIPT mode. However, it leads to highly complex and dynamic
charge-discharge behaviors at R, which poses solid obstacle for tracking
energy state changes mathematically. To tackle this problem, the energy
capacity of PEC is firstly discretized. Then, the MC is invoked to
track the complex transmission procedure among discrete energy states.
Via the stationary distribution of the MC, the probability of satisfied energy requirement is determined.
\vspace{-.3cm}
\subsection{Energy Discretization}

To describe the dynamic charging and discharging behaviors of the PEC,
we need to discretize the battery capacity into discrete energy states
delicately. Each energy state implies the available energy remained
in the PEC, which can be reached by calculating the product of the
corresponding number of energy levels and the unit energy level. In details,
the PEC is quantized into $L+1$ levels, and each energy level characterizes
an energy unit equal to $C_{\text{P}}/L$ where $C_{\text{P}}$ represents
the energy capacity of the PEC. Therefore, the $i$-th energy state
is defined as $E_{i}=iC_{\text{P}}/L,i\in\left\{ 0,1,...,L\right\} $.
In the case of infinite energy discretization, i.e., $L\rightarrow+\infty$,
the proposed discrete battery model can tightly track the behavior
of continuous linear battery which is widely applied in the literature.
Note that $C_{\text{P}}\geq E_{th}$ is considered in this paper,
otherwise R gets no opportunity working in the FD SWIPT mode. It is
also worthy to declare that the analysis of energy discretization
concentrates on arbitrary transmission block.

In the PEH mode, the discretized amount of energy absorbed
by the PEC can be derived as
\begin{equation}
\varXi_{\text{PEH}}\overset{\triangle}{=}\left\lfloor \frac{E_{\text{PEH}}}{C_{\text{P}}/L}\right\rfloor \frac{C_{\text{P}}}{L}=\frac{q_{\text{PEH}}C_{\text{P}}}{L},\label{eq:DiscreEnerPEH}
\end{equation}
where $\left\lfloor \cdot\right\rfloor $ denotes the floor function
and $q_{\text{PEH}}\in\left\{ 1,2,...,L\right\} $ is defined for
notation concision. Here, without loss of generality, we declare that
the $i$-th energy state represents the initial energy amount available
in the PEC. After charging in the PEH mode, if $E_{i}+\varXi_{\text{PEH}}\geq C_{\text{P}}$,
the PEC will be charged to the maximal capacity $E_{L}=C_{\text{P}}$
and any overflowed energy has to be abandoned. Otherwise, the latest
energy state after charging is $E_{i+q_{\text{PEH}}}=E_{i}+\varXi_{\text{PEH}}$
which is guaranteed to be fully accommodated by the PEC.

In the FD SWIPT mode, the harvested energy should be first
stored in the MB and then delivered into the PEC when the FD SWIPT mode terminates. Because the MB is subjected to a predefined
energy capacity $C_{\text{M}}$, the potential amount of energy transferred
into the PEC should be reasonably constrained as $\min\left\{ E_{\text{FS}},C_{\text{M}}\right\} $
where the function $\min\left\{ x,y\right\} $ outputs the smaller
value between $x$ and $y$. practically, energy transfer form the
MB to the PEC suffers from circuitry attenuation. Thus, the actual
amount of energy absorbed by the PEC can be given by
\begin{equation}
\hat{E}_{\text{FS}}=\eta'\times\min\left\{ E_{\text{FS}},C_{\text{M}}\right\} ,
\end{equation}
where $\eta'$ denotes the energy transfer coefficient from the MB
to the PEC, for circuitry attenuation. Furthermore, the discretized
amount of energy absorbed by the PEC should be expressed as
\begin{equation}
\varXi_{\text{FS}}\overset{\triangle}{=}\left\lfloor \frac{\hat{E}_{\text{FS}}}{C_{\text{P}}/L}\right\rfloor \frac{C_{\text{P}}}{L}=\frac{q_{\text{FS}}C_{\text{P}}}{L},\label{eq:DiscreEnerFS}
\end{equation}
where $q_{\text{FS}}\in\left\{ 1,2,...,L\right\} $ is stated for
brief expression. While harvesting energy in the FD SWIPT mode, R
should decode the source signal and forward the recoded information
to D. Invoking the energy requirement, the consumed energy from the
PEC should locates at $E_{\text{FS}}^{\text{C}}\in\left[E_{th},E_{i}\right]$
where we set $E_{\text{FS}}^{\text{C}}=P_{\text{R}}=E_{th}=0.6C_{\text{P}}$ for each
transmission slot for simplicity. After discretization, the amount
of energy consumption at the PEC can be given by
\begin{equation}
\varXi_{\text{FS}}^{\text{C}}=\left\lceil \frac{E_{\text{FS}}^{\text{C}}}{C_{\text{P}}/L}\right\rceil \frac{C_{\text{P}}}{L}=\frac{q_{\text{FS}}^{\text{C}}C_{\text{P}}}{L},\label{eq:DiscreConsuENerFS}
\end{equation}
where $\left\lceil \cdot\right\rceil $ stands as the ceiling function,
and $q_{\text{FS}}^{\text{C}}$ is defined for notation simplicity.
It is worth noting that R may privately broadcast covert massage under
the cover of the legitimate forwarded signal. The energy supporting
convert communication sorely comes from the additional energy supply
unknown by the legitimate party and the transmit power of covert massage
is fixed as $P_{\Delta}$. Similarly, if $E_{i}-\varXi_{\text{FS}}^{\text{C}}+\varXi_{\text{FS}}\geq C_{1}$,
the PEC will be fully charged to $E_{L}=C_{\text{P}}$. On the contrary,
the latest energy state after charging is $E_{i-q_{\text{FS}}^{\text{C}}+q_{\text{FS}}}=E_{i}-\varXi_{\text{FS}}^{\text{C}}+\varXi_{\text{FS}}.$
 
For clarity, we pose the following statement.
At the beginning of the $\left(g+1\right)$-th block, the initial
energy state $E_{i}\left[g+1\right]$ is merely determined by the
transmission mode and energy variation occurred in the adjacently
former block, i.e., the $g$-th block. Note that $E_{i}\left[g+1\right]$
is independent to any transmission block before the $g$-th block,
which implies the Markov property. Specifically, $E_{i}\left[g+1\right]=\min\left\{ E_{i}\left[g\right]+\varXi_{\text{PEH}},C_{\text{P}}\right\} $
and $E_{i}\left[g+1\right]=\min\left\{ E_{i}\left[g\right]-\varXi_{\text{FS}}^{\text{C}}+\varXi_{\text{FS}},C_{\text{P}}\right\} $
correspond respectively to the PEH mode and the FD SWIPT mode applied
at R in the $g$-th block. Hence, energy state transition among different
blocks can be characterized and tracked by the MC. From the aforementioned
analysis, we note that energy state transition process in our proposed
system is time-independent, thus the MC is considered as homogeneous
in time domain.

\subsection{Markov Chain}

Following the energy discretization in a specific transmission block
and the transition relationship between energy states for different
blocks, we are able to track the transition procedure of energy states
at the PEC among multiple transmission blocks as a finite-state time-homogeneous
MC. Note that modeling the energy state transition process is indeed
not necessary for the MB, because it only plays as a temporary energy
storage in the FD SWIPT mode. The transition probability $p_{i,j}$
denotes the probability of energy state transition from $E_{i}$ to
$E_{j}$, which is occurred between the beginning of a transmission
block and that of the next transmission block. The energy state transitions
of the PEC can be stated comprehensively in the following six cases:

1) From $E_{0}$ to $E_{0}$: When initial energy at the PEC is empty,
it surly cannot afford the FD SWIPT mode. After a transmission block, the residual
energy yet remains empty in the considered case. It indicates that
the total harvested energy in this PEH block is discretized to zero,
namely, $\varXi_{\text{PEH}}=0$. Invoking (\ref{eq:EnerPEH}) and
(\ref{eq:DiscreEnerPEH}), the transition probability of states $E_{0}\rightarrow E_{0}$
can be given by
\begin{equation}
p_{0,0}=\Pr\left(q_{\text{PEH}}=0\right)=\Pr\left(\left|h_{\text{SR}}\right|^{2}<\frac{C_{\text{P}}}{\eta P_{s}L}\right).
\end{equation}
For $h_{\text{SR}}$ is subjected to Rayleigh fading and $\mathbb{E}\left\{ \vert h_{\text{SR}}\vert^{2}\right\} =\Omega_{\text{SR}}$,
$\left|h_{\text{SR}}\right|^{2}$ follows the Exponential distribution
with mean $\Omega_{\text{SR}}$. Thus, the cumulative distribution
function (CDF) of $\left|h_{\text{SR}}\right|^{2}$ can be derived
as $F_{\left|h_{\text{SR}}\right|^{2}}\left(x\right)=1-\exp\left(-x/\Omega_{\text{SR}}\right)$.
Furthermore, we get
\begin{equation}
p_{0,0}=F_{\left|h_{\text{SR}}\right|^{2}}\left(\frac{C_{\text{P}}}{\eta P_{s}L}\right).\label{eq:P_00}
\end{equation}

2) From $E_{L}$ to $E_{L}$: In this case, the initial energy certainly
satisfies the energy requirement. Thus, whether R works in the PEH mode or the
FD SWIPT mode depends merely on the SNR requirement. If the PEH mode is turned on, the harvested energy in
this case can be any possible value, since the PEC cannot absorb additional
energy any more. If the FD SWIPT mode is activated, the consumed energy
should be less than or equal to its harvested counterpart. From (\ref{eq:EnerFS}),
(\ref{eq:DiscreEnerFS}) and (\ref{eq:DiscreConsuENerFS}), the transition
probability of states $E_{L}\rightarrow E_{L}$ can be shown as\vspace{-.2cm}
\begin{equation}
p_{L,L}=\Pr\left(\gamma_{\text{SD}}\geq\gamma_{th}\right)+
\Pr\left(\gamma_{\text{SD}}<\gamma_{th}\right)\Pr\left(\varXi_{\text{FS}}^{\text{C}}\leq\varXi_{\text{FS}}\right).\label{eq:P_LLInitial}
\vspace{-.2cm}\end{equation}
Similar to the derivation of (\ref{eq:P_00}), we can obtain $q_{\text{SD}}=\Pr\left(\gamma_{\text{SD}}<\gamma_{th}\right)=F_{\left|h_{\text{SD}}\right|^{2}}\left(\sigma_{\text{D}}^{2}\gamma_{th}/P_{\text{S}}\right)$
and $\Pr\left(\gamma_{\text{SD}}\geq\gamma_{th}\right)=1-F_{\left|h_{\text{SD}}\right|^{2}}\left(\sigma_{\text{D}}^{2}\gamma_{th}/P_{\text{S}}\right)=1-q_{\text{SD}}$.
Regarding $\Pr\left(\varXi_{\text{FS}}^{\text{C}}\leq\varXi_{\text{FS}}\right)$,
we obtain\vspace{-.2cm}
\begin{align}
\Pr&\left(\varXi_{\text{FS}}^{\text{C}}\leq\varXi_{\text{FS}}\right)\nonumber\\
&=\Pr\left[\left(q_{\text{FS}}^{\text{C}}\leq\frac{\eta'E_{\text{FS}}}{C_{\text{P}}/L}\right)\bigcap\left(E_{\text{FS}}<C_{\text{M}}\right)\right]+
\Pr\left[\left(q_{\text{FS}}^{\text{C}}\leq\frac{\eta'C_{\text{M}}}{C_{\text{P}}/L}\right)\bigcap\left(E_{\text{FS}}\geq C_{\text{M}}\right)\right] \nonumber \\
&=\begin{cases}
\Pr\left(E_{\text{FS}}\geq\frac{q_{\text{FS}}^{\text{C}}C_{\text{P}}}{\eta'L}\right), & C_{\text{M}}\geq\frac{q_{\text{FS}}^{\text{C}}C_{\text{P}}}{\eta'L}\\
0, & C_{\text{M}}<\frac{q_{\text{FS}}^{\text{C}}C_{\text{P}}}{\eta'L}
\end{cases}.\label{eq:q_C<q}
\end{align}

Invoking (\ref{eq:EnerFS}), we can gain\vspace{-.2cm}
\begin{equation}
\Pr\left(E_{\text{FS}}\geq\frac{q_{\text{FS}}^{\text{C}}C_{\text{P}}}{\eta'L}\right)=\Pr\left(Z\geq\frac{q_{\text{FS}}^{\text{C}}C_{\text{P}}}{\eta\rho\eta'L}\right),
\vspace{-.2cm}\end{equation}
where $Z=P_{s}\left|h_{\text{SR}}\right|^{2}+kP_{\text{FS}}\left|h_{\text{RR}}\right|^{2}$.
Via convolution of two Exponential distribution variables, we obtain
the CDF of $Z$ as\vspace{-.2cm}
\begin{equation}
F_{Z}\left(x\right)=
\begin{cases}
1-\frac{P_{\text{S}}\Omega_{\text{SR}}}{P_{\text{S}}\Omega_{\text{SR}}-kP_{\text{FS}}\Omega_{\text{RR}}}e^{-\frac{x}{P_{\text{S}}\Omega_{\text{SR}}}}+
\frac{kP_{\text{FS}}\Omega_{\text{RR}}}{P_{\text{S}}\Omega_{\text{SR}}-kP_{\text{FS}}\Omega_{\text{RR}}}e^{-\frac{x}{kP_{\text{FS}}\Omega_{\text{RR}}}}, & P_{\text{S}}\Omega_{\text{SR}}\neq kP_{\text{FS}}\Omega_{\text{RR}}\\
\frac{1}{2}\gamma\left(2,\frac{x}{P_{\text{S}}\Omega_{\text{SR}}}\right), & P_{\text{S}}\Omega_{\text{SR}}=kP_{\text{FS}}\Omega_{\text{RR}}
\end{cases},
\vspace{-.2cm}\end{equation}
where $\gamma\left(\cdot,\cdot\right)$ is the lower incomplete Gamma
function. Furthermore, we get\vspace{-.2cm}
\begin{equation}
\Pr\left(E_{\text{FS}}\geq\frac{q_{\text{FS}}^{\text{C}}C_{\text{P}}}{\eta'L}\right)=1-F_{Z}\left(\frac{q_{\text{FS}}^{\text{C}}C_{\text{P}}}{\eta\rho\eta'L}\right).\label{eq:1-F_Z}
\vspace{-.2cm}\end{equation}
Finally, combining (\ref{eq:P_LLInitial}), (\ref{eq:q_C<q}) and
(\ref{eq:1-F_Z}), we can obtain \vspace{-.2cm}
\begin{equation}
p_{L,L}=\begin{cases}
1-q_{\text{SD}}F_{Z}\left(\frac{q_{\text{FS}}^{\text{C}}C_{\text{P}}}{\eta\rho\eta'L}\right), & C_{\text{M}}\geq\frac{q_{\text{FS}}^{\text{C}}C_{\text{P}}}{\eta'L}\\
1-q_{\text{SD}}, & C_{\text{M}}<\frac{q_{\text{FS}}^{\text{C}}C_{\text{P}}}{\eta'L}
\end{cases}.
\vspace{-.2cm}\end{equation}

3) From $E_{i}$ to $E_{j}$ $\left(0\leq i<j<L\right)$: It is easy
to find that the energy requirement in this case is not always met.
If the initial energy state cannot satisfy the energy requirement,
i.e., $E_{i}<E_{th}$, the PEH mode will be selected. Otherwise, we
need to evaluate whether the SNR requirement is met or not. When $\gamma_{\text{SD}}\geq\gamma_{th}$,
R will choose the PEH mode. On the contrary, R will work in the FD
SWIPT mode. Thus, the transition probability of states $E_{i}\rightarrow E_{j}$
can be expressed as\vspace{-.2cm}
\begin{align}
p_{i,j}
&=q_{\text{SD}}\Pr\left(E_{i}<E_{th}\right)\Pr\left(q_{\text{PEH}}=j-i\right)+
q_{\text{SD}}\Pr\left(E_{i}\geq E_{th}\right)\Pr\left(q_{\text{FS}}-q_{\text{FS}}^{\text{C}}=j-i\right)\nonumber\\
&+\left(1-q_{\text{SD}}\right)\Pr\left(q_{\text{PEH}}=j-i\right) \nonumber\\
&=\begin{cases}
\Pr\left(q_{\text{PEH}}=j-i\right), & i<\varphi\\
\left(1-q_{\text{SD}}\right)\Pr\left(q_{\text{PEH}}=j-i\right)+
q_{\text{SD}}\Pr\left(q_{\text{FS}}-q_{\text{FS}}^{\text{C}}=j-i\right), & i\geq\varphi
\end{cases},\label{eq:p_ijInitial}
\vspace{-.2cm}\end{align}
where $\varphi=\left\lceil \frac{E_{th}}{C_{\text{P}}/L}\right\rceil $
denotes the total number of energy units needed to represent the energy
requirement in the discretized energy regime.

Next, we calculate $\Pr\left(q_{\text{PEH}}=j-i\right)$
and $\Pr\left(q_{\text{FS}}-q_{\text{FS}}^{\text{C}}=j-i\right)$,
shown respectively as (\ref{eq:qPEH=00003Dj-i}) and (\ref{eq:qFS-qFSC=00003Dj-i}) in the following.\vspace{-.2cm}
\begin{align}
\Pr\left(q_{\text{PEH}}=j-i\right)
&=\Pr\left(\frac{\left(j-i\right)C_{\text{P}}}{\eta P_{\text{S}}L}\leq\text{\ensuremath{\vert h_{\text{SR}}\vert^{2}}}<\frac{\left(j-i+1\right)C_{\text{P}}}{\eta P_{\text{S}}L}\right) \nonumber\\
&=F_{\left|h_{\text{SR}}\right|^{2}}\left(\frac{\left(j-i+1\right)C_{\text{P}}}{\eta P_{\text{S}}L}\right)-F_{\left|h_{\text{SR}}\right|^{2}}\left(\frac{\left(j-i\right)C_{\text{P}}}{\eta P_{\text{S}}L}\right).\label{eq:qPEH=00003Dj-i}
\vspace{-.2cm}\end{align}
\begin{align}\vspace{-.2cm}
&\Pr\left(q_{\text{FS}}-q_{\text{FS}}^{\text{C}}=j-i\right)\nonumber\\
&=\begin{cases}
0, & C_{\text{M}}<\frac{\left(j-i+q_{\text{FS}}^{\text{c}}\right)C_{\text{P}}}{\eta'L}\\
1-F_{Z}\left(\frac{\left(j-i+q_{\text{FS}}^{\text{c}}\right)C_{\text{P}}}{\eta\rho\eta'L}\right), & \frac{\left(j-i+q_{\text{FS}}^{\text{c}}\right)C_{\text{P}}}{\eta'L}\leq C_{\text{M}}<\frac{\left(j-i+q_{\text{FS}}^{\text{c}}+1\right)C_{\text{P}}}{\eta'L}\\
F_{Z}\left(\frac{\left(j-i+q_{\text{FS}}^{\text{c}}+1\right)C_{\text{P}}}{\eta\rho\eta'L}\right)-F_{Z}\left(\frac{\left(j-i+q_{\text{FS}}^{\text{c}}\right)C_{\text{P}}}{\eta\rho\eta'L}\right), & C_{\text{M}}\geq\frac{\left(j-i+q_{\text{FS}}^{\text{c}}+1\right)C_{\text{P}}}{\eta'L}
\end{cases}.\label{eq:qFS-qFSC=00003Dj-i}
\vspace{-.2cm}\end{align}

Combining (\ref{eq:p_ijInitial}), (\ref{eq:qPEH=00003Dj-i}) and
(\ref{eq:qFS-qFSC=00003Dj-i}), we get the probability of transition
$E_{i}\rightarrow E_{j}$ as \vspace{-.2cm}
\begin{equation}
p_{i,j}=\begin{cases}
F_{\left|h_{\text{SR}}\right|^{2}}\left(\frac{\left(j-i+1\right)C_{\text{P}}}{\eta P_{\text{S}}L}\right)-F_{\left|h_{\text{SR}}\right|^{2}}\left(\frac{\left(j-i\right)C_{\text{P}}}{\eta P_{\text{S}}L}\right), & i<\varphi\\
\left(1-q_{\text{SD}}\right)\\
\times\left[F_{\left|h_{\text{SR}}\right|^{2}}\left(\frac{\left(j-i+1\right)C_{\text{P}}}{\eta P_{\text{S}}L}\right)-F_{\left|h_{\text{SR}}\right|^{2}}\left(\frac{\left(j-i\right)C_{\text{P}}}{\eta P_{\text{S}}L}\right)\right], & i\geq\varphi\&C_{\text{M}}<\frac{\left(j-i+q_{\text{FS}}^{\text{c}}\right)C_{\text{P}}}{\eta'L}\\
\left(1-q_{\text{SD}}\right)\\
\times\left[F_{\left|h_{\text{SR}}\right|^{2}}\left(\frac{\left(j-i+1\right)C_{\text{P}}}{\eta P_{\text{S}}L}\right)-F_{\left|h_{\text{SR}}\right|^{2}}\left(\frac{\left(j-i\right)C_{\text{P}}}{\eta P_{\text{S}}L}\right)\right]\\
+q_{\text{SD}}\left[1-F_{Z}\left(\frac{\left(j-i+q_{\text{FS}}^{\text{c}}\right)C_{\text{P}}}{\eta\rho\eta'L}\right)\right], & i\geq\varphi\&\frac{\left(j-i+q_{\text{FS}}^{\text{c}}\right)C_{\text{P}}}{\eta'L}\\
                                                                                                                                                                                &\hspace{1cm}\leq C_{\text{M}}<\frac{\left(j-i+q_{\text{FS}}^{\text{c}}+1\right)C_{\text{P}}}{\eta'L}\\
\left(1-q_{\text{SD}}\right)\\
\times\left[F_{\left|h_{\text{SR}}\right|^{2}}\left(\frac{\left(j-i+1\right)C_{\text{P}}}{\eta P_{\text{S}}L}\right)-F_{\left|h_{\text{SR}}\right|^{2}}\left(\frac{\left(j-i\right)C_{\text{P}}}{\eta P_{\text{S}}L}\right)\right]\\
+q_{\text{SD}}\\
\times\left[F_{Z}\left(\frac{\left(j-i+q_{\text{FS}}^{\text{c}}+1\right)C_{\text{P}}}{\eta\rho\eta'L}\right)-F_{Z}\left(\frac{\left(j-i+q_{\text{FS}}^{\text{c}}\right)C_{\text{P}}}{\eta\rho\eta'L}\right)\right], & i\geq\varphi\&C_{\text{M}}\geq\frac{\left(j-i+q_{\text{FS}}^{\text{c}}+1\right)C_{\text{P}}}{\eta'L}
\end{cases}.\label{eq:p_ij}
\end{equation}

4) From $E_{i}$ to $E_{i}$ $\left(0<i<L\right)$: In this case,
it is not certain whether the energy requirement is met or not. If
$E_{i}<E_{th}$, the PEH mode will be invoked and the harvested energy
should be discretized as zero. If $E_{i}\geq E_{th}$ and $\gamma_{\text{SD}}\geq\gamma_{th}$,
the PEH mode is enabled and the harvested energy should also be discretized
as zero, too. If $E_{i}\geq E_{th}$ and $\gamma_{\text{SD}}<\gamma_{th}$,
the FD SWIPT mode will be selected, the discretized amount of consumed
energy should be equal to that of harvested energy. Hence, the transition
probability of states $E_{i}\rightarrow E_{i}$ can be calculated as\vspace{-.2cm}
\begin{align}
p_{i,i}
&=\left(1-q_{\text{SD}}\right)\Pr\left(q_{\text{PEH}}=0\right)\nonumber\\
&+q_{\text{SD}}\Pr\left(E_{i}<E_{th}\right)\Pr\left(q_{\text{PEH}}=0\right)+
q_{\text{SD}}\Pr\left(E_{i}\geq E_{th}\right)\Pr\left(q_{\text{FS}}-q_{\text{FS}}^{\text{C}}=0\right) \nonumber\\
&=\begin{cases}
\Pr\left(q_{\text{PEH}}=0\right), & i<\varphi\\
\left(1-q_{\text{SD}}\right)\Pr\left(q_{\text{PEH}}=0\right)+
q_{\text{SD}}\Pr\left(q_{\text{FS}}-q_{\text{FS}}^{\text{C}}=0\right), & i\geq\varphi
\end{cases},\label{eq:p_iiInitial}
\vspace{-.2cm}\end{align}
where $\Pr\left(q_{\text{PEH}}=0\right)$ and $\Pr\left(q_{\text{FS}}-q_{\text{FS}}^{\text{C}}=0\right)$
are given respectively by (\ref{eq:P_00}) and (\ref{eq:qFS-qFSC=00003D0}) shown
as\vspace{-.2cm}
\begin{align}
\Pr\left(q_{\text{FS}}-q_{\text{FS}}^{\text{C}}=0\right)&=\Pr\left(q_{\text{FS}}=q_{\text{FS}}^{\text{C}}\right)\nonumber\\
&=\begin{cases}
0, & C_{\text{M}}<\frac{q_{\text{FS}}^{\text{c}}C_{\text{P}}}{\eta'L}\\
1-F_{Z}\left(\frac{q_{\text{FS}}^{\text{c}}C_{\text{P}}}{\eta\rho\eta'L}\right), & \frac{q_{\text{FS}}^{\text{c}}C_{\text{P}}}{\eta'L}\leq C_{\text{M}}<\frac{\left(q_{\text{FS}}^{\text{c}}+1\right)C_{\text{P}}}{\eta'L}\\
F_{Z}\left(\frac{\left(q_{\text{FS}}^{\text{c}}+1\right)C_{\text{P}}}{\eta\rho\eta'L}\right)-F_{Z}\left(\frac{q_{\text{FS}}^{\text{c}}C_{\text{P}}}{\eta\rho\eta'L}\right), & C_{\text{M}}\geq\frac{\left(q_{\text{FS}}^{\text{c}}+1\right)C_{\text{P}}}{\eta'L}
\end{cases}.\label{eq:qFS-qFSC=00003D0}
\end{align}

Substituting (\ref{eq:P_00}) and (\ref{eq:qFS-qFSC=00003D0}) into (\ref{eq:p_iiInitial}), we get the transition probability of states $E_{i}\rightarrow E_{i}$ as\vspace{-.2cm}
\begin{equation}
p_{i,i} =\begin{cases}
F_{\left|h_{\text{SR}}\right|^{2}}\left(\frac{C_{\text{P}}}{\eta P_{\text{S}}L}\right), & i<\varphi\\
\left(1-q_{\text{SD}}\right)F_{\left|h_{\text{SR}}\right|^{2}}\left(\frac{C_{\text{P}}}{\eta P_{\text{S}}L}\right), & i\geq\varphi\&C_{\text{M}}<\frac{q_{\text{FS}}^{\text{c}}C_{\text{P}}}{\eta'L}\\
\left(1-q_{\text{SD}}\right)F_{\left|h_{\text{SR}}\right|^{2}}\left(\frac{C_{\text{P}}}{\eta P_{\text{S}}L}\right)\\
+q_{\text{SD}}\left[1-F_{Z}\left(\frac{q_{\text{FS}}^{\text{c}}C_{\text{P}}}{\eta\rho\eta'L}\right)\right], & i\geq\varphi\&\frac{q_{\text{FS}}^{\text{c}}C_{\text{P}}}{\eta'L}\leq C_{\text{M}}<\frac{\left(q_{\text{FS}}^{\text{c}}+1\right)C_{\text{P}}}{\eta'L}\\
\left(1-q_{\text{SD}}\right)F_{\left|h_{\text{SR}}\right|^{2}}\left(\frac{C_{\text{P}}}{\eta P_{\text{S}}L}\right)\\
+q_{\text{SD}}\left[F_{Z}\left(\frac{\left(q_{\text{FS}}^{\text{c}}+1\right)C_{\text{P}}}{\eta\rho\eta'L}\right)-F_{Z}\left(\frac{q_{\text{FS}}^{\text{c}}C_{\text{P}}}{\eta\rho\eta'L}\right)\right], & i\geq\varphi\&C_{\text{M}}\geq\frac{\left(q_{\text{FS}}^{\text{c}}+1\right)C_{\text{P}}}{\eta'L}
\end{cases}.\label{eq:p_ii}
\end{equation}

5) From $E_{i}$ to $E_{j}$ $\left(0\leq j<i\leq L\right)$: Obviously, this circumstance can only occur in the
FD SWIPT mode because the PEH mode can exclusively lead to energy
increasing or energy unchanged. Therefore, the transition probability of states $E_{i}\rightarrow E_{j}$
can be derived as
\begin{align}
p_{i,j}&=
\Pr\left(\gamma_{\text{SD}}<\gamma_{th}\right)\Pr\left(E_{i}\geq E_{th}\right)\Pr\left(q_{\text{FS}}^{\text{C}}-q_{\text{FS}}=i-j\right)\nonumber\\
&=\begin{cases}
0, & i<\varphi\\
q_{\text{SD}}\Pr\left(q_{\text{FS}}^{\text{C}}-q_{\text{FS}}=i-j\right), & i\geq\varphi
\end{cases}.\label{eq:p_jiInitial}
\end{align}
\vspace{-.1cm}
Next, we need to calculate $\Pr\left(q_{\text{FS}}^{\text{C}}-q_{\text{FS}}=i-j\right)$,
shown as 
\begin{align}
&\Pr\left(q_{\text{FS}}^{\text{C}}-q_{\text{FS}}=i-j\right)\nonumber\\
&=
\begin{cases}
0, & C_{\text{M}}<\frac{\left[q_{\text{FS}}^{\text{C}}-\left(i-j\right)\right]C_{\text{P}}}{\eta'L}\\
1-F_{Z}\left(\frac{\left[q_{\text{FS}}^{\text{C}}-\left(i-j\right)\right]C_{\text{P}}}{\eta\rho\eta'L}\right), & \frac{\left[q_{\text{FS}}^{\text{C}}-\left(i-j\right)\right]C_{\text{P}}}{\eta'L}\leq C_{\text{M}}<\frac{\left[q_{\text{FS}}^{\text{C}}-\left(i-j\right)+1\right]C_{\text{P}}}{\eta'L}.\\
F_{Z}\left(\frac{\left[q_{\text{FS}}^{\text{C}}-\left(i-j\right)+1\right]C_{\text{P}}}{\eta\rho\eta'L}\right)-
F_{Z}\left(\frac{\left[q_{\text{FS}}^{\text{C}}-\left(i-j\right)\right]C_{\text{P}}}{\eta\rho\eta'L}\right), & C_{\text{M}}\geq\frac{\left[q_{\text{FS}}^{\text{C}}-\left(i-j\right)+1\right]C_{\text{P}}}{\eta'L}
\end{cases}\label{eq:qFSC-qFS=00003Dj-i}
\end{align}

Invoking (\ref{eq:p_jiInitial}) and (\ref{eq:qFSC-qFS=00003Dj-i}),
we can express the transition probability of state $E_{i}\rightarrow E_{j}$
as 
\begin{equation}
p_{i,j}=\begin{cases}
0, & i<\varphi\|\left(j\geq\varphi\&C_{\text{M}}<\frac{\left[q_{\text{FS}}^{\text{C}}-\left(i-j\right)\right]C_{\text{P}}}{\eta'L}\right)\\
q_{\text{SD}}\left(1-F_{Z}\left(\frac{\left[q_{\text{FS}}^{\text{C}}-\left(i-j\right)\right]C_{\text{P}}}{\eta\rho\eta'L}\right)\right), & i\geq\varphi\&\frac{\left[q_{\text{FS}}^{\text{C}}-\left(i-j\right)\right]C_{\text{P}}}{\eta'L}\leq C_{\text{M}}<\frac{\left[q_{\text{FS}}^{\text{C}}-\left(i-j\right)+1\right]C_{\text{P}}}{\eta'L}\\
q_{\text{SD}}\left[F_{Z}\left(\frac{\left[q_{\text{FS}}^{\text{C}}-\left(i-j\right)+1\right]C_{\text{P}}}{\eta\rho\eta'L}\right)\right.\\
\left.-F_{Z}\left(\frac{\left[q_{\text{FS}}^{\text{C}}-\left(i-j\right)\right]C_{\text{P}}}{\eta\rho\eta'L}\right)\right], & i\geq\varphi\&C_{\text{M}}\geq\frac{\left[q_{\text{FS}}^{\text{C}}-\left(i-j\right)+1\right]C_{\text{P}}}{\eta'L}
\end{cases}.\label{eq:p_ji}
\end{equation}

6) From $E_{i}$ to $E_{L}$ $\left(0\leq i<L\right)$: In this circumstance,
we cannot make sure what mode is applied by R, for whether the initial
energy state can satisfy the energy requirement is not determined.
When $E_{i}<E_{th}$, certainly the PEH mode will be activated, and
the harvested energy should meet $\varXi_{\text{PEH}}\geq E_{L}-E_{i}$.
Otherwise, if $\gamma_{\text{SD}}\geq\gamma_{th}$, the PEH also will
be invoked and the harvested energy is supposed to satisfy $\varXi_{\text{PEH}}\geq E_{L}-E_{i}$.
If $E_{i}\geq E_{th}$ and $\gamma_{\text{SD}}<\gamma_{th}$, the
FD SWIPT mode will be selected and the relationship between the harvested
energy and the released energy should meet $\varXi_{\text{PEH}}-\varXi_{\text{PEH}}^{\text{C}}\geq E_{L}-E_{i}$.
Thus, the transition probability of states $E_{i}\rightarrow E_{L}$
can be expressed as
\begin{align}
p_{i,L}&=\Pr\left(\gamma_{\text{SD}}\geq\gamma_{th}\right)\Pr\left(q_{\text{PEH}}\geq L-i\right)+
\Pr\left(\gamma_{\text{SD}}<\gamma_{th}\right)\Pr\left(E_{i}<E_{th}\right)\Pr\left(q_{\text{PEH}}\geq L-i\right)\nonumber\\
&+\Pr\left(\gamma_{\text{SD}}<\gamma_{th}\right)\Pr\left(E_{i}\geq E_{th}\right)\Pr\left(q_{\text{FS}}-q_{\text{FS}}^{\text{C}}\geq L-i\right)\nonumber\\
&=\begin{cases}
\Pr\left(q_{\text{PEH}}\geq L-i\right), & i<\varphi\\
\left(1-q_{\text{SD}}\right)\Pr\left(q_{\text{PEH}}\geq L-i\right)+
q_{\text{SD}}\Pr\left(q_{\text{FS}}-q_{\text{FS}}^{\text{C}}\geq L-i\right), & i\geq\varphi
\end{cases}.\label{eq:p_iLInitial}
\end{align}

Next, $\Pr\left(q_{\text{PEH}}\geq L-i\right)$ and $\Pr\left(q_{\text{FS}}-q_{\text{FS}}^{\text{C}}\geq L-i\right)$
can be derived respectively as
\begin{equation}
\Pr\left(q_{\text{PEH}}\geq L-i\right)=1-F_{\left|h_{\text{SR}}\right|^{2}}\left(\frac{\left(L-i\right)C_{\text{P}}}{\eta P_{\text{S}}L}\right),\label{eq:PrqPEH>=00003DL-i}
\end{equation}
\begin{equation}
\Pr\left(q_{\text{FS}}-q_{\text{FS}}^{\text{C}}\geq L-i\right)=
\begin{cases}
0, & C_{\text{M}}<\frac{\left(L-i+q_{\text{FS}}^{\text{C}}\right)C_{\text{P}}}{\eta'L}\\
1-F_{Z}\left(\frac{\left(L-i+q_{\text{FS}}^{\text{C}}\right)C_{\text{P}}}{\eta\rho\eta'L}\right), & C_{\text{M}}\geq\frac{\left(L-i+q_{\text{FS}}^{\text{C}}\right)C_{\text{P}}}{\eta'L}
\end{cases}.\label{eq:PrqFS-qFSC>=00003DL-i}
\end{equation}
Invoking (\ref{eq:p_iLInitial}), (\ref{eq:PrqPEH>=00003DL-i}) and
(\ref{eq:PrqFS-qFSC>=00003DL-i}), we get the transition probability
of states $E_{i}\rightarrow E_{L}$, shown as
\begin{equation}
p_{i,L}=\begin{cases}
1-F_{\left|h_{\text{SR}}\right|^{2}}\left(\frac{\left(L-i\right)C_{\text{P}}}{\eta P_{\text{S}}L}\right), & i<\varphi\\
\left(1-q_{\text{SD}}\right)\left[1-F_{\left|h_{\text{SR}}\right|^{2}}\left(\frac{\left(L-i\right)C_{\text{P}}}{\eta P_{\text{S}}L}\right)\right], & i\geq\varphi\&C_{\text{M}}<\frac{\left(L-i+q_{\text{FS}}^{\text{C}}\right)C_{\text{P}}}{\eta'L}\\
\left(1-q_{\text{SD}}\right)\left[1-F_{\left|h_{\text{SR}}\right|^{2}}\left(\frac{\left(L-i\right)C_{\text{P}}}{\eta P_{\text{S}}L}\right)\right]\\
+q_{\text{SD}}\left[1-F_{Z}\left(\frac{\left(L-i+q_{\text{FS}}^{\text{C}}\right)C_{\text{P}}}{\eta\rho\eta'L}\right)\right], & i\geq\varphi\&C_{\text{M}}\geq\frac{\left(L-i+q_{\text{FS}}^{\text{C}}\right)C_{\text{P}}}{\eta'L}
\end{cases}.\label{eq:p_iL}
\end{equation}

\subsection{Stationary Distribution}

We define $\mathbf{M}\overset{\triangle}{=}\left\{ p_{i,j}\right\} $
to denote the $\left(L+1\right)\times\left(L+1\right)$ state transition
matrix. For the convenience of illustration, we provide an example of the possible energy states and the transitions among them in the case of $L=2$. As shown in Figure 1, the state transition diagram and the corresponding transition probability matrix of the MC are clearly depicted.
\begin{figure}
\begin{multicols}{2}

\hspace{2cm}{\centering\includegraphics[scale=0.37]{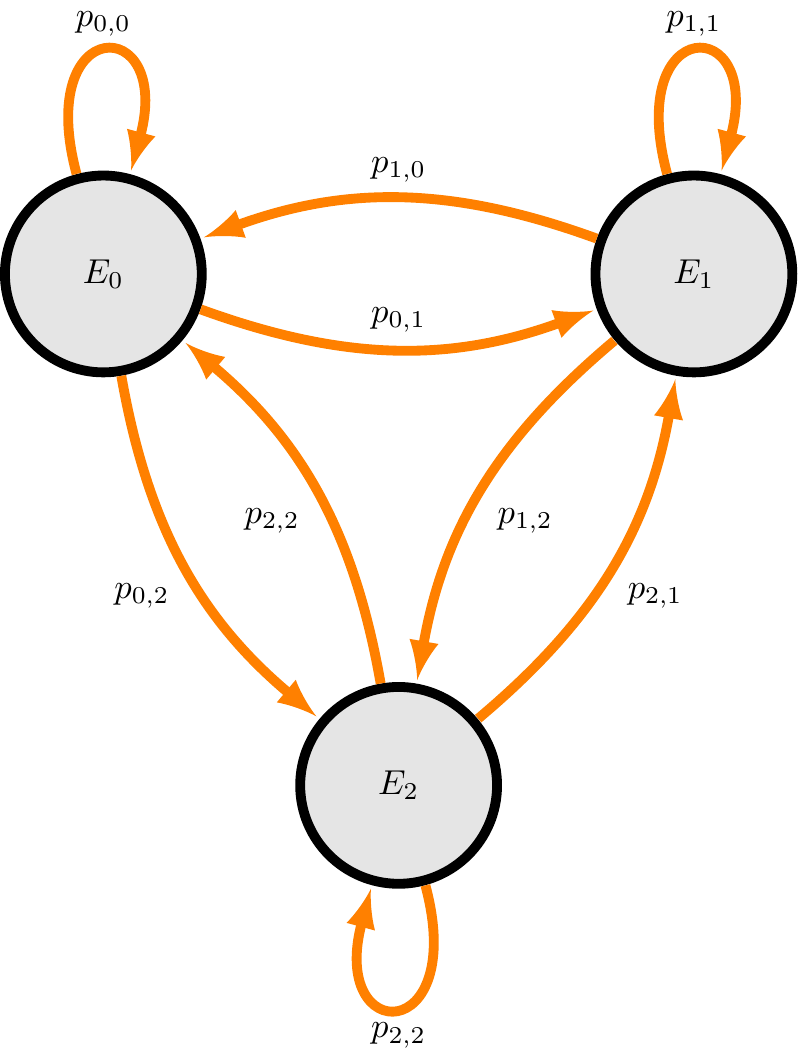}}

\begin{equation*}
\begin{aligned}[b]
\mathbf{M} = 
\begin{bmatrix} 
p_{0,0} & p_{0,1} & p_{0,L}\\ 
p_{1,0} & p_{1,1} & p_{1,L}\\
p_{2,0} & p_{2,1} & p_{L,L}
\end{bmatrix}
\end{aligned}
\end{equation*}

\end{multicols}
\vspace{-1cm}
\caption{The state transition diagram and the corresponding transition probability
matrix of the Markov chain, in the case of $L=2$.}

\end{figure}

\begin{thm}
In this theorem, we derive the probability that the energy status
of arbitrary transmission slot meets the given energy condition. With
the help of stationary distribution $\boldsymbol{\xi}$, for arbitrary
transmission slot, we have 
\begin{equation}
\Pr\left(E_{i}\geq E_{th}\right)=\sum_{i=\varphi}^{L}\xi_{i},
\end{equation}
where $\xi_{i}\in\boldsymbol{\xi}=\left(\xi_{0},\xi_{1},...,\xi_{L}\right)^{T}$,
which can be gained by applying method in the following proof.

Furthermore, we can conclude that
\begin{equation}
\Pr\left(E_{i}<E_{th}\right)=1-\sum_{i=\varphi}^{L}\xi_{i}=\sum_{i=0}^{\varphi-1}\xi_{i}.
\end{equation}
\end{thm}

\begin{IEEEproof}
Using the similar methods in \cite{Krikidis2012Buffer}, we can easily
verify that the transition matrix $\mathbf{M}$ is irreducible$\footnote{In a MC, the transition matrix is said to be irreducible if it is possible to reach any other state form any state in finite number of steps. In our MC analysis, all possible energy states communicate so that the transition matrix $\bf{M}$ is irreducible in this paper.}$ and
row stochastic$\footnote{In a MC, the transition matrix is said to be row stochastic if the sum of all the elements in a row is one and all elements are non-negative. In our MC analysis, the transition probabilities from any energy state to all possible energy states sums up to one and the transition probabilities are definitely non-negative, so we say the transition matrix $\bf{M}$ is row stochastic in this paper. We also note that $\bf{M}$ is asymmetric because $p_{i,j}\ne p_{j,i}$, $\forall i,j$, given the aforementioned analysis.}$,
which can be verified via Figure 1 as an example. Thus, the stationary
distribution $\boldsymbol{\xi}$ must satisfy the following equation
\begin{equation}
\boldsymbol{\xi}=\left(\xi_{0},\xi_{1},...,\xi_{L}\right)^{T}=\mathbf{M}^{T}\boldsymbol{\xi}.
\end{equation}

By solving the above equation, $\boldsymbol{\xi}$ can be derived
as
\begin{equation}
\boldsymbol{\xi}=\left(\mathbf{M}^{T}-\mathbf{I}+\mathbf{B}\right)^{-1}\boldsymbol{b},
\end{equation}
 where $\mathbf{B}_{i,j}=1,\forall i,j,$ $\boldsymbol{b}=\left(1,1,...,1\right)^{T}$
and $\mathbf{I}$ denotes the unit matrix.
\end{IEEEproof}

\begin{remrk}
In $\textbf{\textit{Theorem}}$ \textit{1}, $\xi_{i}$ where $i\in\left\{ 0,1,\dots,L\right\} $
represents the stationary probability of the $i$-th energy state,
on a long-term perspective. The reason why the result $\Pr\left(E_{i}\geq E_{th}\right)=\sum_{i=\varphi}^{L}\xi_{i}$
in $\textbf{\textit{Theorem}}$ \textit{1} holds can be straightly explained as that
$\xi_{i}$ where $i\geq\varphi$ describes the probability of an arbitrary event whose
residual energy is higher than the energy threshold and the probability summation
of all these events makes up the overall probability of $E_{i}\geq E_{th}$.
It is worth noting that $\textbf{\textit{Theorem}}$ \textit{1} serves as the prerequisite
for deriving closed-form expressions of transmission outage probability
which will be shown in Section V.
\end{remrk}

\subsection{Verification and Discussion}

\begin{figure}
\centering
\includegraphics[scale=0.4]{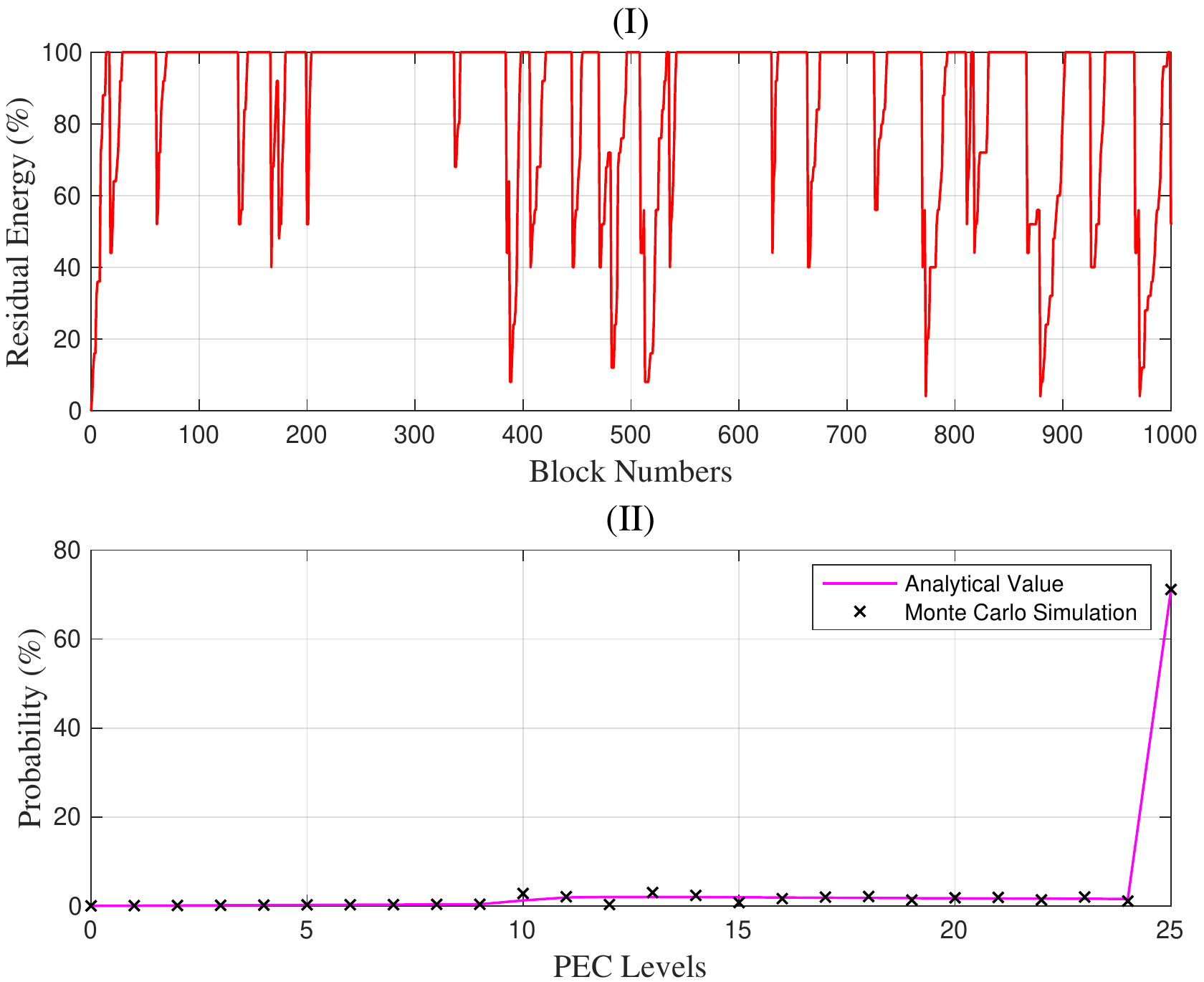}
\includegraphics[scale=0.4]{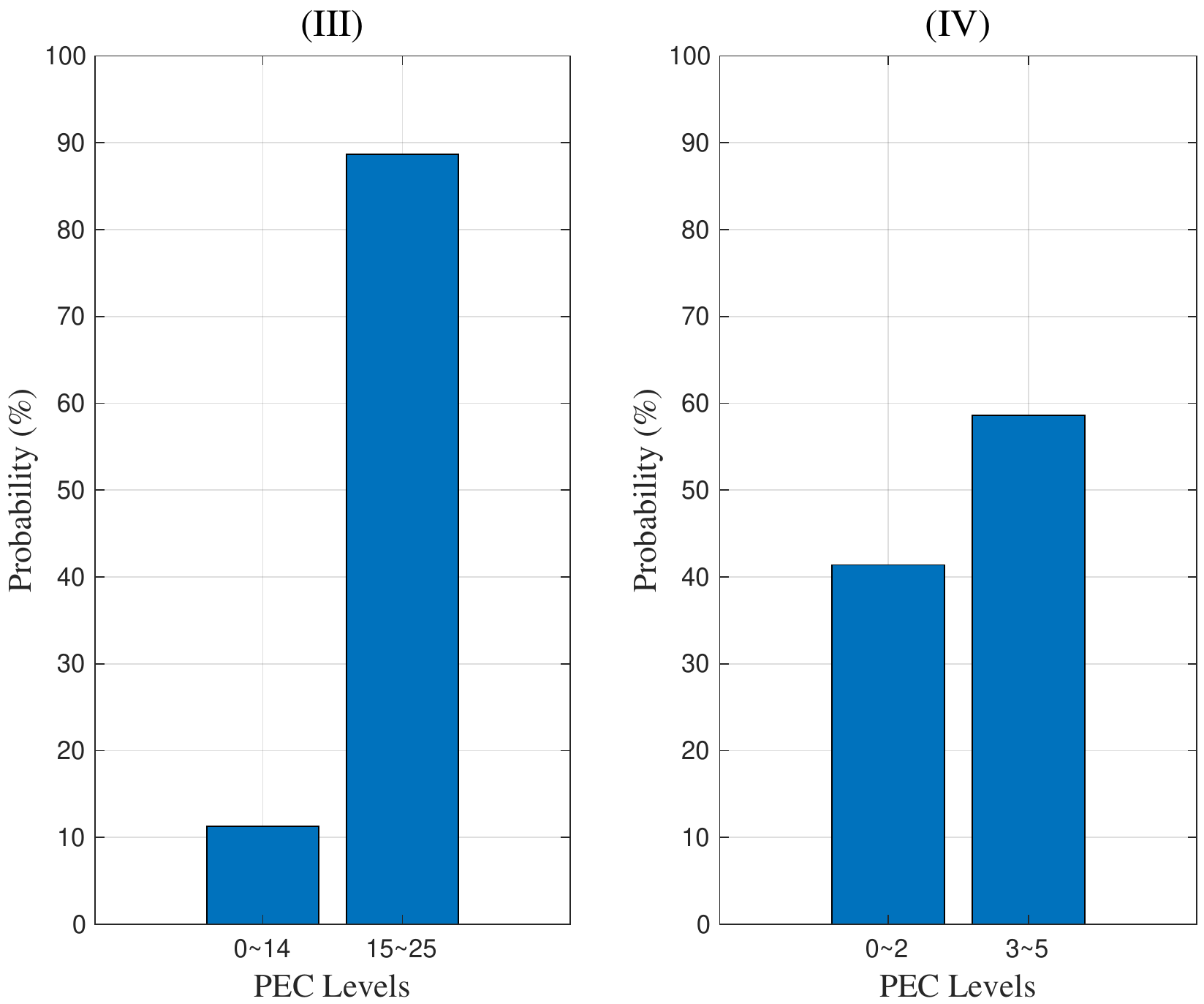}
\caption{Illustration of residual energy fluctuations, validation of the proposed MC analysis and the impact of energy discretisation levels.}
\vspace{-.8cm}
\end{figure}
In Figure 2, we illustrate the dynamic charge-discharge behavior of
the PEC (subfigure (I)) and the comparison of the steady state distribution
gained from the analytical framework in this section against those
generated through Monte Carlo simulation (subfigure (II)).
Note that for subfigure (I), (II) and (III), $L=25$, for subfigure (IV), $L=5$
the other system parameters are all the same among subfigures in Figure
2. The detailed system parameter setups in this figure is in line with that in Section VI. 
\begin{remrk}
The initial energy remained in the PEC is set to be empty, and
as the proposed HOR system runs with respect to (w.r.t.) block numbers,
the complex energy accumulation and consumption process can be clearly
traced as shown in subfigure (I). Observing subfigure (II), it is
confirmed that the proposed analytical model matches the actual distribution
very well, validating the effectiveness of analysis on the MC in this
section. 
\end{remrk}

\begin{remrk}
Comparing subfigures (III) and (IV), one can find that the larger
$L$ (i.e. the PEC levels) is, more likely the residual energy in the PEC can satisfy the ``energy requirement'' which is hereby quantified as that the residual energy in the PEC is greater than or equal to 60$\%$ of the PEC's capacity. This is reasonable for a two-fold reason:
1) the floor function (e.g., formulas (\ref{eq:DiscreEnerPEH}) and
(\ref{eq:DiscreEnerFS})) used to quantify the discretized amount
of energy absorbed by the PEC limits that the proposed energy discretization
model has to abandon the overflow energy assimilated; 2) the ceil
function (e.g., formula (\ref{eq:DiscreConsuENerFS})) applied to
quantify the discretized amount of energy consumed by the PEC restricts
that the proposed energy discretization model should quantify the
underflow amount of discretized energy used up by the PEC as an specific
integer, which means the proposed model consumes extra energy than
its actual counterpart. According to the aforementioned analysis,
we can conclude that the larger $L$ is, i.e., the finer the PEC is
mathematically discretized, the more efficiently manipulating of RF energy can reach. A subsequent influence of $L$ on wireless transmission performance can be found in details in Section VI. However,
there exists the inherent trade-off between the computation complexity
and energy manipulating efficiency of the proposed energy discretization model so that
the value of $L$ should be chosen carefully and delicately in the
practical application scenarios. 
\end{remrk}

Based on the MC analysis, we can mathematically track wireless transmission
performance of the proposed HOR protocol, like, the connection outage
probability, which will be clearly stated and analyzed in  Section
V. Besides, the inherent SNR requirement when the FD SWIPT is invoked
restricts that $\gamma_{\text{SD}}$ has to be less $\gamma_{th}$,
which puts congenital influences on the covert performance analysis
in Section IV.

\section{Covert Communication Performance Analysis}

In Section III, we investigated the stationary distribution of energy
states discretized at R's PEC, via energy discretization and finite-state
homogeneous MC. In this section, we will analyze the covert performance
of the proposed HOR protocol. Note that R only intends to broadcast
covert massage in the case of working in the FD SWIPT mode, because
there exists no solid cover in the PEH mode so that D which also plays
the role as\textit{ warden} can detect the arising of covert communication
easily. Thus, this paper focuses on the circumstance
in which D performs detection regarding covert communication only
in the case of FD SWIPT mode. In the PEH mode, R will not broadcast
covert massage and D ceases the detection. This consideration is reasonable
because the exact work mode R applies is an open consensus among all
nodes at the beginning of each specific transmission block. Note that in this section, the constraint $\gamma_{\text{SD}} \textless \gamma_{th}$ holds due to the nature of the FD SWIPT mode.

\subsection{Channel Uncertainty Model}

To investigate the impact of channel uncertainty on covert detection
performance at D, it is assumed that D gets an imperfect estimation
of the wireless channel R$\rightarrow$D and the imperfect channel
estimation model of D is formulated as
\begin{equation}
h_{\text{RD}}=\hat{h}_{\text{RD}}+\tilde{h}_{\text{RD}},
\end{equation}
where $\hat{h}_{\text{RD}}\sim\mathcal{CN}\left(0,\left(1-\beta\right)\Omega_{\text{RD}}\right)$
and $\tilde{h}_{\text{RD}}\sim\mathcal{CN}\left(0,\beta\Omega_{\text{RD}}\right)$
are independent complex Gaussian random variables (RVs) which represent
D's channel estimation and the corresponding estimation error, respectively.
It is worth noting that $\beta\in\left(0,1\right)$ measures the degree
of channel uncertainty and the aforementioned assumption of Gaussian
estimation error comes from the minimum mean square error (MMSE) estimation
method. Although the instantaneous knowledge of $h_{\text{RD}}$ that
D gains is incomplete and contains estimation error, we assume that
D does know the fading distribution to which $h_{\text{RD}}$ is subjected.

\subsection{Binary Detection at the Destination}

The source node S transmits wireless energy to charge R for gaining
assistance helping the main wireless transmission between S and D.
As one part of the main party, D performs detection regarding whether
R emits illegitimate information, i.e., covert message, under the
cover of legal forwarded version of source messages. Hence, apart
form receiving desired information form S and R, D also needs to perform
simple (binary) hypothesis test in which $\mathcal{H}_{0}$ means
the null hypothesis indicating that R does not transmit covert information
while $\mathcal{H}_{1}$ represents the alternative hypothesis implicating
that R does emit the covert message. In a specific transmission slot,
we define the False Alarm (i.e., type I error) probability by $\mathbb{P}_{\text{FA}}\ensuremath{\triangleq}\Pr\left(\mathcal{D}_{1}|\mathcal{H}_{0}\right)$
and the Missed Detection (i.e., type II error) probability by $\mathbb{P}_{\text{MD}}\ensuremath{\triangleq}\Pr\left(\mathcal{D}_{0}|\mathcal{H}_{1}\right)$,
where $\mathcal{D}_{1}$ and $\mathcal{D}_{0}$ represent the binary
decisions in favor of the occurrence of covert transmission or not,
respectively. Besides, the $a$ $priori$ probabilities of hypotheses
$\mathcal{H}_{0}$ and $\mathcal{H}_{1}$ are assumed to be equal
(i.e., both are 0.5) in this paper$\footnote{Note that the equal $a$ $priori$
probability assumption corresponds to the circumstance in which D
has no $a$ $priori$ knowledge on whether R emits covert message
or not and completely ignores R's covert transmission possibility.}$, which is a widely adopted assumption
in the field of covert communication. 
Following this assumption, the detection performance of D is measured
by the detection error probability $\mathbb{P}_{\text{E}}$, defined as
\begin{equation}
\mathbb{P}_{\text{E}}\ensuremath{\triangleq}\mathbb{P}_{\text{FA}}+\mathbb{P}_{\text{MD}}.\label{eq:Pr_E_Definition}
\end{equation}

For arbitrary $\epsilon>0$, we define R achieving covert communication
if any communication scheme exists satisfying $\mathbb{P}_{\text{E}}\geq1-\epsilon$.
Note that the lower bound on $\mathbb{P}_{\text{E}}$ characterizes
the necessary trade-off between the false alarms and missed detections
in a simple hypothesis test. Specifically, $\mathbb{P}_{\text{E}}\geq1-\epsilon$
represents the covert communication constraint and $\epsilon$ signifies
the covert requirement, cause a sufficiently small $\epsilon$ renders
any detector employed at D to be ineffective.

\subsection{Derivation and Analytics}

When a transmission block is determined to adopt the FD SWIPT mode,
D would like to keep an eye on whether R broadcasts covert massage
under the cover of the forwarded version of source information. In
the case of FD SWIPT mode, the received signals at D in the $\omega$-th
channel use within a transmission block can be expressed as 
\begin{equation}
\boldsymbol{y}_{\text{D}}\left[\omega\right]=\begin{cases}
\sqrt{P_{\text{S}}}h_{\text{SD}}\boldsymbol{x}_{\text{S}}\left[\omega\right]+\sqrt{P_{\text{R}}}h_{\text{RD}}\boldsymbol{x}_{\text{R}}\left[\omega\right]+
\boldsymbol{n}_{\text{D}}\left[\omega\right], & \mathcal{H}_{0}\\
\sqrt{P_{\text{S}}}h_{\text{SD}}\boldsymbol{x}_{\text{S}}\left[\omega\right]+\sqrt{P_{\text{R}}}h_{\text{RD}}\boldsymbol{x}_{\text{R}}\left[\omega\right]+
\sqrt{P_{\Delta}}h_{\text{RD}}\boldsymbol{x}_{\text{c}}\left[\omega\right]+\boldsymbol{n}_{\text{D}}\left[\omega\right], & \mathcal{H}_{1}
\end{cases}.
\end{equation}

\begin{lemma}
\label{Lemma_OptimalityofRadiometer}
A radiometer is utilized by D to perform the detection test monitoring
potential covert communications launched by R. In the case of availability
of noise power at D, it is approved that radiometer is the optimal
detector for covert communication detection.
\end{lemma}

\begin{IEEEproof}
See Appendix \ref{AppendixofLemmaOptimalityofRadiometer}.
\end{IEEEproof}

Applying a radiometer as the optimal detection strategy at D, closed-form
expressions of false alarm, missed detection and detection error probabilities
for any given threshold $\tau$ will be derived in the following Theorem.
Then, closed-form expressions of the optimal detection threshold $\tau$
and minimum detection error probability will be derived and given.
The impacts of imperfect channel estimation on minimum detection error
probability will be analyzed via discussion of its monotonicity w.r.t.
$\beta$.

\begin{thm}
\label{TheoremofClosedFormDEP}
For arbitrary threshold $\tau$, closed-form expressions of false
alarm and missed detection probabilities can be respectively given
by
\begin{equation}
\mathbb{P}_{\text{FA}}=\begin{cases}
\exp\left(\frac{j_{0}-\tau}{\beta P_{\text{R}}\Omega_{\text{RD}}}\right), & \tau\geq j_{0}\\
1, & \text{otherwise}
\end{cases},\label{eq:Pr_FA}
\end{equation}
\begin{equation}
\mathbb{P}_{\text{MD}}=\begin{cases}
1-\exp\left(\frac{j_{1}-\tau}{\beta\left(P_{\text{R}}+P_{\Delta}\right)\Omega_{\text{RD}}}\right), & \tau\geq j_{1}\\
0, & \text{otherwise}
\end{cases},\label{eq:Pr_MD}
\end{equation}
where $j_{0}=P_{\text{S}}\vert h_{\text{SD}}\vert^{2}+P_{\text{R}}\vert\hat{h}_{\text{RD}}\vert^{2}+\sigma_{\text{D}}^{2}$
and $j_{1}=P_{\text{S}}\vert h_{\text{SD}}\vert^{2}+\left(P_{\text{R}}+P_{\Delta}\right)\vert\hat{h}_{\text{RD}}\vert^{2}+\sigma_{\text{D}}^{2}$.
Furthermore, invoking (\ref{eq:Pr_E_Definition}), (\ref{eq:Pr_FA})
and (\ref{eq:Pr_MD}), we can derive closed-form expression of the
detection error probability, shown as 
\begin{equation}
\mathbb{P}_{\text{E}}=\begin{cases}
1, & \tau<j_{0}\\
\exp\left(\frac{j_{0}-\tau}{\beta P_{\text{R}}\Omega_{\text{RD}}}\right), & j_{0}\le\tau<j_{1}\\
1+\exp\left(\frac{j_{0}-\tau}{\beta P_{\text{R}}\Omega_{\text{RD}}}\right)-
\exp\left(\frac{j_{1}-\tau}{\beta\left(P_{\text{R}}+P_{\Delta}\right)\Omega_{\text{RD}}}\right), & \tau\geq j_{1}
\end{cases}.\label{eq:Pr_E}
\end{equation}
\end{thm}

\begin{IEEEproof}
See Appendix \ref{AppendixofTheoremofClosedFormDEP}.
\end{IEEEproof}
\begin{thm}
\label{OptimalDetectionThreshold}
The optimal detection threshold of D's radiometer, which is supposed
to minimize $\mathbb{P}_{\text{E}}$, is given by
\begin{equation}
\tau^{*}=\begin{cases}
j_{1}, & j_{1}\geq\tau_{k_{1}=0}\\
\tau_{k_{1}=0}, & j_{1}<\tau_{k_{1}=0}
\end{cases},\label{eq:Optimal_Tau}
\end{equation}
where
\begin{equation}
\tau_{k_{1}=0}=-\frac{\beta P_{\text{R}}\left(P_{\text{R}}+P_{\Delta}\right)\Omega_{\text{RD}}}{P_{\Delta}}\ln\frac{P_{\text{R}}}{P_{\text{R}}+P_{\Delta}}+
P_{\text{S}}\vert h_{\text{SD}}\vert^{2}+\sigma_{\text{D}}^{2}.\label{eq:tqu_k1=00003D0}
\end{equation}
\end{thm}

\begin{IEEEproof}
See Appendix \ref{AppendixofOptimalDetectionThreshold}.
\end{IEEEproof}
\begin{figure}
\centering
\includegraphics[scale=0.4]{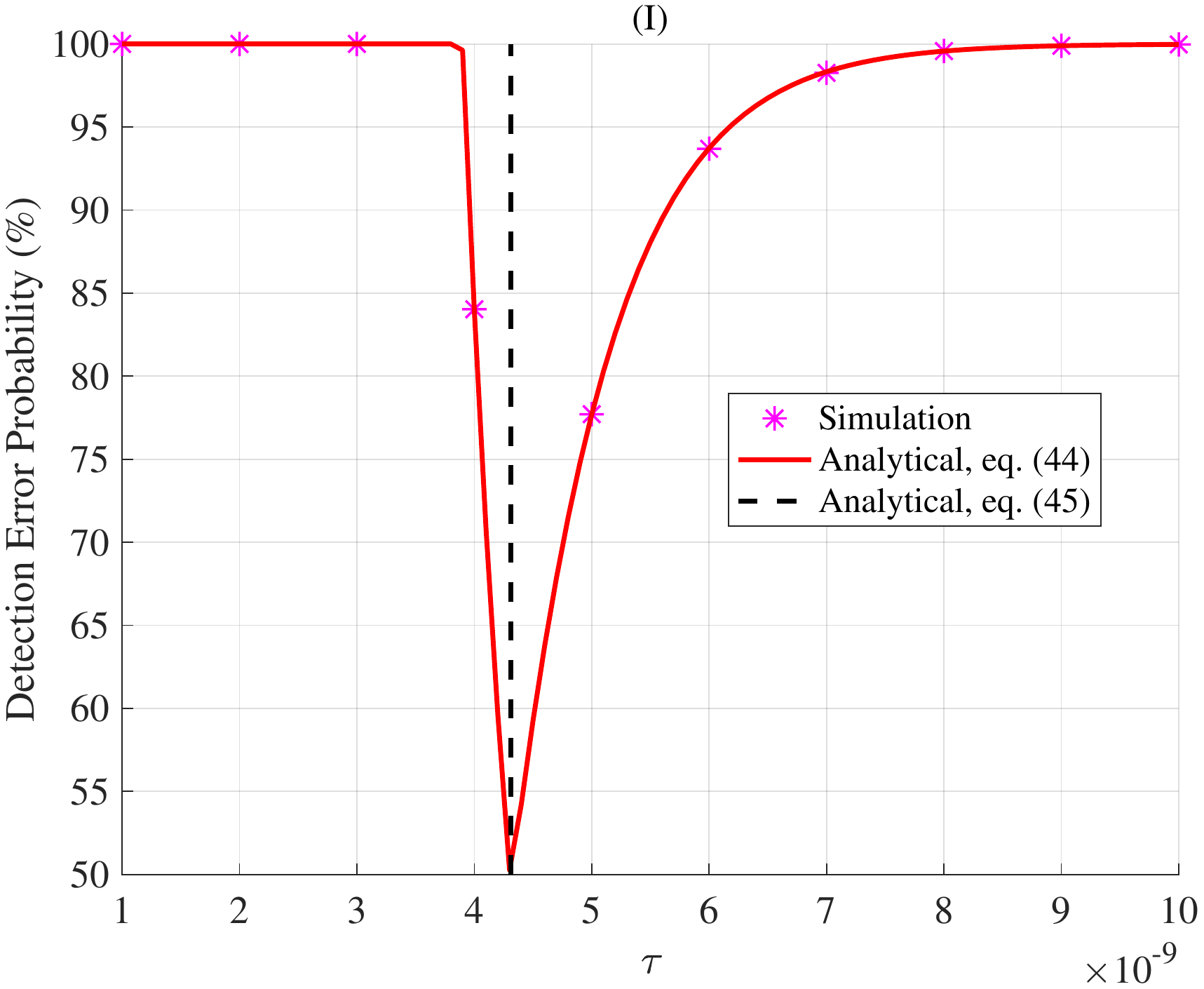}
\includegraphics[scale=0.42]{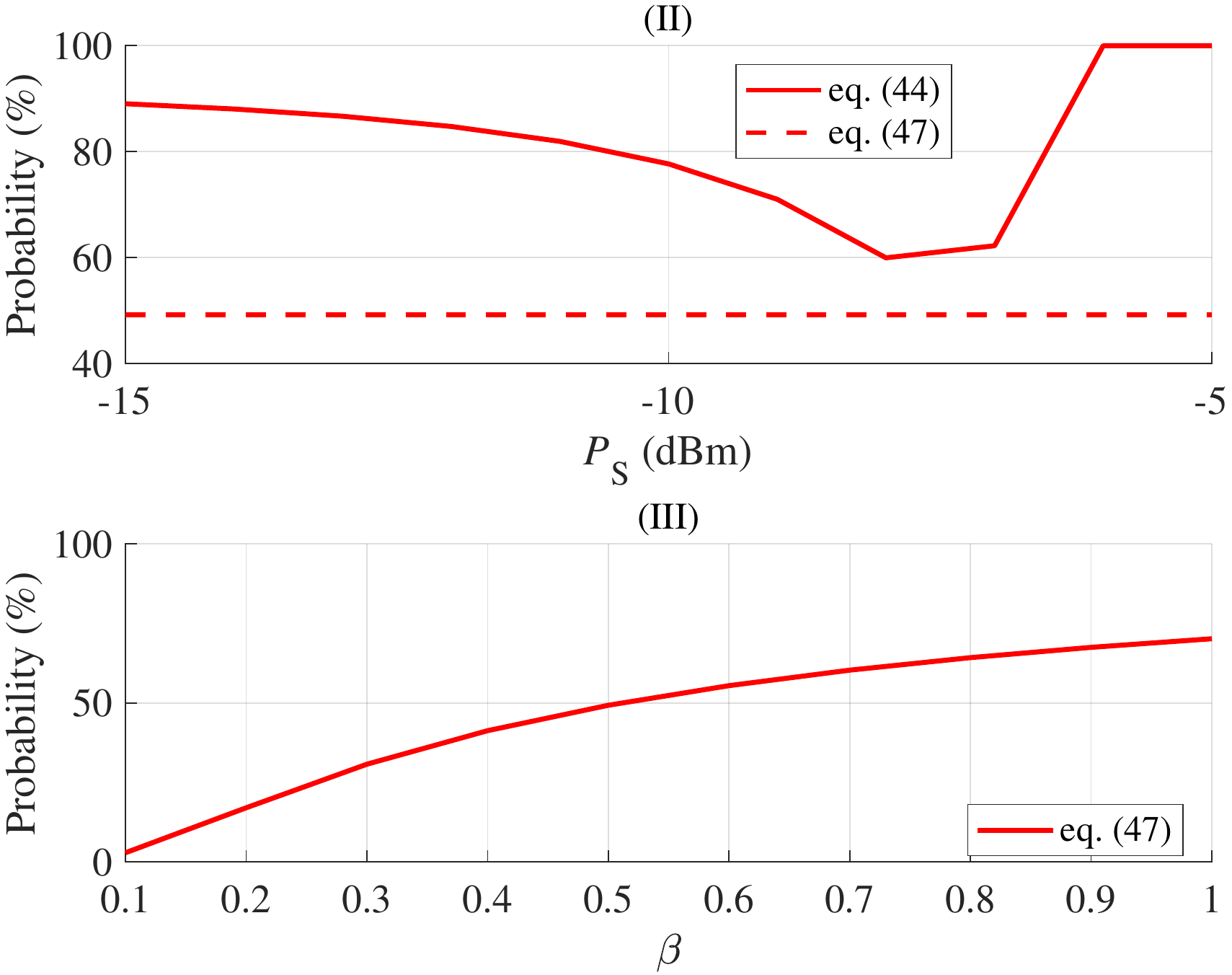}
\caption{Validation of the derived closed-from expressions of detection error probability and the optimal detection threshold, illustration of performance superiority of the proposed minimum detection error probability and its monotonicity w.r.t. $\beta$.}
\vspace{-.8cm}
\end{figure}

\begin{cor}
To achieve the best detection performance, D will always select the
optimal detection threshold as per (\ref{eq:Optimal_Tau}). Thus,
closed-form expression of minimum detection error probability can
be calculated as
\begin{equation}
\mathbb{P}_{\text{E}}^{*}=\begin{cases}
\exp\left(\frac{j_{0}-j_{1}}{\beta P_{\text{R}}\Omega_{\text{RD}}}\right), & j_{1}\geq\tau_{k_{1}=0}\\
1+\exp\left(\frac{j_{0}-\tau_{k_{1}=0}}{\beta P_{\text{R}}\Omega_{\text{RD}}}\right)-
\exp\left(\frac{j_{1}-\tau_{k_{1}=0}}{\beta\left(P_{\text{R}}+P_{\Delta}\right)\Omega_{\text{RD}}}\right), & j_{1}<\tau_{k_{1}=0}
\end{cases}.\label{eq:Minimum_P_E}
\end{equation}
\end{cor}

\begin{remrk}
According to \textbf{\textit{Theorem}} \textit{2}, \textbf{\textit{Theorem}} \textit{3} and \textbf{\textit{Corellary}} \textit{1}, it is confirmed that $\mathbb{P}_{\text{E}}$, $\tau^{*}$ and $\mathbb{P}_{\text{E}}^{*}$ are independent to parameters $k$, $L$, $\gamma_{th}$, $C_{\text{M}}$, $\eta$, $\eta'$, $\sigma_{\text{R}}^{2}$, $h_{\text{RR}}$ and $h_{\text{SR}}$. This is because, concisely speaking, covert communication is constrained to be possible only within the FD SWIPT mode, and parameters $C_{\text{P}}$ and $E_{th}$ can affect covert metrics in the manner of the aforementioned $P_{\text{R}}=E_{th}=0.6C_{\text{P}}$. Moreover, $\mathbb{P}_{\text{E}}^{*}$ is not subjected to $P_{\text{S}}$ and $h_{\text{SD}}$ either, because of subtractions of $j_{0}-j_{1}$, $j_{0}-\tau_{k1=0}$ and $j_{1}-\tau_{k1=0}$. This finding can guide designers to understand clearly what parameters are valid to pose impacts on covert communication detection performance. 
\end{remrk}

With the help of \textbf{\textit{Corollary}} \textit{1}, it is mathematically guaranteed
that the detection error probability at D is minimized on the perspective
of imperfect channel estimation. However, how does the factor $\beta$
influence the performance of minimum detection error probability?
This question motivates us to provide the following Corollary.
\begin{cor}
\label{CorIncreaingFunofBeta}
Minimum detection error probability $\mathbb{P}_{\text{E}}^{*}$ is monotonically increasing function w.r.t. $\beta$.
\end{cor}

\begin{IEEEproof}
See Appendix \ref{AppendixofCorIncreaingFunofBeta}.
\end{IEEEproof}

\begin{remrk}
Based on \textbf{\textit{Corollary}} \textit{2}, the imperfect channel estimation
is proved to be an important factor posing significant impacts on
$\mathbb{P}_{\text{E}}^{*}$. A smaller $\beta$, i.e., the better
channel estimation method, is desired to enhance the covert communication
detection performance at D.
\end{remrk}

To better show the covert communication performance analysis and verify the correctness of the corresponding analytical expressions, Figure 3 is illustrated in which $\beta=0.5$ stands unless otherwise specified and the other system parameters are set in line with those in Section VI. Note that in Figure 3, we evaluate covert metrics for arbitrarily selected transmission block pair in which $h_{\text{SD}}=-0.0010 - 0.0027j$ and $h_{\text{RD}}= 0.0261 + 0.0526j$ stand. From subfigure (I), the Monte Carlo simulation nodes mach perfectly with the analytical curve of (\ref{eq:Pr_E}) and the dash line generated from (\ref{eq:Optimal_Tau}) coincides tightly with the simulated optimal $\tau$'s coordinate, validating the correctness of the derived analytical expressions in \textbf{\textit{Theorem}} \textit{2}, \textbf{\textit{Theorem}} \textit{3}. Subfigure (II) depicts clearly that applying \textbf{\textit{Corellary}} \textit{1} can significantly reduce the detection error probability, compared with its counterpart without the optimal detection threshold. It can also be observed from subfigure (II) that the curve of $\mathbb{P}_{\text{E}}^{*}$ holds instant w.r.t. $P_{\text{S}}$, the reason was explained in  \textbf{\textit{Remark}} \textit{4}. Last but not least, subfigure (III) shows that $\mathbb{P}_{\text{E}}^{*}$ is a monotonically increasing function w.r.t. $\beta$, justifying the effectiveness of  \textbf{\textit{Corollary}} \textit{2} and  \textbf{\textit{Remark}} \textit{5}.

In this section, we analyzed the covert communication performance
by proving the optimality of radiometer on detection of potential
covert communication and deriving closed-form expressions of detection
error probability. Based on the mathematical analysis, we calculated
and stated closed-form expressions of the optimal detection threshold
and minimum detection error probability. Note that in this section,
we focused on the situation where the FD SWIPT mode is invoked at
R. For each particular transmission block which is located in the
domain of FD SWIPT, we provided closed-form expressions of the optimal
detection threshold and minimum detection probability from the perspective
of imperfect channel estimation of instantaneous wireless channel
between S and D, which means the values of (\ref{eq:Optimal_Tau})
and (\ref{eq:Minimum_P_E}) stand for this specific transmission block
and vary among different transmission blocks when the FD SWIPT mode
is activated. Besides, we would like to emphasize hereby that the
optimality of our analysis in this section is valid for any particular
wireless channel applications when the FD SWIPT mode is active. It
is also worth noting that the proposed HOR model inherently limits
$\gamma_{\text{SD}}<\gamma_{th}$ for the analysis in this section,
due to the SNR requirement.

\section{Transmission Outage Performance Analysis}
In this section, a typical transmission performance metrics, namely,
transmission outage probability (TOP) is derived and analyzed in details.
In this paper, we consider the circumstance in which D applies Maximum
Ratio Combination (MRC) protocol to combine the received signals from
S and R, when the FD SWIPT mode stands.

In the FD SWIPT mode, invoking (\ref{eq:y_DH0}) and (\ref{eq:y_DH1}),
the received SINR at D can be given by
\begin{equation}
\gamma_{\text{D}}=\begin{cases}
\gamma_{\text{SD}}+Y_{\mathcal{H}_{0}}, & \mathcal{H}_{0}\\
\gamma_{\text{SD}}+Y_{\mathcal{H}_{1}}, & \mathcal{H}_{1}
\end{cases},\label{eq:gamma_D}
\end{equation}
where
\begin{equation}
Y_{\mathcal{H}_{0}}=\min\left\{\frac{\left(1-\rho\right)P_{\text{S}}\vert h_{\text{SR}}\vert^{2}}{\left(1-\rho\right)kP_{\text{R}}\vert h_{\text{RR}}\vert^{2}+\sigma_{\text{R}}^{2}},\frac{P_{\text{R}}\vert h_{\text{RD}}\vert^{2}}{\sigma_{\text{D}}^{2}}\right\} ,\label{eq:Y_H0}
\end{equation}
\begin{equation}
Y_{\mathcal{H}_{1}}=\min\left\{ \frac{\left(1-\rho\right)P_{\text{S}}\vert h_{\text{SR}}\vert^{2}}{\left(1-\rho\right)k\left(P_{\text{R}}+P_{\Delta}\right)\vert h_{\text{RR}}\vert^{2}+\sigma_{\text{R}}^{2}},\frac{P_{\text{R}}\vert h_{\text{RD}}\vert^{2}}{P_{\Delta}\vert h_{\text{RD}}\vert^{2}+\sigma_{\text{D}}^{2}}\right\} .\label{eq:Y_H1}
\end{equation}
Note that the term $\min\left\{ \cdot,\cdot\right\} $ in (\ref{eq:Y_H0})
and (\ref{eq:Y_H1}) is introduced by the fixed DF relaying policy
applied at R \cite{zhao2016secrecy}. Knowing $\vert h_{\text{SR}}\vert^{2}\sim E\left(\Omega_{\text{SR}}\right)$,
$\vert h_{\text{RR}}\vert^{2}\sim E\left(\Omega_{\text{RR}}\right)$
and $\vert h_{\text{RD}}\vert^{2}\sim E\left(\Omega_{\text{RD}}\right)$,
closed-form CDF expressions of $Y_{\mathcal{H}_{0}}$ and $Y_{\mathcal{H}_{1}}$
can be calculated as

\begin{equation}
F_{Y_{\mathcal{H}_{\phi}}}\left(x\right)=\begin{cases}
1-\frac{P_{\text{S}}\Omega_{\text{SR}}\exp\left(-\left(\frac{\sigma_{\text{R}}^{2}}{\left(1-\rho\right)P_{\text{S}}\Omega_{\text{SR}}}+\frac{\sigma_{\text{D}}^{2}}{P_{\text{R}}\Omega_{\text{RD}}}\right)x\right)}{P_{\text{S}}\Omega_{\text{SR}}+kP_{\text{R}}\Omega_{\text{RR}}x}, & \phi=0\\
1-\frac{P_{\text{S}}\Omega_{\text{SR}}\exp\left(-\left(\frac{\sigma_{\text{R}}^{2}}{\left(1-\rho\right)P_{\text{S}}\Omega_{\text{SR}}}+\frac{\sigma_{\text{D}}^{2}}{\left(P_{\text{R}}-P_{\Delta}x\right)\Omega_{\text{RD}}}\right)x\right)}{P_{\text{S}}\Omega_{\text{SR}}+k\left(P_{\text{R}}+P_{\Delta}\right)\Omega_{\text{RR}}x}, & \phi=1\&\&x<\frac{P_{\text{R}}}{P_{\Delta}}\\
1, & \phi=1\&\&x\geq\frac{P_{\text{R}}}{P_{\Delta}}
\end{cases}.\label{eq:ClosedForm_CDF_Y_H_Phi}
\end{equation}

\begin{lemma}\label{LemmaofClosedFormCDFofGammadH0}
Closed-form expression of CDF of $\gamma_{\text{D}}|\mathcal{H}_{0}$ can be derived as
\begin{align}
F_{\gamma_{\text{D}}|\mathcal{H}_{0}}\left(x\right)&=q_{\text{SD}}-v_{1}\left[\text{Ei}\left(v_{3}\right)-\text{Ei}\left(v_{4}\right)\right]\nonumber \\
&\times\exp\left(\frac{P_{\text{S}}\Omega_{\text{SR}}\left(\frac{\sigma_{\text{R}}^{2}}{\left(1-\rho\right)P_{\text{S}}\Omega_{\text{SR}}}+\frac{\sigma_{\text{D}}^{2}}{P_{\text{R}}\Omega_{\text{RD}}}\right)-\frac{\sigma_{\text{D}}^{2}}{P_{\text{S}}\Omega_{\text{SD}}}\left(P_{\text{S}}\Omega_{\text{SR}}+kP_{\text{R}}\Omega_{\text{RR}}x\right)}{kP_{\text{R}}\Omega_{\text{RR}}}\right),\label{eq:ClosedForm_CDF_Gamma_D_H_0}
\end{align}
where $\text{Ei}\left(\cdot\right)$
represents the one-argument Exponential Integral function. For concise
expression, we define the following variables in (\ref{eq:ClosedForm_CDF_Gamma_D_H_0})
as 
\begin{equation}
v_{1}=\frac{\sigma_{\text{D}}^{2}\Omega_{\text{SR}}}{kP_{\text{R}}\Omega_{\text{RR}}\Omega_{\text{SD}}},
\end{equation}
\begin{equation}
v_{2}=\frac{\frac{\sigma_{\text{D}}^{2}}{P_{\text{S}}\Omega_{\text{SD}}}-\frac{\sigma_{\text{R}}^{2}}{\left(1-\rho\right)P_{\text{S}}\Omega_{\text{SR}}}-\frac{\sigma_{\text{D}}^{2}}{P_{\text{R}}\Omega_{\text{RD}}}}{kP_{\text{R}}\Omega_{\text{RR}}},
\end{equation}
\begin{equation}
v_{3}=v_{2}\left(P_{\text{S}}\Omega_{\text{SR}}+kP_{\text{R}}\Omega_{\text{RR}}x\right),
\end{equation}
\begin{equation}
v_{4}=v_{2}\left(P_{\text{S}}\Omega_{\text{SR}}+kP_{\text{R}}\Omega_{\text{RR}}\left(x-\gamma_{th}\right)\right).
\end{equation}
\end{lemma}

\begin{IEEEproof}
See Appendix \ref{AppendixofLemmaofClosedFormCDFofGammadH0}.
\end{IEEEproof}

\begin{lemma}
\label{LemmaofApproxCDFofGammadH1}
Closed-form CDF expression of $\gamma_{\text{D}}|\mathcal{H}_{1}$
in the case of FD SWIPT mode can be derived approximately as
\begin{equation}
F_{\gamma_{\text{D}}|\mathcal{H}_{1}}\left(x\right)\thickapprox\text{quadgk}\left(\text{fun}\left(y\right),0,\gamma_{th}\right),\label{eq:ClosedForm_Appromy_CDF_Gamma_D_H_1}
\end{equation}
where the definitions of $\text{quadgk}(\cdot,\cdot,\cdot)$ and $\text{fun}\left(y\right)$
can be found in the following proof.
\end{lemma}

\begin{IEEEproof}
See Appendix \ref{AppendixofLemmaofApproxCDFofGammadH1}.
\end{IEEEproof}
\begin{remrk}
In \textbf{\textit{Lemma}} \textit{3}, the approximation of $F_{\gamma_{\text{D}}|\mathcal{H}_{1}}$ is achieved by converting infinite integral to finite summation. The accuracy of this approximation is mainly affected by the amount of nodes used within the finite summation, the more nodes is applied, the more complex the summation is, though the preciser approximation it can achieve.
\end{remrk}
\begin{thm}
\label{TheoremofTOPinFS}
Closed-form expression of the TOP in the FD SWIPT mode can be given
by
\begin{equation}
TOP_{\text{FS}}=
\frac{1}{2}\sum_{i=\varphi}^{L}\xi_{i}\left[F_{\gamma_{\text{D}}|\mathcal{H}_{0}}\left(2^{R_{th}}-1\right)+F_{\gamma_{\text{D}}|\mathcal{H}_{1}}\left(2^{R_{th}}-1\right)\right].\label{eq:ClosedForm_TOP_FS}
\end{equation}
\end{thm}

\begin{IEEEproof}
See Appendix \ref{AppendixofTheoremofTOPinFS}.
\end{IEEEproof}

\begin{thm}
\label{TheoremofTOPinPEH}
Closed-form expression of the TOP in the PEH mode can be given as
\begin{equation}
TOP_{\text{PEH}}=q_{\text{SD}}\sum_{i=0}^{\varphi-1}\xi_{i}+F_{\gamma_{\text{SD}}|\gamma_{\text{SD}}\geq\gamma_{th}}\left(2^{R_{th}}-1\right),\label{eq:ClosedForm_TOP_PEH}
\end{equation}
where the concept of $F_{\gamma_{\text{SD}}|\gamma_{\text{SD}}\geq\gamma_{th}}\left(x\right)$
can be found in the following proof.
\end{thm}

\begin{IEEEproof}
See Appendix \ref{AppendixofTheoremofTOPinPEH}.
\end{IEEEproof}

\begin{cor}
Finally, invoking (\ref{eq:ClosedForm_TOP_FS}) and (\ref{eq:ClosedForm_TOP_PEH}),
closed-form expression of overall TOP for our proposed HOR model can
be derived as
\begin{align}
TOP=q_{\text{SD}}\sum_{i=0}^{\varphi-1}\xi_{i}&+F_{\gamma_{\text{SD}}|\gamma_{\text{SD}}\geq\gamma_{th}}\left(2^{R_{th}}-1\right)\nonumber \\
&+\frac{1}{2}\sum_{i=\varphi}^{L}\xi_{i}\left[F_{\gamma_{\text{D}}|\mathcal{H}_{0}}\left(2^{R_{th}}-1\right)+F_{\gamma_{\text{D}}|\mathcal{H}_{1}}\left(2^{R_{th}}-1\right)\right].
\end{align}
\end{cor}

In this section, the developed closed-form expression of the TOP is indeed in a form of complicated composition, from which the impacts of system parameters on the TOP performance are impossible to be unveiled and discussed thoroughly. To analyse the TOP performance of the proposed HOR system and thus highlight the superiority of the HOR scheme as well as the impacts of various system parameters on the TOP performance, we pose detailed investigation via showing numerical results in Section VI.

\section{Numerical Results}

In this section, applying the analytical expressions derived in the
previous contents, numerical results will be performed and the impact
of key parameters on the performance will also be investigated. The simulation is deployed in a 2-dimensional (2D) topology where all the nodes are placed with the same altitude, i.e., terrestrial relaying scenario. Unless otherwise specified, the simulation results are based on the following parameter setups. The distances among nodes are allocated as $d_{{\text{SD}}}=15$ m, $d_{{\text{SR}}}=8$ m, $d_{\text{RR}}=0.1$ m and $d_{{\text{RD}}}=8$ m, where it is reasonable to consider that the distance between R's dual antennas is relatively near. We set the average wireless channel gains as $\Omega_{ij}=1/(1+d^{\alpha}_{ij}), \left\{i.j\right\}\in\left\{\text{S, R, D}\right\}$ where the path loss exponent is predefined as $\alpha=3$, the AWGN powers $\sigma_{\text{R}}^{{2}}=\sigma_{\text{D}}^{{2}}=-60$ dBm, the target transmission rate $R_{th}=1$ bps/Hz, the SNR threshold $\gamma_{th}=1$, the energy threshold $E_{th}=0.6C_{\text{P}}$, the transmit power of S $P_{\text{S}}=-10$ dBm, the PS factor $\rho=0.5$ and the covert transmitting power $P_{\Delta}=0.2P_{\text{R}}$. Regarding parameters of the hybrid energy storage, we set $C_{\text{P}}=C_{\text{M}}=10^{-6}$ Joule, the energy conversion efficiency $\eta=0.4$, the energy transfer coefficient $\eta'=0.9$ and the discretisation level $L=25$.

\begin{figure}[!htb]
   \centering \includegraphics[width = 0.48\linewidth]{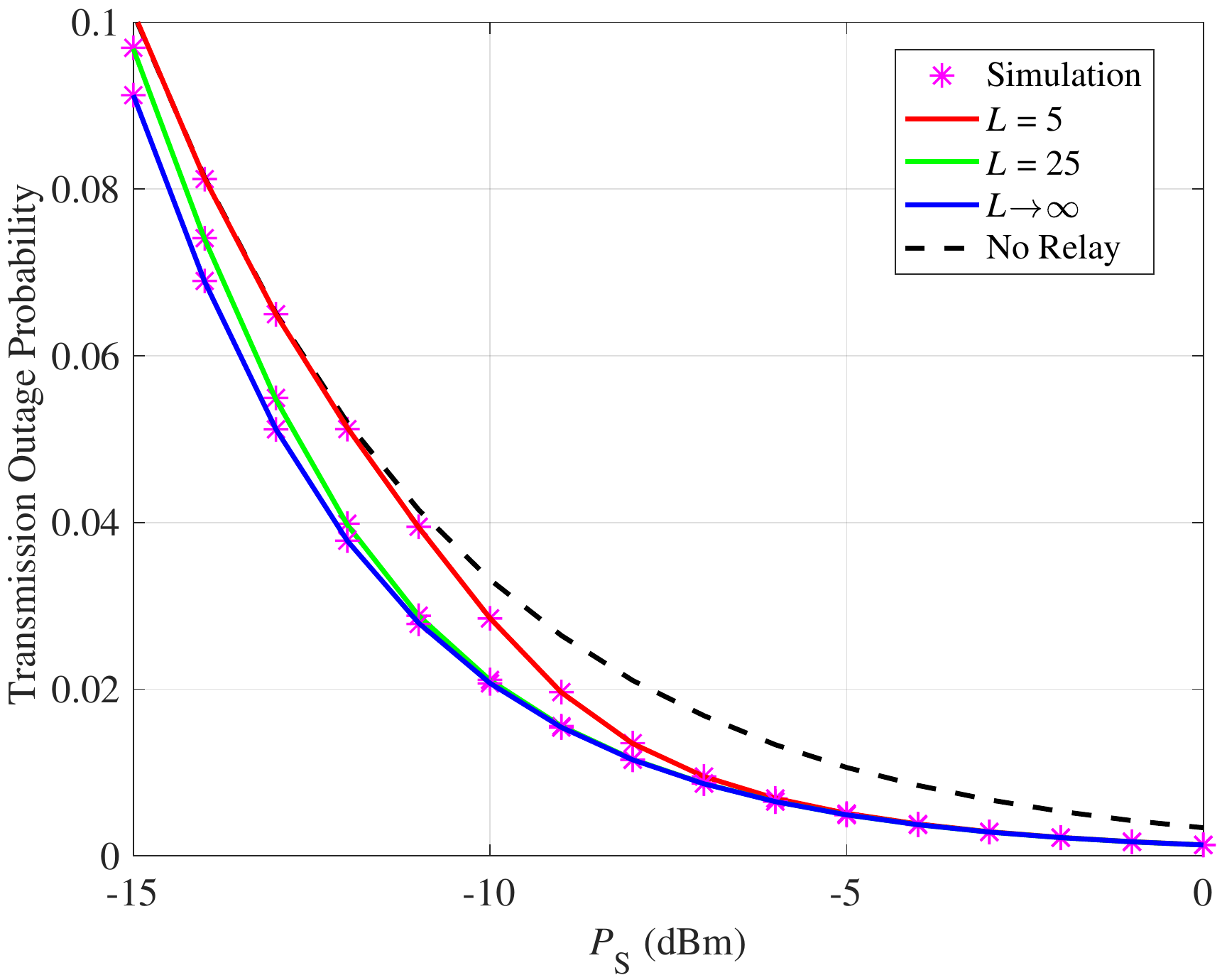}
    \caption{Transmission outage probability versus $P_{\text{S}}$ with various $L$ values.}
\end{figure}
\subsection{Validation of The Proposed Energy Discretization Method}
In this part, we validate the feasibility and accuracy of the proposed discrete energy model described in Section III, by showing curves generated from the MC based TOP analysis and the corresponding Monte Carlo simulation points. Figure 4 depicts curves of the TOP versus $P_{\text{S}}$ with different energy discretisation levels. Note that $L\rightarrow\infty$ serves as upper bound of the TOP performance, in the case of a massive energy discretisation. It can be observed from Figure 4 that even a small energy discretisation level ($L=5$) is enough to provide considerable TOP performance gain for majority of the simulated $P_{\text{S}}$ regime, compared to the circumstance in which no relay assists wireless communication between S and D. Comparing the TOP performance curves of various $L$ values, one can conclude that the TOP performance approaches the upper bound gradually as the value of $L$ increases. The reason why $L$ can affect the HOR system has been explained in details in \textbf{\textit{Remark}} \textit{3}. Specifically, the TOP performance curve when $L$'s value is not so large, i.e., $L=25$ can coincide with the upper bound in the most region of simulated $P_{\text{S}}$. The aforementioned observations validates the effectiveness of the proposed HOR system on helping wireless transmissions between devices, even with practical energy discretisation levels ($L=5,L=25$).
\begin{figure}[!htb]
\begin{tabular}{cc}
\begin{minipage}[t]{0.48\linewidth}
    \includegraphics[width = 1\linewidth]{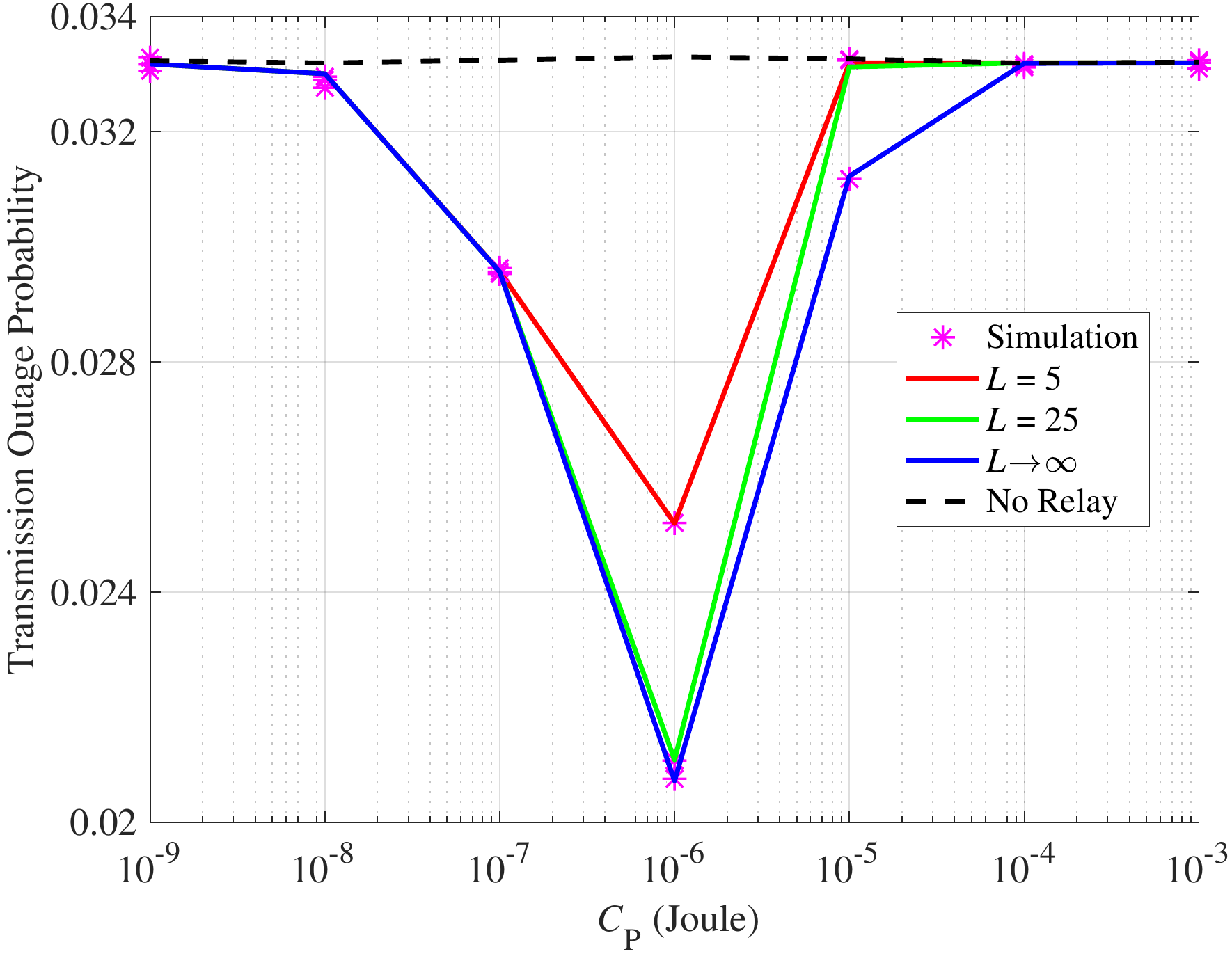}
    \caption{Transmission outage probability versus $C_{\text{P}}$ with various $L$ values.}
\end{minipage}
\begin{minipage}[t]{0.48\linewidth}
    \includegraphics[width = 1\linewidth]{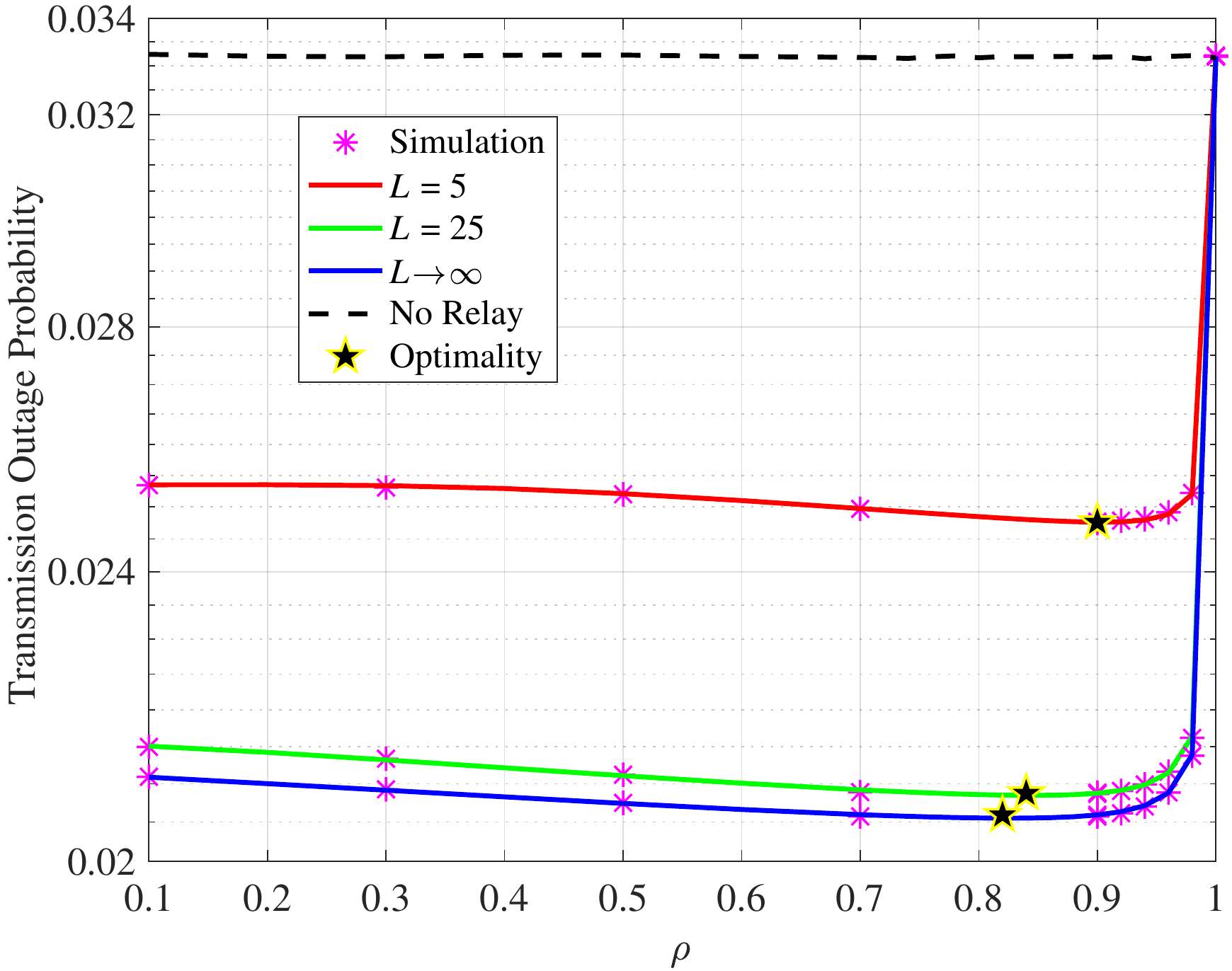}
    \caption{Transmission outage probability versus $\rho$ with various $L$ values.}
\end{minipage}
\end{tabular}
\end{figure}
\subsection{The Impact of Capacity of The PEC}
In this subsection, we examine that how $C_{\text{P}}$ influences the TOP performance. Figure 5 shows the TOP curves versus $C_{\text{P}}$ with various $L$ values. It is straightforward to find that for specific HOR system parameter setup, there exists optimal value of $C_{\text{P}}$ to minimise the TOP performance. The existence of the optimal $C_{\text{P}}$ is because, briefly speaking, it influences the values of $P_{\text{R}}$ and $E_{th}$ by the means of $P_{\text{R}}=E_{th}=0.6C_{\text{P}}$. Under the system parameter setup of this example, the values of $L$ does not pose any impact on value of the optimal $C_{\text{P}}$. It can be observed that $L=25$ can almost act as a feasible alternative of the TOP performance's upper bound, revealing the efficiency of the proposed energy discretisation model. The observation of this example allows the system designer to determine an optimal $C_{\text{P}}$ while reducing computation by selecting a small but sufficient $L$, for various system parameter setups.
\subsection{The Impact of The PS Factor}
In this part, we investigate the impact of $\rho$ on the TOP performance. Figure 6 demonstrates the TOP curves versus $\rho$ with various $L$ values. Alongside all the possible values of $\rho$ towards $\rho=1$, we can find that the TOP curves first decreases, reach the optimality and then rapidly rocket to the worst case at which performance gain offered by the proposed HOR protocol does not exist any more. The existence of the optimality is because the inherent trade-off at R between harvesting more energy and gaining stronger received SNR of the signals from S. Also, one can find that the energy discretisation levels does pose impact on the value of the optimality. Specifically, a larger $L$ leads to a smaller value of the optimality. It does make sense because a larger $L$ can reduce the energy loss in the proposed energy discretisation model based on the discussion in \textbf{\textit{Remark}} \textit{3} so that R has the space to pose more efforts on information processing.
\begin{figure}[!htb]
\begin{tabular}{cc}
\begin{minipage}[t]{0.48\linewidth}
    \includegraphics[width = 1\linewidth]{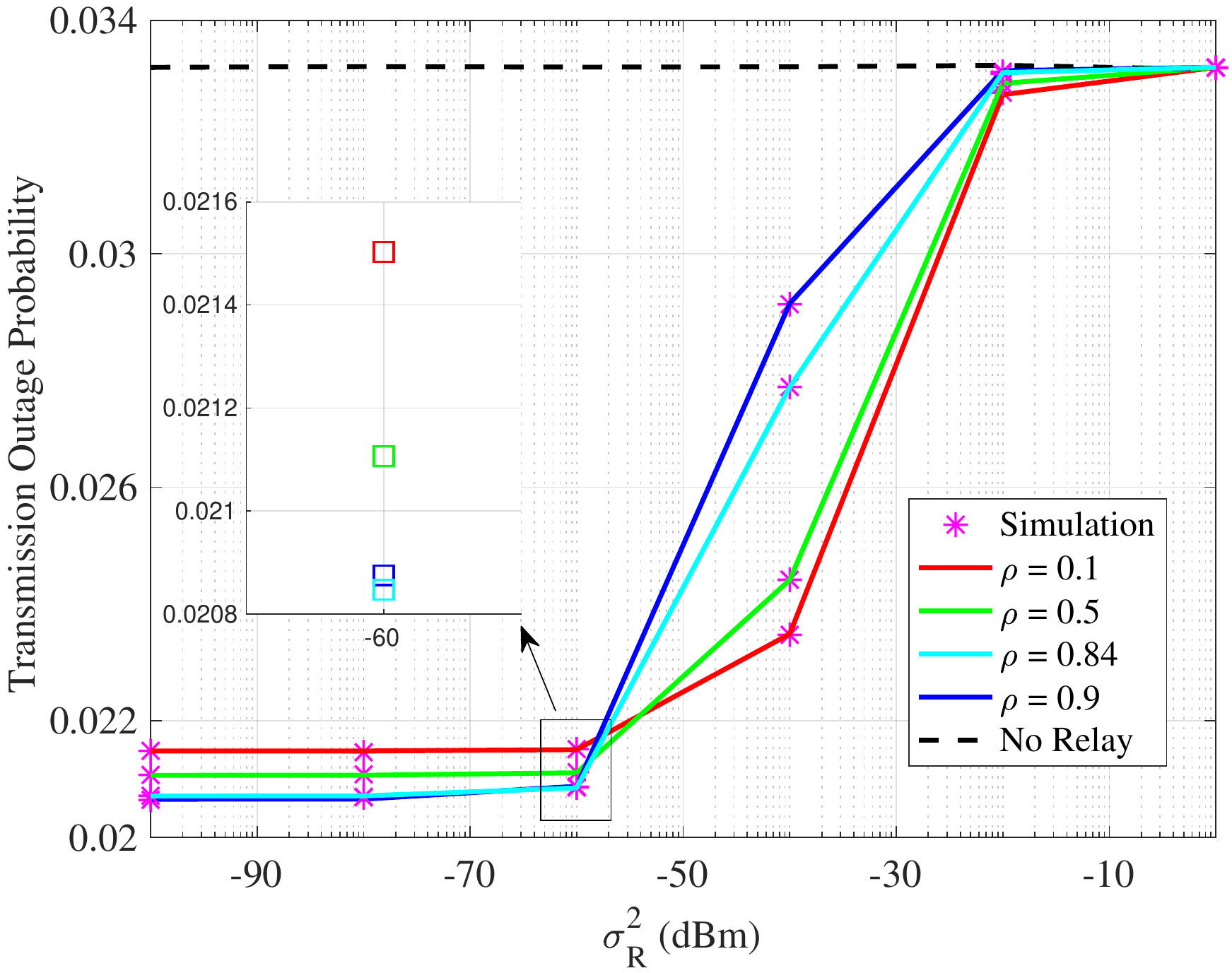}
    \caption{Transmission outage probability versus $\sigma^{2}_{\text{R}}$ with various $\rho$ values.}
\end{minipage}
\begin{minipage}[t]{0.48\linewidth}
    \includegraphics[width = 1\linewidth]{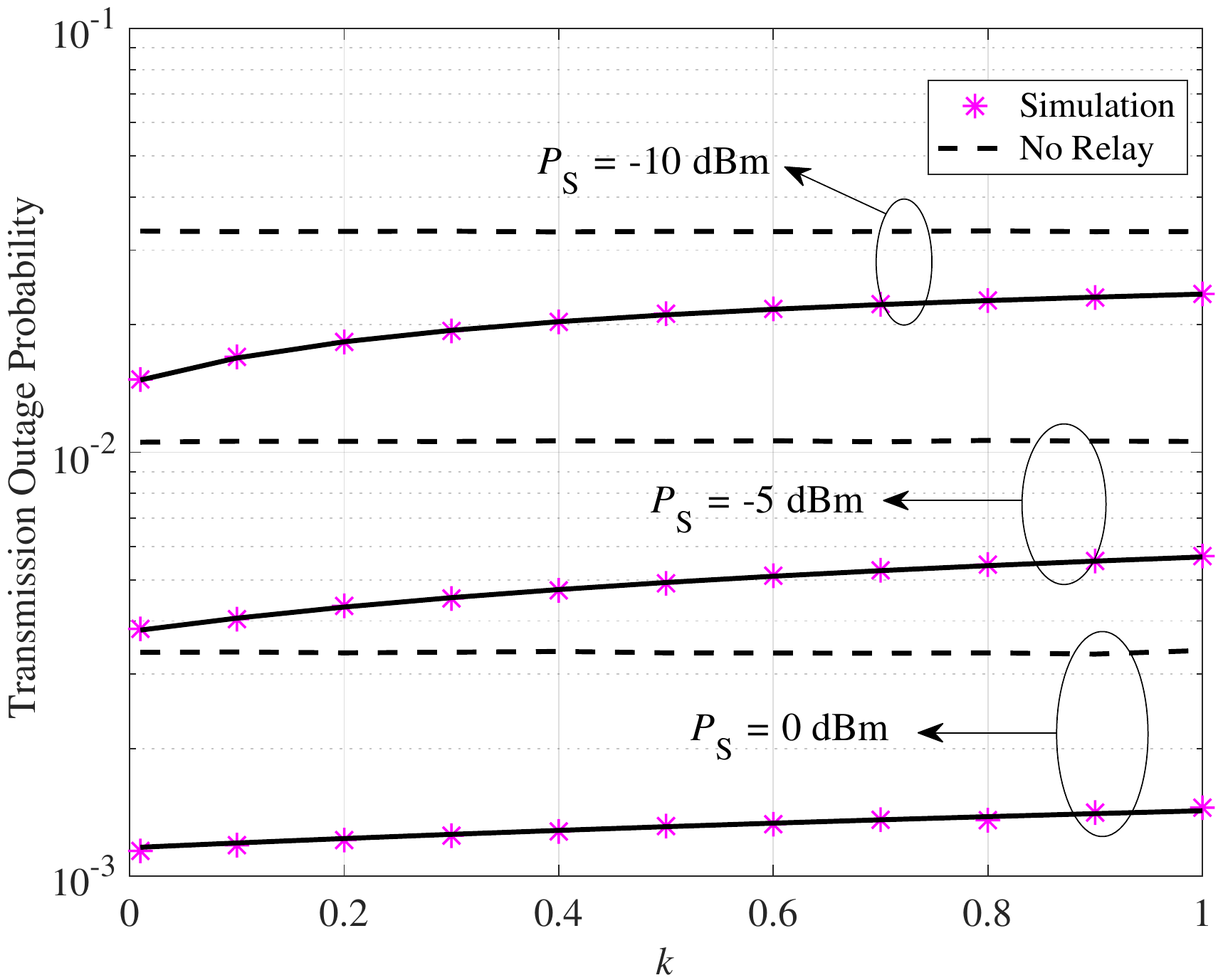}
    \caption{Transmission outage probability versus $k$ with various $P_{\text{S}}$ values.}
\end{minipage}
\end{tabular}
\end{figure}
\subsection{The Impact of R's AWGN Power}
In this subsection, we show the influence of $\sigma^{2}_{\text{R}}$ on the TOP performance. Figure 7 depicts the TOP curves versus $\sigma^{2}_{\text{R}}$ with various values of $\rho$. From the figure, it is straightforward to conclude that the TOP performance is getting worse with the increasing of $\sigma^{2}_{\text{R}}$. Specifically, when R is less or equally ``noisy'' than D, i.e., in the case of $\sigma^{2}_{\text{R}} \leq \sigma^{2}_{\text{D}}$, the TOP performance remains static at the minimum value. On the contrary, a ``noisier'' R will lead to the loss of performance gain offered by the proposed HOR system. This is because, in short, the min function introduced by the DF relaying strategy in formulas (\ref{eq:Y_H0}) and (\ref{eq:Y_H1}) forces the overall received SNR $\gamma_{\text{D}}$ to behave the segmentation feature. Besides, with the increasing of $\sigma^{2}_{\text{R}}$, the impact of $\rho$ on the TOP performance gradually becomes negligible, e.g., in the case of $\rho\in\left[-20,0\right]$ dBm. This is because, at this moment, $Y_{\mathcal{H}_{i}}, i\in\left\{0,1\right\}$ is way too small compared with $\gamma_{\text{SD}}$. Moreover, we give the detailed illustration in the case of $\sigma^{2}_{\text{R}}=-60$ dBm. At this point, the TOP performance of $\rho=0.84$ (the empirical optimal PS factor from Figure 6) is superior to that of $\rho=0.9$, validating the existing of the optimal $\rho$ which was found and discussed in the aforementioned Subsection $C$.
\subsection{The Impact of SIC Strength}
In this part, we examine how $k$ can affect the TOP performance. Figure 8 shows the TOP curves versus $k$ with various $P_{\text{S}}$ values. It is direct to find from this figure that the TOP performance is becoming worse with the increasing of $k$, for all simulated $P_{\text{S}}$ setups. The reason is that a larger $k$ means a stronger SI which suppresses the received SNR of R more. Although a larger $k$ can lead R to harvest more energy from the loop SI channel, from Figure 8, it is still better to pursue a good SIC efficiency, i.e., a smaller value of $k$, when implementing the proposed HOR system. Besides, with a higher $P_{\text{S}}$, the impact of $k$ becomes less obvious. This is because the strengths of both energy harvested from the loop SI channel and the interference caused by the SI link become minor, in the front of a high value of $P_{\text{S}}$, which is determined by formulas (\ref{eq:EnerFS}), (\ref{eq:Y_H0}) and (\ref{eq:Y_H1}).
\begin{figure}[!htb]
\begin{tabular}{cc}
\begin{minipage}[t]{0.48\linewidth}
    \includegraphics[width = 1\linewidth]{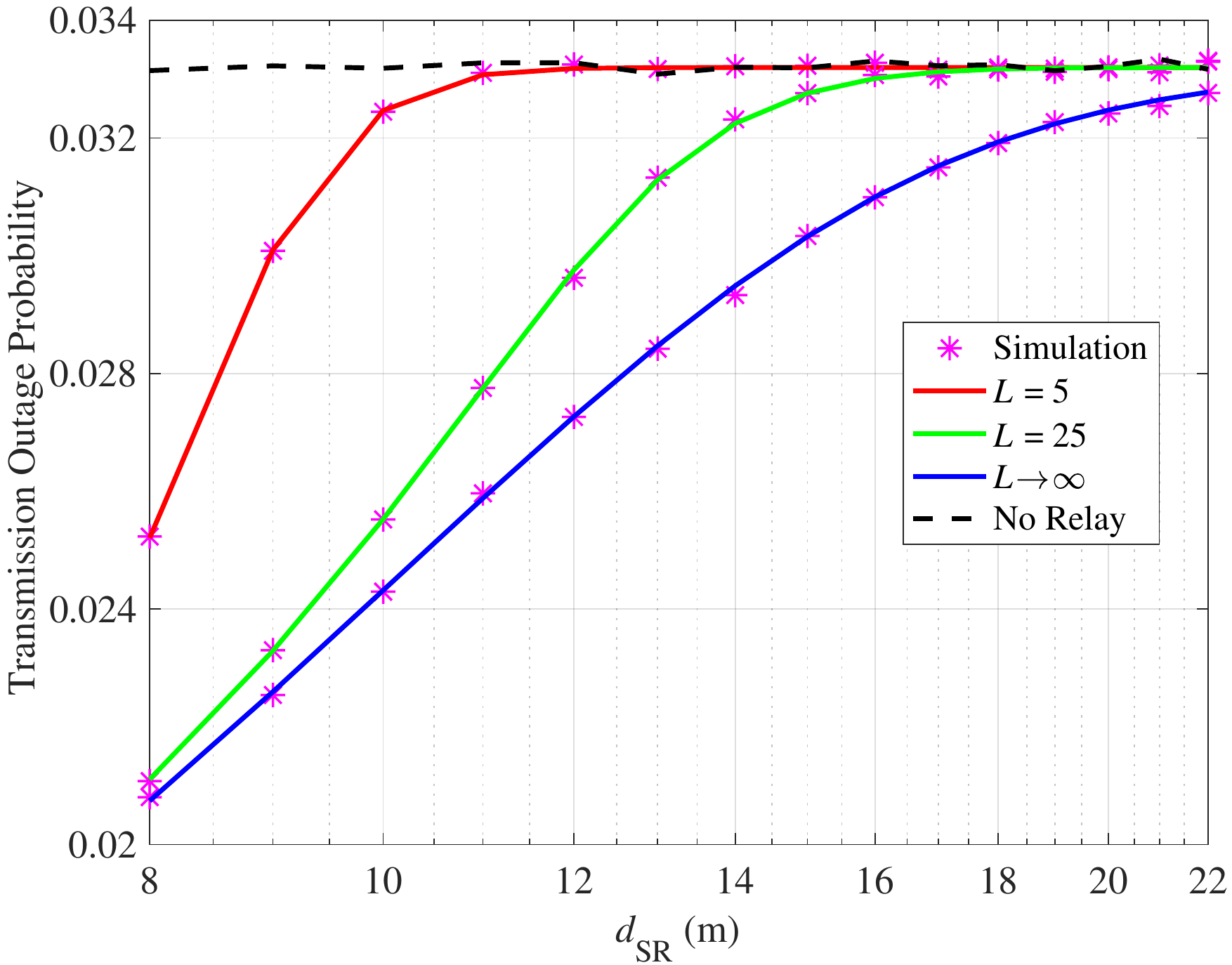}
    \caption{Transmission outage probability versus $d_{\text{SR}}$ with various $L$ values.}
\end{minipage}
\begin{minipage}[t]{0.48\linewidth}
    \includegraphics[width = 1\linewidth]{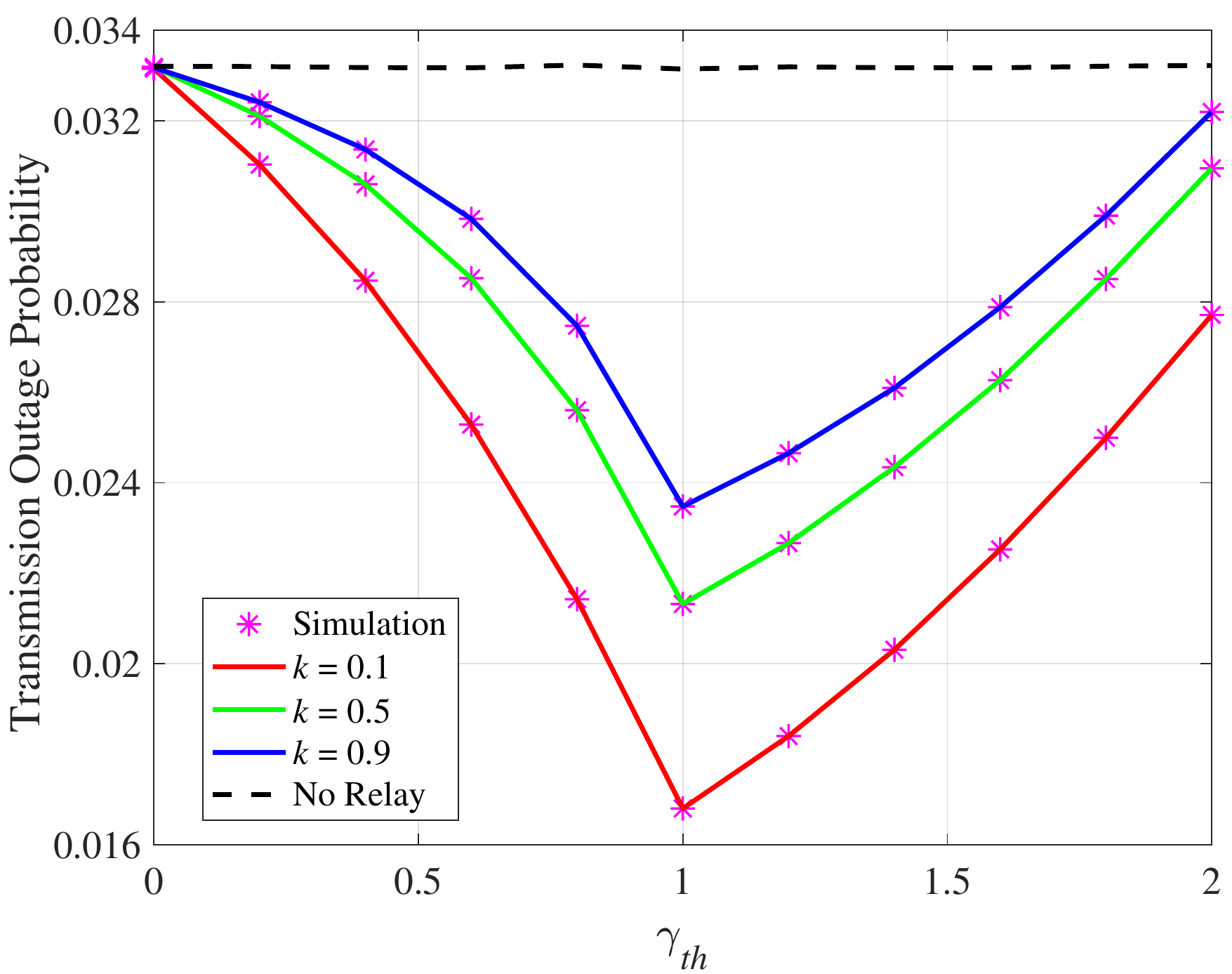}
    \caption{Transmission outage probability versus $\gamma_{th}$ with various $k$ values.}
\end{minipage}
\end{tabular}
\end{figure}
\subsection{The Impact of The Distance Between S and R}
In this subsection, we discuss the impact of $d_{\text{SR}}$ on the TOP performance. Figure 9 illustrates the TOP curves versus $d_{\text{SR}}$ with various values of $L$. Under the subjective of the Triangle Side Length Rule, the possible length of $d_{\text{SR}}$ should locates in $d_{\text{SR}}\in\left(7,23\right)$ m. From Figure 9, it is straightforward to find that no matter what value $L$ is, a reasonable shorter distance between S and R is always preferred for achieving more TOP performance gain. The reason is simply because the amount of harvested energy is very sensitive to $d_{\text{SR}}$, which can be found in the assumption of $\Omega_{\text{SR}}=1/(1+d^{3}_{\text{SR}})$. From this figure, the approaching speed of the TOP curves to the ``No Relay'' line is much slower for a larger $L$, validating the discussion in \textbf{\textit{Remark}} \textit{3}.
\subsection{The Impact of The SNR Threshold}
In this part, we analyse how the value of $\gamma_{th}$ affects the TOP performance. Figure 10 depicts the TOP curves versus $\gamma_{th}$ with different $k$ values. From this figure, one can observe that there exists an optimal value of $\gamma_{th}$ which can minimise the TOP curves. This is because, concisely speaking, the value of $\gamma_{th}$ directly influences the occurrence frequency of the FD SWIPT mode, which is determined by the activation condition as $\left\{\gamma_{\textnormal{SD}}<\gamma_{th}\right\}\cap \left\{E_{i}\geq E_{th}\right\}$. The dilemma of ``never or less frequently using R'' and ``using R too much'' makes the optimal $\gamma_{th}$ standing. Besides, the optimal value of $\gamma_{th}$ is independent to $k$. However, a more solid SIC degree, i.e., a smaller $k$, is still preferable, which is consistent with the discussion in Subsection $E$.
\begin{figure}[!htb]
\begin{tabular}{cc}
\begin{minipage}[t]{0.48\linewidth}
    \includegraphics[width = 1\linewidth]{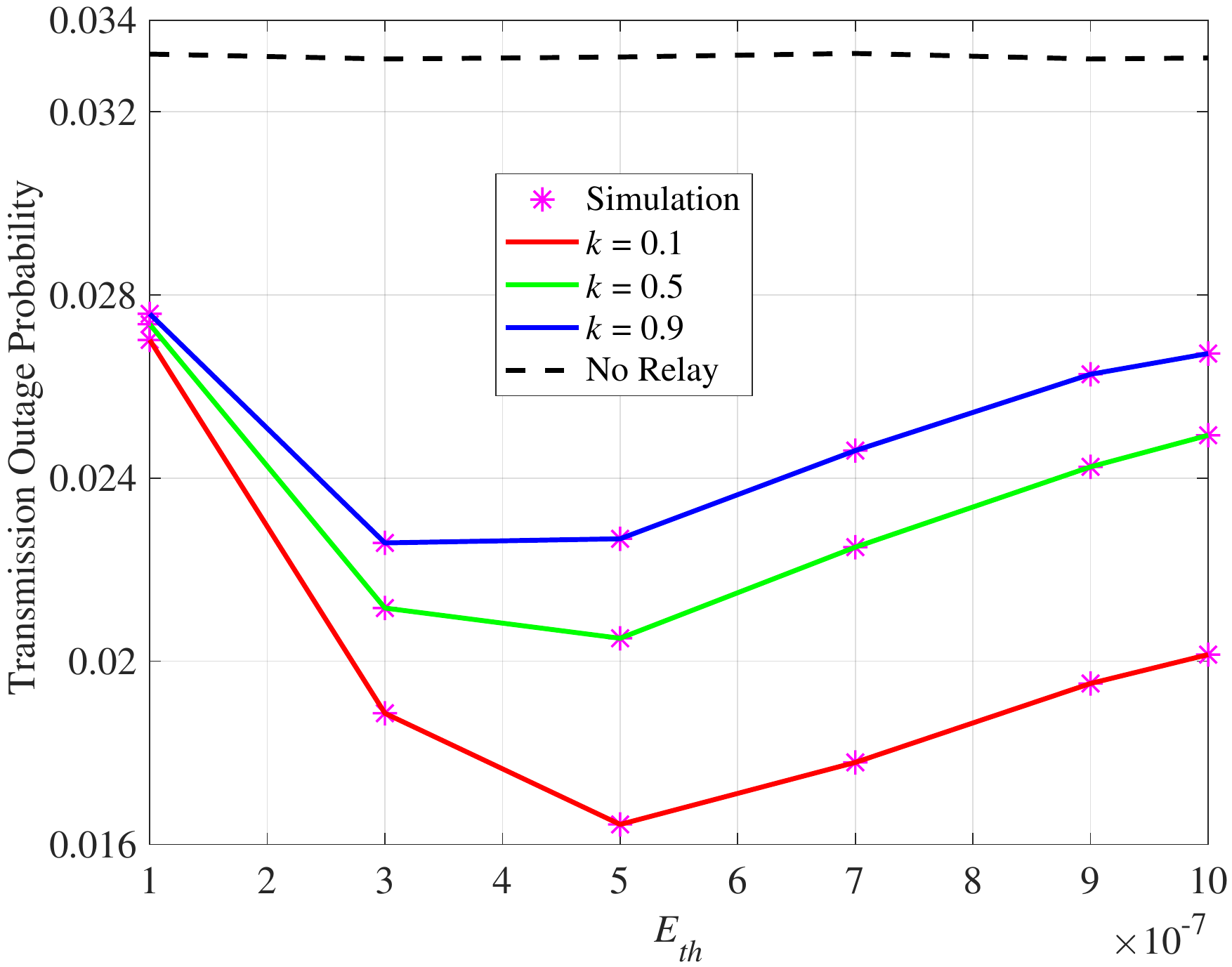}
    \caption{Transmission outage probability versus $E_{th}$ with various $k$ values.}
\end{minipage}
\begin{minipage}[t]{0.48\linewidth}
    \includegraphics[width = 1\linewidth]{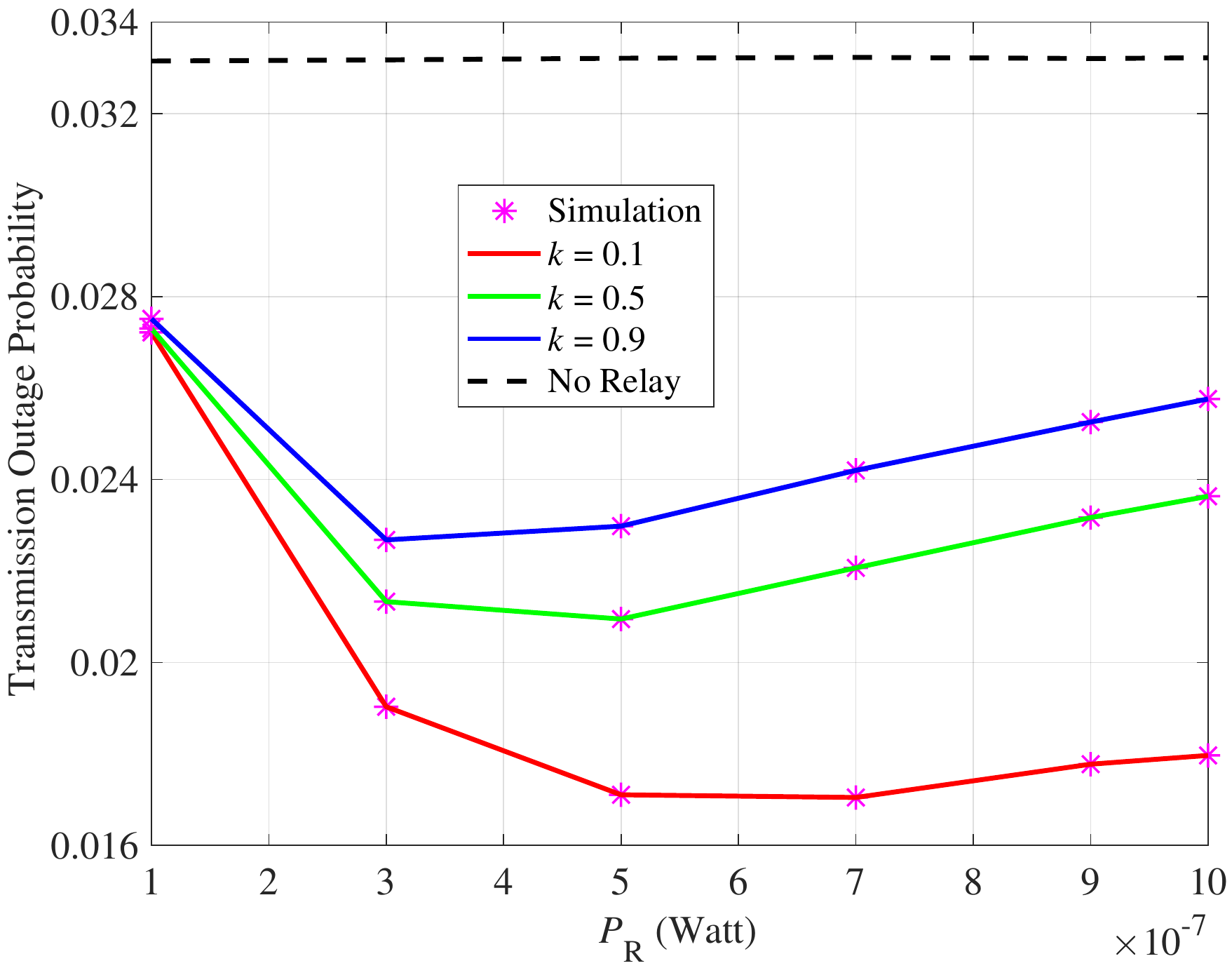}
    \caption{Transmission outage probability versus $P_{\text{R}}$ with various $k$ values.}
\end{minipage}
\end{tabular}
\end{figure}
\subsection{The Impact of The Energy Threshold}
In this subsection, we discuss the impact of $E_{th}$ on the TOP performance. Figure 11 illustrates the TOP curves versus $E_{th}$ with various values of $k$. It is easy to conclude that an optimal choice of $E_{th}$ which can minimise the TOP performance does exist for a specific $k$. The reason is similar to that discussed in Subsection $G$, which can be explained by the activation condition of the FD SWIPT mode, i.e., $\left\{\gamma_{\textnormal{SD}}<\gamma_{th}\right\}\cap \left\{E_{i}\geq E_{th}\right\}$. This observation can help the designer to determine a feasible setup of $E_{th}$ in practical applications.
\subsection{The Impact of R's Transmit Power}
In this part, we discuss the impact of $P_{\text{R}}$ on the TOP performance. Figure 12 depicts the TOP curves versus $P_{\text{R}}$ with various values of $k$. The overall appearance of this figure is similar to that of Figure 11, however the subtle differences can be found by comparing these two figures, illustrating the different influence strengths of $E_{th}$ and $P_{\text{R}}$ on the TOP performance. The existence of the optimal $P_{\text{R}}$ is due to the following two trade-offs: 1) a larger $P_{\text{R}}$ will consume more stored energy at the PEC but also lead the PEC to absorb more energy from the SI channel. 2) the min function introduced by the DF relaying strategy limits that $\gamma_{\text{D}}$ is not always increasing with the increasing of $P_{\text{R}}$. This two kinds of dilemma cause that simply enlarging $P_{\text{R}}$ does not lead to a better TOP performance, and also make the optimal value of $P_{\text{R}}$ existing. This finding is beneficial for designer to choose a feasible value of $P_{\text{R}}$ in implement of the proposed HOR system.

\section{Conclusion}
In this paper, we initiated a novel wireless relaying transmission scheme termed HOR, via both listing the necessary hardware devices and designing the essential transmission protocol. To realise the SWIPT and true FD functionalities of the proposed HOR system, a practical finite-capacity hybrid energy storage model  is applied, which is composed of two independent energy containers. The relay can work opportunistically in either PEH or FD SWIPT mode according to the proposed HOR scheme, not only providing a better way to manipulate available wireless energy but also improving the overall wireless transmission performance within the end-to-end wireless communication scenario. To track the dynamic charge-discharge behaviour of the PEC, a discrete-state MC method is adopted, based on which the long-term stationary distribution of energy states is quantified. Furthermore, covert communication and transmission performances of the proposed HOR system were analysed via deriving closed-form expressions of minimum detection error probability and transmission outage probability. Numerical results validated the correctness of the aforementioned analyses and the impacts of key system parameters were investigated. The proposed HOR scheme can enhance wireless energy manipulating efficiency, wireless transmission performance and privacy level of the end-to-end wireless transmission system, which has been proved throughout this paper.

\appendices
\section{Proof of Lemma~\ref{Lemma_OptimalityofRadiometer}}
\label{AppendixofLemmaOptimalityofRadiometer}

Under the assumption that D has complete knowledge of his noise power,
similar to the proof of Lemma 3 in \cite{sobers2017covert}, by applying
Fisher-Neyman factorization theorem and Likelihood Ratio ordering,
we can prove that radiometer is indeed the optimal choice for D to
perform the detection test. The proof details can be stated as follows.

Because each symbol of the received message vector $\boldsymbol{y}_{\text{D}}$
in a specific transmission slot follows i.i.d. complex Gaussian distribution,
$\boldsymbol{y}_{\text{D}}\left[\omega\right]$ is ruled by the distribution
shown as 
\begin{equation}
\begin{cases}
\mathcal{CN}\left(0,P_{\text{S}}\vert h_{\text{SD}}\vert^{2}+P_{\text{R}}\vert\hat{h}_{\text{RD}}\vert^{2}+P_{\text{R}}\vert\tilde{h}_{\text{RD}}\vert^{2}+\sigma_{\text{D}}^{2}\right), & \text{\ensuremath{\mathcal{H}_{0}}}\\
\mathcal{CN}\left(0,P_{\text{S}}\vert h_{\text{SD}}\vert^{2}+\left(P_{\text{R}}+P_{\Delta}\right)\vert\hat{h}_{\text{RD}}\vert^{2}+\left(P_{\text{R}}+P_{\Delta}\right)\vert\tilde{h}_{\text{RD}}\vert^{2}+\sigma_{\text{D}}^{2}\right), & \text{\ensuremath{\mathcal{H}_{1}}}
\end{cases}.
\end{equation}

Denote the observation conditioned on $\psi$ by
$\boldsymbol{y}_{\text{D}}\left(\psi\right)=\left[y_{\text{D}}\left[1\right]\left(\psi\right),y_{\text{D}}\left[2\right]\left(\psi\right),\dots,y_{\text{D}}\left[n\right]\left(\psi\right)\right]$
in which $y_{\text{D}}\left[\omega\right]\left(\psi\right)\sim\mathcal{CN}\left(0,\sigma_{\text{D}}^{2}+\text{\ensuremath{\psi}}\right)$.
Note that $\psi$ represents the sum variance of D's received signals
from S and R. To distinguish the null hypothesis $\mathcal{H}_{0}$
from the alternative hypothesis $\mathcal{H}_{1}$, we here introduce
a couple of non-negative and real-value RVs $\Psi_{0}$ and $\Psi_{1}$,
whose PDFs are integratedly given by
\begin{equation}
f_{\Psi_{q}}\left(\psi\right)=\begin{cases}
\frac{\exp\left(-\frac{\psi-\phi_{0}}{\beta P_{\text{R}}\Omega_{\text{RD}}}\right)}{\beta P_{\text{R}}\Omega_{\text{RD}}}, & x>\phi_{0},q=0\\
\frac{\exp\left(-\frac{\psi-\phi_{1}}{\beta\left(P_{\text{R}}+P_{\Delta}\right)\Omega_{\text{RD}}}\right)}{\beta\left(P_{\text{R}}+P_{\Delta}\right)\Omega_{\text{RD}}}, & x>\phi_{1},q=1\\
0, & \text{otherwise}
\end{cases},\label{eq:PDF_Psi_q}
\end{equation}
where $\phi_{0}=P_{\text{S}}\Omega_{\text{SD}}+\left(1-\beta\right)P_{\text{R}}\Omega_{\text{RD}}$
and $\phi_{1}=P_{\text{S}}\Omega_{\text{SD}}+\left(1-\beta\right)\left(P_{\text{R}}+P_{\Delta}\right)\Omega_{\text{RD}}$.

Furthermore, the PDF of vector $\boldsymbol{y}_{\text{D}}$ given
$\psi$ can be calculated as
\begin{equation}
f_{\boldsymbol{y}_{\text{D}}\left(\psi\right)}\left(\boldsymbol{y}\right)=\prod_{\omega=1}^{n}\frac{\exp\left(-\frac{\vert\boldsymbol{y}_{\text{D}}\left[\omega\right]\left(\psi\right)\vert^{2}}{\sigma_{\text{D}}^{2}+\text{\ensuremath{\psi}}}\right)}{\pi\left(\sigma_{\text{D}}^{2}+\text{\ensuremath{\psi}}\right)}
=\left(\frac{1}{\pi\left(\sigma_{\text{D}}^{2}+\text{\ensuremath{\psi}}\right)}\right)^{n}\exp\left(-\frac{\sum_{\omega=1}^{n}\vert\boldsymbol{y}_{\text{D}}\left[\omega\right]\left(\psi\right)\vert^{2}}{\sigma_{\text{D}}^{2}+\text{\ensuremath{\psi}}}\right).
\end{equation}
Here, invoking the Fisher-Neyman Fctorization Theorem, the total received
power in a transmission slot $\sum_{\omega=1}^{n}\vert\boldsymbol{y}_{\text{D}}\left[\omega\right]\left(\psi\right)\vert^{2}$
is a sufficient statistic for D's hypothesis test. It is worth noting
that $\sum_{\omega=1}^{n}\vert\boldsymbol{y}_{\text{D}}\left[\omega\right]\left(\psi\right)\vert^{2}=\left(\sigma_{\text{D}}^{2}+\text{\ensuremath{\psi}}\right)\mathcal{X}_{2n}^{2}$
where $\mathcal{X}_{2n}^{2}$ denotes chi-squared RV with $2n$ degrees
of freedom. Because D performs testing between two simple hypotheses
and he knows the statistical knowledge of his received signals when
either hypothesis stands, with help of the Neyman-Pearson Lemma, the
best testing rule for D to decide which hypothesis stands is the likelihood
ratio test (LRT), given by
\begin{equation}
\Lambda\left(\boldsymbol{y}_{\text{D}}\right)=\frac{f_{\boldsymbol{y}_{\text{D}}\mid\mathcal{H}_{1}}\left(\boldsymbol{y}\right)}{f_{\boldsymbol{y}_{\text{D}}\mid\mathcal{H}_{0}}\left(\boldsymbol{y}\right)}\stackrel[\mathcal{D}_{0}]{\mathcal{D}_{1}}{\gtrless}\Gamma,
\end{equation}
where $\Gamma=\Pr\left(\mathcal{H}_{1}\right)/\Pr\left(\mathcal{H}_{0}\right)=1$
due to the application of equal $a$ $priori$ assumption. D does
not have instantaneous knowledge of either $\Psi_{0}$ or $\Psi_{1}$,
so he modifies his LRT as
\begin{equation}
\Lambda\left(\boldsymbol{y}_{\text{D}}\right)=\frac{\mathbb{E}_{\Psi_{1}}\left[f_{\boldsymbol{y}_{\text{D}}\left(\psi\right)}\left(\boldsymbol{y}\right)\right]}{\mathbb{E}_{\Psi_{0}}\left[f_{\boldsymbol{y}_{\text{D}}\left(\psi\right)}\left(\boldsymbol{y}\right)\right]}\stackrel[\mathcal{D}_{0}]{\mathcal{D}_{1}}{\gtrless}\Gamma.\label{eq:LRT}
\end{equation}
We introduce here that RV $X$ is smaller than RV $Y$ in the likelihood
ratio order, i.e., $X\leq_{\text{lr}}Y$, when $f_{Y}\left(x\right)/f_{X}\left(x\right)$
is an non-decreasing function over the union of their supports.

Invoking (\ref{eq:PDF_Psi_q}), we have
\begin{equation}
\frac{f_{\Psi_{1}}\left(\psi\right)}{f_{\Psi_{0}}\left(\psi\right)}=
\frac{P_{\text{R}}}{P_{\text{R}}+P_{\Delta}}\exp\left(\frac{P_{\Delta}\psi-\left(P_{\text{R}}+P_{\Delta}\right)\phi_{0}+P_{\text{R}}\phi_{1}}{\beta P_{\text{R}}\left(P_{\text{R}}+P_{\Delta}\right)\Omega_{\text{RD}}}\right).\label{eq:PDFRatio_Psi1Psi0}
\end{equation}
It is straightforward to find that (\ref{eq:PDFRatio_Psi1Psi0}) is
non-decreasing over the union of supports of $\Psi_{0}$ and $\Psi_{1}$,
hence $\Psi_{0}\leq_{\text{lr}}\Psi_{1}$. From the statistical nature
of chi-squared RVs, for any $\psi_{1}\leq\psi_{2}$, we have $\boldsymbol{y}_{\text{D}}\left(\psi_{1}\right)\leq_{\text{lr}}\boldsymbol{y}_{\text{D}}\left(\psi_{2}\right)$.
Then, according to Theorem 1, Chapter 11 in \cite{shakedstochastic},
the monotonicity of $\Lambda\left(\boldsymbol{y}_{\text{D}}\right)$
is ruled by Stochastic Ordering and $\Lambda\left(\boldsymbol{y}_{\text{D}}\right)$
is non-decreasing w.r.t. $\sum_{\omega=1}^{n}\vert\boldsymbol{y}_{\text{D}}\left[\omega\right]\left(\psi\right)\vert^{2}$.
Hence, the LRT (\ref{eq:LRT}) is equivalent to a received power threshold
test. Since any one-to-one transformation of a sufficient statistic
remains the sufficiency, the term $\sum_{\omega=1}^{n}\vert\boldsymbol{y}_{\text{D}}\left[\omega\right]\vert^{2}/n$
is also a sufficient statistic. From the strong law of large numbers,
we have $\mathcal{X}_{2n}^{2}/n\rightarrow1$ when infinite blocklength
($n\rightarrow\infty$) assumption is considered. Invoking the Lebesgue's
Dominated Convergence Theorem, it is allowed to replace $\mathcal{X}_{2n}^{2}/n$
by 1, when $n\rightarrow\infty$. Thus, we get
\begin{align}
T &=\lim_{n\rightarrow\infty}\frac{1}{n}\sum_{\omega=1}^{n}\vert\boldsymbol{y}_{\text{D}}\left[\omega\right]\vert^{2}\nonumber \\
&=
\begin{cases}
P_{\text{S}}\vert h_{\text{SD}}\vert^{2}+P_{\text{R}}\vert\hat{h}_{\text{RD}}\vert^{2}+P_{\text{R}}\vert\tilde{h}_{\text{RD}}\vert^{2}+\sigma_{\text{D}}^{2}, & \mathcal{H}_{0}\\
P_{\text{S}}\vert h_{\text{SD}}\vert^{2}+\left(P_{\text{R}}+P_{\Delta}\right)\vert\hat{h}_{\text{RD}}\vert^{2}+
\left(P_{\text{R}}+P_{\Delta}\right)\vert\tilde{h}_{\text{RD}}\vert^{2}+\sigma_{\text{D}}^{2}, & \mathcal{H}_{1}
\end{cases}.\label{eq:T}
\end{align}
Then, the optimal decision rule at D can be expressed as
\[
T\stackrel[\mathcal{D}_{0}]{\mathcal{D}_{1}}{\gtrless}\tau,
\]
where $\tau$ denotes the threshold which will be optimized to minimize
$\mathbb{P}_{\text{E}}$.

After all, a radiometer which is able to detect the total power of
received messages at D is proved to be optimal. Besides, radiometers
are also beneficial for D due to its low complexity and ease of implementation.
So, it is sufficient and optimal for D to apply a radiometer to perform
hypothesis test regarding covert communication detection.

\section{Proof of Theorem~\ref{TheoremofClosedFormDEP}}
\label{AppendixofTheoremofClosedFormDEP}
Based on (\ref{eq:T}), we can calculate the false alarm and missed
detection probabilities, given respectively by
\begin{equation}
\mathbb{P}_{\text{FA}} =\Pr\left(T>\tau|\mathcal{H}_{0}\right)=\Pr\left(P_{\text{R}}\vert\tilde{h}_{\text{RD}}\vert^{2}+j_{0}>\tau\right)
=\begin{cases}
\Pr\left(\vert\tilde{h}_{\text{RD}}\vert^{2}>\frac{\tau-j_{0}}{P_{\text{R}}}\right), & \tau\geq j_{0}\\
1, & \text{otherwise}
\end{cases},\label{eq:Pr_FA_Proof}
\end{equation}

\begin{align}
\mathbb{P}_{\text{MD}}&=\Pr\left(T<\tau|\mathcal{H}_{1}\right)\nonumber \\
&=\Pr\left(\left(P_{\text{R}}+P_{\Delta}\right)\vert\tilde{h}_{\text{RD}}\vert^{2}+j_{1}<\tau\right)
=\begin{cases}
\Pr\left(\vert\tilde{h}_{\text{RD}}\vert^{2}<\frac{\tau-j_{1}}{P_{\text{R}}+P_{\Delta}}\right), & \tau\geq j_{1}\\
0, & \text{otherwise}
\end{cases}.\label{eq:Pr_MD_Proof}
\end{align}
Because the uncertain part of channel R$\rightarrow$D follows the
distribution $\tilde{h}_{\text{RD}}\sim\mathcal{CN}\left(0,\beta\Omega_{\text{RD}}\right)$,
it is straightforward to know that $\vert\tilde{h}_{\text{RD}}\vert^{2}$
obeys the Exponential distribution. Thus, the CDF of $\vert\tilde{h}_{\text{RD}}\vert^{2}$
can be gained as $F_{\vert\tilde{h}_{\text{RD}}\vert^{2}}\left(x\right)=1-\exp\left(-x/\left(\beta\Omega_{\text{RD}}\right)\right)$.
Then, after some simple algebra calculation, we gain closed-form expressions
of false alarm and missed detection probabilities, expressed respectively
as (\ref{eq:Pr_FA}) and (\ref{eq:Pr_MD}). Invoking (\ref{eq:Pr_E_Definition}),
(\ref{eq:Pr_FA}) and (\ref{eq:Pr_MD}), closed-form expression of
detection error probability can be gained after simple derivation
as (\ref{eq:Pr_E}).

\section{Proof of Theorem~\ref{OptimalDetectionThreshold}}
\label{AppendixofOptimalDetectionThreshold}
To determine the optimal detection threshold of D's radiometer, it
is supposed to solve the following optimization problem, shown as
\begin{equation}
\tau^{*}=\underset{\tau}{\arg\min}\thinspace\thinspace\mathbb{P}_{\text{E}}.
\end{equation}

In the case of $\tau<j_{0}$, the detection error probability at D
remains 1. This is the worst case for D and D will never choose any
value satisfying $\tau<j_{0}$. Thus, the optimization problem did
not stand in this case.

In the case of $j_{0}\le\tau<j_{1}$, it is easy to find that $\mathbb{P}_{\text{E}}$
monotonically decreases w.r.t. $\tau$. Besides, the piecewise function
$\mathbb{P}_{\text{E}}$ is a continuous function along side the whole
feasible domain of $\tau$. Thus, D will choose $j_{1}$ to minimize
$\mathbb{P}_{\text{E}}$, leading to $\mathbb{P}_{\text{E}}=\exp\left(\left(j_{0}-j_{1}\right)/\left(\beta P_{\text{R}}\Omega_{\text{RD}}\right)\right)$.

In the case of $\tau\geq j_{1}$, to determine the optimal value of
$\tau$, the first derivative of function $\mathbb{P}_{\text{E}}$
w.r.t. $\tau$ is calculated as
\begin{equation}
\frac{\partial\mathbb{P}_{\text{E}}}{\partial\tau}=\frac{k}{\beta P_{\text{R}}\left(P_{\text{R}}+P_{\Delta}\right)\Omega_{\text{RD}}},\label{eq:Partial_P_E}
\end{equation}
where $k=P_{\text{R}}\exp\left[\left(j_{1}-\tau\right)/\left(\beta\left(P_{\text{R}}+P_{\Delta}\right)\Omega_{\text{RD}}\right)\right]-\left(P_{\text{R}}+P_{\Delta}\right)\exp\left[\left(j_{0}-\tau\right)/\left(\beta P_{\text{R}}\Omega_{\text{RD}}\right)\right]$.
It is easy to find that whether (\ref{eq:Partial_P_E})
is positive or not depends only on the value of $k$. After simple
manipulations, we can modify $k$ as
\begin{equation}
k=\exp\left(\ln P_{\text{R}}+\frac{j_{1}-\tau}{\beta\left(P_{\text{R}}+P_{\Delta}\right)\Omega_{\text{RD}}}\right)-\exp\left(\ln\left(P_{\text{R}}+P_{\Delta}\right)+\frac{j_{0}-\tau}{\beta P_{\text{R}}\Omega_{\text{RD}}}\right).
\end{equation}
Besides, the Exponential function $\exp$ is monotonically increasing
w.r.t. the feasible independent variable region. Thus, we can determine
whether $k$ is positive or not by 
\begin{align}
k_{1} & =\ln\frac{P_{\text{R}}}{P_{\text{R}}+P_{\Delta}}+\frac{P_{\text{R}}\left(j_{1}-\tau\right)-\left(P_{\text{R}}+P_{\Delta}\right)\left(j_{0}-\tau\right)}{\beta P_{\text{R}}\left(P_{\text{R}}+P_{\Delta}\right)\Omega_{\text{RD}}}\nonumber \\
 & =\ln\frac{P_{\text{R}}}{P_{\text{R}}+P_{\Delta}}+\frac{P_{\Delta}\left(\tau-P_{\text{S}}\vert h_{\text{SD}}\vert^{2}-\sigma_{\text{D}}^{2}\right)}{\beta P_{\text{R}}\left(P_{\text{R}}+P_{\Delta}\right)\Omega_{\text{RD}}}.\label{eq:k_1}
\end{align}
Because $\tau\geq j_{1}$ stands in this considered case, the right
hand of (\ref{eq:k_1}) is absolutely positive. However, the left
hand of (\ref{eq:k_1}) is negative due to $P_{\text{R}}<P_{\text{R}}+P_{\Delta}$.
Most importantly, from (\ref{eq:k_1}), we can find that $k_{1}$
is a monotonically increasing function w.r.t. $\tau$. Let $k_{1}=0$,
we can get the solution as (\ref{eq:tqu_k1=00003D0}). From (\ref{eq:tqu_k1=00003D0}),
we can conclude that $k_{1}\geq0$ in the case of $\tau\geq\tau_{k_{1}=0}$
and $k_{1}<0$ otherwise. If $j_{1}\geq\tau_{k_{1}=0}$ holds, in
the case of $\tau\geq j_{1}$, we can determine that $k>0$ and furthermore
$\partial\mathbb{P}_{\text{E}}/\partial\tau>0$ which means $\mathbb{P}_{\text{E}}$
monotonically increases w.r.t. $\tau$ when $\tau\geq j_{1}$. Here,
it is the optimal choice for D to choose $j_{1}$ as the optimal threshold
which is able to minimize $\mathbb{P}_{\text{E}}$. If $j_{1}<\tau_{k_{1}=0}$,
we know that for $\tau\in\left(j_{1},\tau_{k_{1}=0}\right)$, $\partial\mathbb{P}_{\text{E}}/\partial\tau<0$
and for $\tau\in\left(\tau_{k_{1}=0},+\infty\right)$, $\partial\mathbb{P}_{\text{E}}/\partial\tau>0$.
Thus, the optimal detection threshold for D is $\tau_{k_{1}=0}$ in
this case.

\section{Proof of Corollary~\ref{CorIncreaingFunofBeta}}
\label{AppendixofCorIncreaingFunofBeta}
In the case of $j_{1}\geq\tau_{k_{1}=0}$, i.e., $\beta\geq-P_{\Delta}\vert\hat{h}_{\text{RD}}\vert^{2}/\left(P_{\text{R}}\Omega_{\text{RD}}\ln\frac{P_{\text{R}}}{P_{\text{R}}+P_{\Delta}}\right)$,
the first derivative of $\mathbb{P}_{\text{E}}^{*}$ w.r.t. $\beta$
can be calculated as
\begin{equation}
\frac{\partial\mathbb{P}_{\text{E}}^{*}}{\partial\beta}\mid_{j_{1}\geq\tau_{k_{1}=0}}=-\frac{j_{0}-j_{1}}{\beta^{2}P_{\text{R}}\Omega_{\text{RD}}}\exp\left(\frac{j_{0}-j_{1}}{\beta P_{\text{R}}\Omega_{\text{RD}}}\right),
\end{equation}
whose value is positive due to $j_{0}<j_{1}$. For $j_{1}<\tau_{k_{1}=0}$, i.e., $\beta<-P_{\Delta}\vert\hat{h}_{\text{RD}}\vert^{2}/\left(P_{\text{R}}\Omega_{\text{RD}}\ln\frac{P_{\text{R}}}{P_{\text{R}}+P_{\Delta}}\right)$,
the first derivative of $\mathbb{P}_{\text{E}}^{*}$ w.r.t. $\beta$
can be calculated as
\begin{align}
\frac{\partial\mathbb{P}_{\text{E}}^{*}}{\partial\beta}&\mid_{j_{1}<\tau_{k_{1}=0}}=
\frac{\vert\hat{h}_{\text{RD}}\vert^{2}}{\beta^{2}\Omega_{\text{RD}}}\nonumber \\
&\times\left[\exp\left(\frac{\vert\hat{h}_{\text{RD}}\vert^{2}}{\beta^{2}\Omega_{\text{RD}}}+\frac{P_{\text{R}}}{P_{\Delta}}\ln\frac{P_{\text{R}}}{P_{\text{R}}+P_{\Delta}}\right)-\exp\left(\frac{\vert\hat{h}_{\text{RD}}\vert^{2}}{\beta^{2}\Omega_{\text{RD}}}+\frac{P_{\text{R}}+P_{\Delta}}{P_{\Delta}}\ln\frac{P_{\text{R}}}{P_{\text{R}}+P_{\Delta}}\right)\right],
\end{align}
whose value is also positive due to the truth of $P_{\text{R}}>P_{\Delta}>0$.
Thus, we can conclude that $\mathbb{P}_{\text{E}}^{*}$ monotonically
increases as $\beta$ increases.

\section{Proof of Lemma~\ref{LemmaofClosedFormCDFofGammadH0}}
\label{AppendixofLemmaofClosedFormCDFofGammadH0}
The CDF of $\gamma_{\text{D}}|\mathcal{H}_{0}$ can be constructed
as 
\begin{equation}
F_{\gamma_{\text{D}}|\mathcal{H}_{0}}\left(x\right)=\Pr\left(\gamma_{\text{SD}}+Y_{\mathcal{H}_{0}}<x\bigcap\gamma_{\text{SD}}<\gamma_{th}\right).\label{eq:CDF_Constructed_GammaD_H0}
\end{equation}
Note that the limitation of variable $\gamma_{\text{SD}}$ should
be constrained as $\gamma_{\text{SD}}<\gamma_{th}$ due to the nature
of FD SWIPT mode. Invoking (\ref{eq:ClosedForm_CDF_Y_H_Phi}) and
after some simple mathematical computation, we can earn closed-form
expression of (\ref{eq:CDF_Constructed_GammaD_H0}) as (\ref{eq:ClosedForm_CDF_Gamma_D_H_0}).

\section{Proof of Lemma~\ref{LemmaofApproxCDFofGammadH1}}
\label{AppendixofLemmaofApproxCDFofGammadH1}
Closed-form CDF expression of $\gamma_{\text{D}}|\mathcal{H}_{1}$
should be calculated in the way similar to the derivation of (\ref{eq:ClosedForm_CDF_Gamma_D_H_0}).
However, we found that it is mathematically intractable. To tackle
this problem, we resort to Gauss-Kronrod Quadrature (GKQ) method to
approximately solve it, shown as
\begin{align}
F_{\gamma_{\text{D}}|\mathcal{H}_{1}}\left(x\right) & =\Pr\left[\gamma_{\text{SD}}+Y_{\mathcal{H}_{1}}<x\bigcap\gamma_{\text{SD}}<\gamma_{th}\right]\nonumber \\
 & =\int_{0}^{\gamma_{th}}\underset{\text{fun}}{\underbrace{\frac{\sigma_{\text{D}}^{2}}{P_{\text{S}}\Omega_{\text{SD}}}F_{Y_{\mathcal{H}_{1}}}\left(x-y\right)\exp\left(-\frac{\sigma_{\text{D}}^{2}y}{P_{\text{S}}\Omega_{\text{SD}}}\right)}dy}\nonumber \\
 & \thickapprox\sum_{i=1}^{n}\varrho_{i}\text{fun}\left(y_{i}\right),\label{eq:ClosedForm_CDF_Gamma_D_H_1}
\end{align}
where $\varrho_{i}$ and $y_{i}$ denote the weights and points which
are essential to evaluate the function $\text{fun\ensuremath{\left(y\right)}}$.
Note that the GKQ formula is an adaptive method for numerical integration,
which is a variant of Gaussian quadrature. In this paper, we use the
built-in function of Matlab named $\text{quadgk}(\cdot,\cdot,\cdot)$
to calculate (\ref{eq:ClosedForm_CDF_Gamma_D_H_1}), which implements
adaptive quadrature based on a Gauss-Kronrod pair ($15^{th}$ and
$7^{th}$ order formulas). Applying $\text{quadgk}(\cdot,\cdot,\cdot)$,
we can derive closed-form approximate CDF expression of $\gamma_{\text{D}}|\mathcal{H}_{1}$
as (\ref{eq:ClosedForm_Appromy_CDF_Gamma_D_H_1}).

\vspace{-1cm}
\section{Proof of Theorem~\ref{TheoremofTOPinFS}}
\label{AppendixofTheoremofTOPinFS}
In our considered HOR model, the TOP in the case of FD SWIPT should
be constructed as
\begin{align}
TOP_{\text{FS}} & =\Pr\left[\log_{2}\left(1+\gamma_{\text{D}}\right)<R_{th}\bigcap\mathcal{H}_{0}\bigcap\text{FS}\right]+\Pr\left[\log_{2}\left(1+\gamma_{\text{D}}\right)<R_{th}\bigcap\mathcal{H}_{1}\bigcap\text{FS}\right]\nonumber \\
 & \overset{a}{=}\Pr\left[\log_{2}\left(1+\gamma_{\text{D}}\right)<R_{th}\bigcap\mathcal{H}_{0}\bigcap\gamma_{\text{SD}}<\gamma_{th}\right]\sum_{i=\varphi}^{L}\xi_{i}+\nonumber\\
 &\Pr\left[\log_{2}\left(1+\gamma_{\text{D}}\right)<R_{th}\bigcap\mathcal{H}_{1}\bigcap\gamma_{\text{SD}}<\gamma_{th}\right]\sum_{i=\varphi}^{L}\xi_{i}\nonumber \\
 & =\frac{1}{2}\sum_{i=\varphi}^{L}\xi_{i}\nonumber\\
 &\times\left\{ \underset{f_{1}}{\underbrace{\Pr\left[\gamma_{\text{D}}|\mathcal{H}_{0}<2^{R_{th}}-1\bigcap\gamma_{\text{SD}}<\gamma_{th}\right]}}+\underset{f_{2}}{\underbrace{\Pr\left[\gamma_{\text{D}}|\mathcal{H}_{1}<2^{R_{th}}-1\bigcap\gamma_{\text{SD}}<\gamma_{th}\right]}}\right\},\label{eq:TOP_FS}
\end{align}
where the factor $1/2$ is due to the assumption of equal $a$ $priori$,
$R_{th}$ is the target rate under which the transmission outage occurs.
Note that step (a) in (\ref{eq:TOP_FS}) holds, because of the fact
that the energy requirement is independent with other factors. With
the help of  \textbf{\textit{Lemma}} \textit{2} and  \textbf{\textit{Lemma}} \textit{3}, we are able
to derive closed-form expressions of $f_{1}$ and $f_{2}$, which
can be achieved by simply replacing variable $x$ in (\ref{eq:ClosedForm_CDF_Gamma_D_H_0})
and (\ref{eq:ClosedForm_Appromy_CDF_Gamma_D_H_1}) with factor $2^{R_{th}}-1$.
Substituting $f_{1}$ and $f_{2}$ into (\ref{eq:TOP_FS}), we can
derive closed-form expression of the TOP in the FD SWIPT mode as (\ref{eq:ClosedForm_TOP_FS})
and this completes the proof.

\section{Proof of Theorem~\ref{TheoremofTOPinPEH}}
\label{AppendixofTheoremofTOPinPEH}
Similar to the derivation of (\ref{eq:ClosedForm_TOP_FS}), in the PEH mode, the TOP should be constructed as
\begin{align}
TOP_{\text{PEH}} & =\Pr\left[\log_{2}\left(1+\gamma_{\text{D}}\right)<R_{th}\bigcap\text{PEH}\right]\nonumber \\
 & =\underset{f_{3}}{\underbrace{\Pr\left(\gamma_{\text{SD}}<2^{R_{th}}-1\cap\gamma_{\text{SD}}<\gamma_{th}\right)}}\sum_{i=0}^{\varphi-1}\xi_{i}+\underset{f_{4}}{\underbrace{\Pr\left(\gamma_{\text{SD}}<2^{R_{th}}-1\cap\gamma_{\text{SD}}\geq\gamma_{th}\right)}}.\label{eq:TOP_PEH}
\end{align}

In the case of $\gamma_{\text{SD}}<\gamma_{th}\cap E_{i}<E_{th}$,
we have $\Pr\left(\gamma_{\text{SD}}<2^{R_{th}}-1\right)=1$. It is
worth noting that hereby $\Pr\left(\gamma_{\text{SD}}<2^{R_{th}}-1\right)$
and $\Pr\left(\gamma_{\text{SD}}<\gamma_{th}\right)$ are independent
with each other, because D cases signal-processing and forces $\Pr\left(\gamma_{\text{SD}}<2^{R_{th}}-1\right)=1$,
leading $f_{3}=\Pr\left(\gamma_{\text{SD}}<\gamma_{th}\right)=q_{\text{SD}}$.
In the case of $\gamma_{\text{SD}}\geq\gamma_{th}$, the main wireless
channel is good enough, closed-form expression of CDF of $\gamma_{\text{SD}}|\gamma_{\text{SD}}\geq\gamma_{th}$
can be derive as
\begin{equation}
F_{\gamma_{\text{SD}}|\gamma_{\text{SD}}\geq\gamma_{th}}\left(x\right)=
\begin{cases}
\exp\left(-\frac{\sigma_{\text{D}}^{2}\gamma_{th}}{P_{\text{S}}\Omega_{\text{SD}}}\right)-\exp\left(-\frac{\sigma_{\text{D}}^{2}x}{P_{\text{S}}\Omega_{\text{SD}}}\right), & x>\gamma_{th}\\
0, & x\le\gamma_{th}
\end{cases}.\label{eq:ClosedForm_CDF_Gamma_SD_geq_Gamma_th}
\end{equation}
Hence, closed-form expression of $f_{4}$ can be given by $f_{4}=F_{\gamma_{\text{SD}}|\gamma_{\text{SD}}\geq\gamma_{th}}\left(2^{R_{th}}-1\right)$.
Substituting $f_{3}$ and $f_{4}$ into (\ref{eq:TOP_PEH}), we can
derive closed-form expression of the TOP in the PEH mode as (\ref{eq:ClosedForm_TOP_PEH})
and this completes the proof.

\bibliographystyle{IEEEtran}
\bibliography{reference-icc}

\begin{thebibliography}{10}
\providecommand{\url}[1]{#1}
\csname url@samestyle\endcsname
\providecommand{\newblock}{\relax}
\providecommand{\bibinfo}[2]{#2}
\providecommand{\BIBentrySTDinterwordspacing}{\spaceskip=0pt\relax}
\providecommand{\BIBentryALTinterwordstretchfactor}{4}
\providecommand{\BIBentryALTinterwordspacing}{\spaceskip=\fontdimen2\font plus
\BIBentryALTinterwordstretchfactor\fontdimen3\font minus
  \fontdimen4\font\relax}
\providecommand{\BIBforeignlanguage}[2]{{%
\expandafter\ifx\csname l@#1\endcsname\relax
\typeout{** WARNING: IEEEtran.bst: No hyphenation pattern has been}%
\typeout{** loaded for the language `#1'. Using the pattern for}%
\typeout{** the default language instead.}%
\else
\language=\csname l@#1\endcsname
\fi
#2}}
\providecommand{\BIBdecl}{\relax}
\BIBdecl

\bibitem{bi2016accumulate}
Y.~Bi and H.~Chen, ``Accumulate and jam: Towards secure communication via a
  wireless-powered full-duplex jammer,'' \emph{IEEE J. Sel. Signal Process.},
  vol.~10, no.~8, pp. 1538--1550, 2016.

\bibitem{chu2018resource}
Z.~Chu, F.~Zhou, P.~Xiao, Z.~Zhu, D.~Mi, N.~Al-Dhahir, and R.~Tafazolli,
  ``Resource allocation for secure wireless powered integrated multicast and
  unicast services with full duplex self-energy recycling,'' \emph{IEEE Trans.
  Wireless Commun.}, vol.~18, no.~1, pp. 620--636, 2018.

\bibitem{Varshney2008Transporting}
L.~R. Varshney, ``Transporting information and energy simultaneously,'' in
  \emph{Proc. IEEE Int. Symp. Inf. Theory (ISIT)}, Toronto, Canada, Jul. 2008.

\bibitem{Zhang2013MIMO}
R.~Zhang and C.~K. Ho, ``{MIMO} broadcasting for simultaneous wireless
  information and power transfer,'' \emph{IEEE Trans. Wireless Commun.},
  vol.~12, no.~5, pp. 1989--2001, May 2013.

\bibitem{krikidis2014simultaneous}
I.~Krikidis, ``Simultaneous information and energy transfer in large-scale
  networks with/without relaying,'' \emph{IEEE Trans. Commun.}, vol.~62, no.~3,
  pp. 900--912, Mar. 2014.

\bibitem{yan2017dynamic}
J.~Yan and Y.~Liu, ``A dynamic {SWIPT} approach for cooperative cognitive radio
  networks,'' \emph{IEEE Trans. Veh. Technol.}, vol.~66, no.~12, pp.
  11\,122--11\,136, Dec. 2017.

\bibitem{zhang2018incomplete}
H.~Zhang, J.~Du, J.~Cheng, K.~Long, and V.~C. Leung, ``Incomplete {CSI} based
  resource optimization in {SWIPT} enabled heterogeneous networks: A
  non-cooperative game theoretic approach,'' \emph{IEEE Trans. Wireless
  Commun.}, vol.~17, no.~3, pp. 1882--1892, Mar. 2018.

\bibitem{rostampoor2017energy}
J.~Rostampoor, S.~M. Razavizadeh, and I.~Lee, ``Energy efficient precoding
  design for {SWIPT} in {MIMO} two-way relay networks,'' \emph{IEEE Trans. Veh.
  Technol.}, vol.~66, no.~9, pp. 7888--7896, Sep. 2017.

\bibitem{chen2017spectral}
Z.~Chen, T.~Q. Quek, and Y.-C. Liang, ``Spectral efficiency and relay energy
  efficiency of full-duplex relay channel,'' \emph{IEEE Trans. Wireless
  Commun.}, vol.~16, no.~5, pp. 3162--3175, May 2017.

\bibitem{xing2017multipair}
P.~Xing, J.~Liu, C.~Zhai, X.~Wang, and X.~Zhang, ``Multipair two-way
  full-duplex relaying with massive array and power allocation,'' \emph{IEEE
  Trans. Veh. Technol.}, vol.~66, no.~10, pp. 8926--8939, Oct. 2017.

\bibitem{li2018antenna}
Y.~Li, R.~Zhao, L.~Fan, and A.~Liu, ``Antenna mode switching for full-duplex
  destination-based jamming secure transmission,'' \emph{IEEE Access}, vol.~6,
  pp. 9442--9453, Mar. 2018.

\bibitem{ahmed2015all}
E.~Ahmed and A.~M. Eltawil, ``All-digital self-interference cancellation
  technique for full-duplex systems,'' \emph{IEEE Trans. Wireless Commun.},
  vol.~14, no.~7, pp. 3519--3532, Jul. 2015.

\bibitem{li2017outage}
C.~Li, Z.~Chen, Y.~Wang, Y.~Yao, and B.~Xia, ``Outage analysis of the
  full-duplex decode-and-forward two-way relay system,'' \emph{IEEE Trans. Veh.
  Technol.}, vol.~66, no.~5, pp. 4073--4086, May 2017.

\bibitem{Sohaib2017Full}
S.~Sohaib and M.~Uppal, ``Full duplex compress-and-forward relaying under
  residual self-interference,'' \emph{IEEE Trans. Veh. Technol.}, vol.~67,
  no.~3, pp. 2776--2780, Mar. 2017.

\bibitem{Razlighi2017Buffer}
M.~M. Razlighi and N.~Zlatanov, ``Buffer-aided relaying for the two-hop
  full-duplex relay channel with self-interference,'' \emph{IEEE Trans.
  Wireless Commun.}, vol.~17, no.~1, pp. 477--491, Jan. 2018.

\bibitem{Wang2015Outage}
Q.~Wang, Y.~Dong, X.~Xu, and X.~Tao, ``Outage probability of full-duplex {AF}
  relaying with processing delay and residual self-interference,'' \emph{IEEE
  Commun. Lett.}, vol.~19, no.~5, pp. 783--786, May 2015.

\bibitem{li2017Secrecy}
Y.~Li, R.~Zhao, X.~Tan, and Z.~Nie, ``Secrecy performance analysis of
  artificial noise aided precoding in full-duplex relay systems,'' in
  \emph{Proc. IEEE GLOBECOM}, Dec. 2017, pp. 1--6.

\bibitem{bash2013limits}
B.~A. Bash, D.~Goeckel, and D.~Towsley, ``Limits of reliable communication with
  low probability of detection on awgn channels,'' \emph{IEEE J. Sel. Areas
  Commun}, vol.~31, no.~9, pp. 1921--1930, 2013.

\bibitem{shahzad2018achieving}
K.~Shahzad, X.~Zhou, S.~Yan, J.~Hu, F.~Shu, and J.~Li, ``Achieving covert
  wireless communications using a full-duplex receiver,'' \emph{IEEE Trans.
  Wireless Commun.}, vol.~17, no.~12, pp. 8517--8530, 2018.

\bibitem{zheng2019multi}
T.-X. Zheng, H.-M. Wang, D.~W.~K. Ng, and J.~Yuan, ``Multi-antenna covert
  communications in random wireless networks,'' \emph{IEEE Trans. Wireless
  Commun.}, vol.~18, no.~3, pp. 1974--1987, 2019.

\bibitem{zhou2019joint}
X.~Zhou, S.~Yan, J.~Hu, J.~Sun, J.~Li, and F.~Shu, ``Joint optimization of a
  uav's trajectory and transmit power for covert communications,'' \emph{IEEE
  Trans. Signal Process.}, vol.~67, no.~16, pp. 4276--4290, 2019.

\bibitem{Wang2017Relay}
D.~Wang, R.~Zhang, X.~Cheng, L.~Yang, and C.~Chen, ``Relay selection in
  full-duplex energy-harvesting two-way relay networks,'' \emph{IEEE Trans.
  Green Commun. Netw.}, vol.~1, no.~2, pp. 182--191, Jun. 2017.

\bibitem{Wen2016Joint}
Z.~Wen, X.~Liu, N.~C. Beaulieu, and R.~Wang, ``Joint source and relay
  beamforming design for full-duplex {MIMO} {AF} relay {SWIPT} systems,''
  \emph{IEEE Commun. Lett.}, vol.~20, no.~2, pp. 320--323, 2016.

\bibitem{zeng2015full}
Y.~Zeng and R.~Zhang, ``Full-duplex wireless-powered relay with self-energy
  recycling,'' \emph{IEEE Wireless Commun. Lett.}, vol.~4, no.~2, pp. 201--204,
  2015.

\bibitem{liu2016power}
H.~Liu, K.~J. Kim, K.~S. Kwak, and H.~V. Poor, ``Power splitting-based swipt
  with decode-and-forward full-duplex relaying,'' \emph{IEEE Trans. Wireless
  Commun.}, vol.~15, no.~11, pp. 7561--7577, 2016.

\bibitem{hu2019covert}
J.~Hu, S.~Yan, F.~Shu, and J.~Wang, ``Covert transmission with a self-sustained
  relay,'' \emph{IEEE Trans. Wireless Commun.}, vol.~18, no.~8, pp. 4089--4102,
  2019.

\bibitem{wang2018covert}
J.~Wang, W.~Tang, Q.~Zhu, X.~Li, H.~Rao, and S.~Li, ``Covert communication with
  the help of relay and channel uncertainty,'' \emph{IEEE Wireless Commun.
  Lett.}, vol.~8, no.~1, pp. 317--320, 2018.

\bibitem{shahzad2018relaying}
K.~Shahzad, ``Relaying via cooperative jamming in covert wireless
  communications,'' in \emph{2018 12th International Conference on Signal
  Processing and Communication Systems (ICSPCS)}.\hskip 1em plus 0.5em minus
  0.4em\relax IEEE, 2018, pp. 1--6.

\bibitem{Krikidis2012Buffer}
I.~Krikidis, T.~Charalambous, and J.~S. Thompson, ``Buffer-aided relay
  selection for cooperative diversity systems without delay constraints,''
  \emph{IEEE Trans. Wireless Commun.}, vol.~11, no.~5, pp. 1957--1967, 2012.

\bibitem{zhao2016secrecy}
R.~Zhao, Y.~Yuan, L.~Fan, and Y.-C. He, ``Secrecy performance analysis of
  cognitive decode-and-forward relay networks in nakagami-$ m $ fading
  channels,'' \emph{IEEE Trans. Commun.}, vol.~65, no.~2, pp. 549--563, 2016.

\bibitem{sobers2017covert}
T.~V. Sobers, B.~A. Bash, S.~Guha, D.~Towsley, and D.~Goeckel, ``Covert
  communication in the presence of an uninformed jammer,'' \emph{IEEE Trans.
  Wireless Commun.}, vol.~16, no.~9, pp. 6193--6206, 2017.

\bibitem{shakedstochastic}
M.~Shaked and J.~G. Shanthikumar, ``Stochastic orders and their applications.
  1994,'' \emph{Acad-emic Press, New York}.

\end{thebibliography}

\end{document}